

\def\unlockat{\catcode`\@=11}

\def\lockat{\catcode`\@=12}

\unlockat

\global\newcount\secno \global\secno=0
\global\newcount\prono \global\prono=0
\def\newsec#1{\vfill\eject\global\advance\secno by1\message{(\the\secno. #1)}
\global\subsecno=0\global\subsubsecno=0
\global\deno=0\global\teno=0
\eqnres@t\noindent
{\bf\the\secno. #1}
\writetoca{{\bf\secsym} {\rm #1}}\par\nobreak\medskip\nobreak}
\global\newcount\subsecno \global\subsecno=0
\def\subsec#1{\global\advance\subsecno
by1\message{(\secsym\the\subsecno. #1)}
\ifnum\lastpenalty>9000\else\bigbreak\fi
\global\subsubsecno=0
\global\deno=0
\global\teno=0
\noindent{\bf\secsym\the\subsecno. #1}
\writetoca{\bf \string\quad {\secsym\the\subsecno.} {\it #1}}
\par\nobreak\medskip\nobreak}
\global\newcount\subsubsecno \global\subsubsecno=0
\def\subsubsec#1{\global\advance\subsubsecno by1
\message{(\secsym\the\subsecno.\the\subsubsecno. #1)}
\ifnum\lastpenalty>9000\else\bigbreak\fi
\noindent\quad{\bf \secsym\the\subsecno.\the\subsubsecno.}{\ \bf #1}
\writetoca{\string\qquad\bf { \secsym\the\subsecno.\the\subsubsecno.}{\sl  #1}}
\par\nobreak\medskip\nobreak}

\global\newcount\deno \global\deno=0
\def\de#1{\global\advance\deno by1
\message{(\bf Definition\quad\secsym\the\subsecno.\the\deno #1)}
\ifnum\lastpenalty>9000\else\bigbreak\fi
\noindent{\bf Definition\quad\secsym\the\subsecno.\the\deno}{#1}
\writetoca{\string\qquad{\secsym\the\subsecno.\the\deno}{#1}}}

\global\newcount\prono \global\prono=0
\def\pro#1{\global\advance\prono by1
\message{(\bf Proposition\quad\secsym\the\subsecno.\the\prono 
)}
\ifnum\lastpenalty>9000\else\bigbreak\fi
\noindent{\bf Proposition\quad
\the\prono\quad}{\ninepoint #1}
}

\global\newcount\teno \global\prono=0
\def\te#1{\global\advance\teno by1
\message{(\bf Theorem\quad\secsym\the\subsecno.\the\teno #1)}
\ifnum\lastpenalty>9000\else\bigbreak\fi
\noindent{\bf Theorem\quad\secsym\the\subsecno.\the\teno}{#1}
\writetoca{\string\qquad{\secsym\the\subsecno.\the\teno}{#1}}}
\def\subsubseclab#1{\DefWarn#1\xdef #1{\noexpand\hyperref{}{subsubsection}%
{\secsym\the\subsecno.\the\subsubsecno}%
{\secsym\the\subsecno.\the\subsubsecno}}%
\writedef{#1\leftbracket#1}\wrlabeL{#1=#1}}

\input epsf
\def\Figx#1#2#3{
\bigskip
\vbox{\centerline{\epsfxsize=#1 cm \epsfbox{NOpic#2.eps}}
\centerline{{\bf Fig.\the\figno} #3}}\bigskip\global\advance\figno by1}

\def\Figy#1#2#3{
\bigskip
\vbox{\centerline{\epsfysize=#1 cm \epsfbox{NOpic#2.eps}}
\centerline{{\bf Fig.\the\figno} #3}}\bigskip\global\advance\figno by1}

\def\unredoffs{} \def\redoffs{\voffset=-.40truein\hoffset=-.40truein}
\def\speclscape{}

\newbox\leftpage \newdimen\fullhsize \newdimen\hstitle \newdimen\hsbody
\tolerance=1000\hfuzz=2pt

\catcode`\@=11
\def\bigans{b }
\def\answ{b }

\ifx\answ\bigans\message{(This will come out unreduced.}
\magnification=1200\unredoffs\baselineskip=16pt plus 2pt minus 1pt
\hsbody=\hsize \hstitle=\hsize

\else\message{(This will be reduced.} \let\l@r=L
\magnification=1200\baselineskip=16pt plus 2pt minus 1pt \vsize=7truein
\redoffs \hstitle=8truein\hsbody=4.75truein\fullhsize=10truein\hsize=\hsbody
\output={\ifnum\pageno=0

   \shipout\vbox{{\hsize\fullhsize\makeheadline}
     \hbox to \fullhsize{\hfill\pagebody\hfill}}\advancepageno
   \else
   \almostshipout{\leftline{\vbox{\pagebody\makefootline}}}\advancepageno
   \fi}
\def\almostshipout#1{\if L\l@r \count1=1 \message{[\the\count0.\the\count1]}
       \global\setbox\leftpage=#1 \global\let\l@r=R
  \else \count1=2
   \shipout\vbox{\speclscape{\hsize\fullhsize\makeheadline}
       \hbox to\fullhsize{\box\leftpage\hfil#1}}  \global\let\l@r=L\fi}
\fi

\newcount\yearltd\yearltd=\year\advance\yearltd by -2000

\def\Title#1#2{\nopagenumbers
\abstractfont\hsize=\hstitle\rightline{#1}%
\vskip 1in\centerline{\titlefont #2}\abstractfont\vskip .5in\pageno=0}
\def\Date#1{\vfill\leftline{#1}\tenpoint\supereject\global\hsize=\hsbody%
\footline={\hss\tenrm\folio\hss}}


\def\draftmode{\message{ DRAFTMODE }\def\draftdate{{\rm preliminary draft:
\number\month/\number\day/\number\yearltd\ \ \hourmin}}%

\writelabels\baselineskip=20pt plus 2pt minus 2pt
  {\count255=\time\divide\count255 by 60 \xdef\hourmin{\number\count255}
   \multiply\count255 by-60\advance\count255 by\time
   \xdef\hourmin{\hourmin:\ifnum\count255<10 0\fi\the\count255}}}

\def\nolabels{\def\wrlabeL##1{}\def\eqlabeL##1{}\def\reflabeL##1{}}
\def\writelabels{\def\wrlabeL##1{\leavevmode\vadjust{\rlap{\smash%
{\line{{\escapechar=` \hfill\rlap{\sevenrm\hskip.03in\string##1}}}}}}}%
\def\eqlabeL##1{{\escapechar-1\rlap{\sevenrm\hskip.05in\string##1}}}%
\def\reflabeL##1{\noexpand\llap{\noexpand\sevenrm\string\string\string##1}}}
\nolabels
%


\global\newcount\secno \global\secno=0
\global\newcount\meqno
\global\meqno=1
\def\eqnres@t{\xdef\secsym{\the\secno.}\global\meqno=1
\bigbreak\bigskip}
\def\sequentialequations{\def\eqnres@t{\bigbreak}}
\def\appendix#1#2{\vfill\eject\global\meqno=1\global\subsecno=0\xdef\secsym{\hbox{#1.}}
\bigbreak\bigskip\noindent{\bf Appendix #1. #2}\message{(#1. #2)}
\writetoca{Appendix {#1.} {#2}}\par\nobreak\medskip\nobreak}

\def\eqnn#1{\xdef #1{(\secsym\the\meqno)}\writedef{#1\leftbracket#1}%
\global\advance\meqno by1\wrlabeL#1}
\def\eqna#1{\xdef #1##1{\hbox{$(\secsym\the\meqno##1)$}}
\writedef{#1\numbersign1\leftbracket#1{\numbersign1}}%
\global\advance\meqno by1\wrlabeL{#1$\{\}$}}
\def\eqn#1#2{\xdef #1{(\secsym\the\meqno)}\writedef{#1\leftbracket#1}%
\global\advance\meqno by1$$#2\eqno#1\eqlabeL#1$$}

\newskip\footskip\footskip14pt plus 1pt minus 1pt

\def\footnotefont{\ninepoint}\def\f@t#1{\footnotefont #1\@foot}
\def\f@@t{\baselineskip\footskip\bgroup\footnotefont\aftergroup\@foot\let\next}
\setbox\strutbox=\hbox{\vrule height9.5pt depth4.5pt width0pt}
\global\newcount\ftno \global\ftno=0
\def\foot{\global\advance\ftno by1\footnote{$^{\the\ftno}$}}

\newwrite\ftfile
\def\footend{\def\foot{\global\advance\ftno by1\chardef\wfile=\ftfile
$^{\the\ftno}$\ifnum\ftno=1\immediate\openout\ftfile=foots.tmp\fi%
\immediate\write\ftfile{\noexpand\smallskip%
\noexpand\item{f\the\ftno:\ }\pctsign}\findarg}%
\def\footatend{\vfill\eject\immediate\closeout\ftfile{\parindent=20pt
\centerline{\bf Footnotes}\nobreak\bigskip\input foots.tmp }}}
\def\footatend{}

\global\newcount\refno \global\refno=1
\newwrite\rfile
\def\ref{[\the\refno]\nref}
\def\nref#1{\xdef#1{[\the\refno]}\writedef{#1\leftbracket#1}%
\ifnum\refno=1\immediate\openout\rfile=refs.tmp\fi \global\advance\refno
by1\chardef\wfile=\rfile\immediate \write\rfile{\noexpand\item{#1\
}\reflabeL{#1\hskip.31in}\pctsign}\findarg}

\def\findarg#1#{\begingroup\obeylines\newlinechar=`\^^M\pass@rg}
{\obeylines\gdef\pass@rg#1{\writ@line\relax #1^^M\hbox{}^^M}%
\gdef\writ@line#1^^M{\expandafter\toks0\expandafter{\striprel@x #1}%
\edef\next{\the\toks0}\ifx\next\em@rk\let\next=\endgroup\else\ifx\next\empty%
\else\immediate\write\wfile{\the\toks0}\fi\let\next=\writ@line\fi\next\relax}}
\def\striprel@x#1{} \def\em@rk{\hbox{}}
\def\lref{\begingroup\obeylines\lr@f}
\def\lr@f#1#2{\gdef#1{\ref#1{#2}}\endgroup\unskip}
\def\semi{;\hfil\break}
\def\addref#1{\immediate\write\rfile{\noexpand\item{}#1}}

\def\startrefs#1{\immediate\openout\rfile=refs.tmp\refno=#1}
\def\xref{\expandafter\xr@f}\def\xr@f[#1]{#1}
\def\refs#1{\count255=1[\r@fs #1{\hbox{}}]}
\def\r@fs#1{\ifx\und@fined#1\message{reflabel \string#1 is undefined.}%
\nref#1{need to supply reference \string#1.}\fi%
\vphantom{\hphantom{#1}}\edef\next{#1}\ifx\next\em@rk\def\next{}%
\else\ifx\next#1\ifodd\count255\relax\xref#1\count255=0\fi%
\else#1\count255=1\fi\let\next=\r@fs\fi\next}


\def\writetoc{\immediate\openout\tfile=NOI.tmp
    \def\writetoca##1{{\edef\next{\write\tfile{\noindent  ##1
    \string\leaderfill {\noexpand\number\pageno} \par}}\next}}}


%
\catcode`\@=12 
%
\edef\tfontsize{\ifx\answ\bigans scaled\magstep3\else scaled\magstep4\fi}
\font\titlerm=cmr10 \tfontsize \font\titlerms=cmr7 \tfontsize
\font\titlermss=cmr5 \tfontsize \font\titlei=cmmi10 \tfontsize
\font\titleis=cmmi7 \tfontsize \font\titleiss=cmmi5 \tfontsize
\font\titlesy=cmsy10 \tfontsize \font\titlesys=cmsy7 \tfontsize
\font\titlesyss=cmsy5 \tfontsize \font\titleit=cmti10 \tfontsize
\skewchar\titlei='177 \skewchar\titleis='177 \skewchar\titleiss='177
\skewchar\titlesy='60 \skewchar\titlesys='60 \skewchar\titlesyss='60
\def\titlefont{\def\rm{\fam0\titlerm}
\textfont0=\titlerm \scriptfont0=\titlerms \scriptscriptfont0=\titlermss
\textfont1=\titlei \scriptfont1=\titleis \scriptscriptfont1=\titleiss
\textfont2=\titlesy \scriptfont2=\titlesys \scriptscriptfont2=\titlesyss
\textfont\itfam=\titleit
\def\it{\fam\itfam\titleit}\rm}
\font\authorfont=cmcsc10 \ifx\answ\bigans\else scaled\magstep1\fi
\ifx\answ\bigans\def\abstractfont{\tenpoint}\else \font\abssl=cmsl10 scaled
\magstep1 \font\absrm=cmr10 scaled\magstep1 \font\absrms=cmr7
scaled\magstep1 \font\absrmss=cmr5 scaled\magstep1 \font\absi=cmmi10
scaled\magstep1 \font\absis=cmmi7 scaled\magstep1 \font\absiss=cmmi5
scaled\magstep1 \font\abssy=cmsy10 scaled\magstep1 \font\abssys=cmsy7
scaled\magstep1 \font\abssyss=cmsy5 scaled\magstep1 \font\absbf=cmbx10
scaled\magstep1 \skewchar\absi='177 \skewchar\absis='177
\skewchar\absiss='177 \skewchar\abssy='60 \skewchar\abssys='60
\skewchar\abssyss='60
\def\abstractfont{\def\rm{\fam0\absrm}
\textfont0=\absrm \scriptfont0=\absrms \scriptscriptfont0=\absrmss
\textfont1=\absi \scriptfont1=\absis \scriptscriptfont1=\absiss
\textfont2=\abssy \scriptfont2=\abssys \scriptscriptfont2=\abssyss
\textfont\itfam=\bigit \def\it{\fam\itfam\bigit}\def\footnotefont{\tenpoint}%
\textfont\slfam=\abssl \def\sl{\fam\slfam\abssl}%
\textfont\bffam=\absbf \def\bf{\fam\bffam\absbf}\rm}\fi
\def\tenpoint{\def\rm{\fam0\tenrm}
\textfont0=\tenrm \scriptfont0=\sevenrm \scriptscriptfont0=\fiverm
\textfont1=\teni  \scriptfont1=\seveni  \scriptscriptfont1=\fivei
\textfont2=\tensy \scriptfont2=\sevensy \scriptscriptfont2=\fivesy
\textfont\itfam=\tenit \def\it{\fam\itfam\tenit}\def\footnotefont{\ninepoint}%
\textfont\bffam=\tenbf
\def\bf{\fam\bffam\tenbf}\def\sl{\fam\slfam\tensl}\rm}
\font\ninerm=cmr9 \font\sixrm=cmr6 \font\ninei=cmmi9 \font\sixi=cmmi6
\font\ninesy=cmsy9 \font\sixsy=cmsy6 \font\ninebf=cmbx9 \font\nineit=cmti9
\font\ninesl=cmsl9 \skewchar\ninei='177 \skewchar\sixi='177
\skewchar\ninesy='60 \skewchar\sixsy='60
\def\ninepoint{\def\rm{\fam0\ninerm}
\textfont0=\ninerm \scriptfont0=\sixrm \scriptscriptfont0=\fiverm
\textfont1=\ninei \scriptfont1=\sixi \scriptscriptfont1=\fivei
\textfont2=\ninesy \scriptfont2=\sixsy \scriptscriptfont2=\fivesy
\textfont\itfam=\ninei \def\it{\fam\itfam\nineit}\def\sl{\fam\slfam\ninesl}%
\textfont\bffam=\ninebf \def\bf{\fam\bffam\ninebf}\rm}
%
%

\hyphenation{anom-aly anom-alies coun-ter-term coun-ter-terms}
\def\inv{^{\raise.15ex\hbox{${\scriptscriptstyle -}$}\kern-.05em 1}}

\def\Dsl{\,\raise.15ex\hbox{/}\mkern-13.5mu D} 
\def\dsl{\raise.15ex\hbox{/}\kern-.57em\partial}

 \def\Tr{{\rm Tr}}
\font\bigit=cmti10 scaled \magstep1
\def\lspace{\ifx\answ\bigans{}\else\qquad\fi}
\def\lbspace{\ifx\answ\bigans{}\else\hskip-.2in\fi} 
\def\boxeqn#1{\vcenter{\vbox{\hrule\hbox{\vrule\kern3pt\vbox{\kern3pt
     \hbox{${\displaystyle #1}$}\kern3pt}\kern3pt\vrule}\hrule}}}
\def\mbox#1#2{\vcenter{\hrule \hbox{\vrule height#2in
         \kern#1in \vrule} \hrule}}  


\newwrite\ffile\global\newcount\figno \global\figno=1
\def\fig{fig.~\the\figno\nfig}
\def\nfig#1{\xdef#1{fig.~\the\figno}%
\writedef{#1\leftbracket fig.\noexpand~\the\figno}%
\ifnum\figno=1\immediate\openout\ffile=figs.tmp\fi\chardef\wfile=\ffile%
\immediate\write\ffile{\noexpand\medskip\noexpand\item{Fig.\ \the\figno. }
\reflabeL{#1\hskip.55in}\pctsign}\global\advance\figno by1\findarg}
\def\xfig{\expandafter\xf@g}
\def\xf@g fig.\penalty\@M\ {}
\def\figs#1{figs.~\f@gs #1{\hbox{}}}
\def\f@gs#1{\edef\next{#1}\ifx\next\em@rk\def\next{}\else
\ifx\next#1\xfig #1\else#1\fi\let\next=\f@gs\fi\next}
\newwrite\lfile
{\escapechar-1\xdef\pctsign{\string\%}\xdef\leftbracket{\string\{}
\xdef\rightbracket{\string\}}\xdef\numbersign{\string\#}}

\def\writestop{\def\writestoppt{\immediate\write\lfile{\string\pageno%
\the\pageno\string\startrefs\leftbracket\the\refno\rightbracket%
\string\def\string\secsym\leftbracket\secsym\rightbracket%
\string\secno\the\secno\string\meqno\the\meqno}\immediate\closeout\lfile}}
\def\writestoppt{}\def\writedef#1{}
\def\seclab#1{\xdef #1{\the\secno}\writedef{#1\leftbracket#1}\wrlabeL{#1=#1}}
\def\subseclab#1{\xdef #1{\secsym\the\subsecno}%
\writedef{#1\leftbracket#1}\wrlabeL{#1=#1}}
\newwrite\tfile \def\writetoca#1{}
\def\leaderfill{\leaders\hbox to 1em{\hss.\hss}\hfill}

\def\tilde{\widetilde}
\def\bar{\overline}
\def\hat{\widehat}

\def\darr#1{\raise1.5ex\hbox{$\leftrightarrow$}\mkern-16.5mu #1}

\def\half{{\textstyle{1\over2}}}

\def\roughly#1{\raise.3ex\hbox{$#1$\kern-.75em\lower1ex\hbox{$\sim$}}}

\def\IA{\relax\hbox{$\inbar\kern-.3em{\rm A}$}}
\def\IB{\relax\hbox{$\inbar\kern-.3em{\rm B}$}}
\def\IC{\relax\hbox{$\inbar\kern-.3em{\rm C}$}}
\def\ID{\relax\hbox{$\inbar\kern-.3em{\rm D}$}}
\def\IE{\relax\hbox{$\inbar\kern-.3em{\rm E}$}}
\def\IF{\relax\hbox{$\inbar\kern-.3em{\rm F}$}}
\def\IG{\relax\hbox{$\inbar\kern-.3em{\rm G}$}}
\def\IGa{\relax\hbox{${\rm I}\kern-.18em\Gamma$}}
\def\IH{\relax{\rm I\kern-.18em H}}
\def\IK{\relax{\rm I\kern-.18em K}}
\def\IL{\relax{\rm I\kern-.18em L}}
\def\IP{\relax{\rm I\kern-.18em P}}
\def\IR{\relax{\rm I\kern-.18em R}}
\def\IS{\relax\hbox{$\inbar\kern-.3em{\rm S}$}}

\def\IZ{\relax\ifmmode\mathchoice{
\hbox{\cmss Z\kern-.4em Z}}{\hbox{\cmss Z\kern-.4em Z}}
{\lower.9pt\hbox{\cmsss Z\kern-.4em Z}} {\lower1.2pt\hbox{\cmsss Z\kern-.4em
Z}} \else{\cmss Z\kern-.4em Z}\fi}

\def\II{\relax{\rm I\kern-.18em I}}


\lockat

\def\square{{\mbox{.07}{.07}}}
\def\squarel{{\mbox{.04}{.04}}}

\def\frac1#1{{1 \over{#1}}}
\def\sinv#1{{\scriptstyle{1\over{#1}}}}
\def\sfrac#1#2{{\scriptstyle{{#1}\over{#2}}}}
\def\frac#1#2{{{#1}\over{#2}}}
\def\lg#1{{{\rm log}\left(#1\right)}}
\def\lgm#1{{{\rm log}\Biggl| #1\Biggr|}}

\def\const{{\rm const}}
\def\supp{{\rm supp}}
\def\mod{\ {\rm mod} \ }

\def\fs{f_\star}

\def\zu#1{{{\g}_{\hbar}\left( #1 ; {\Lambda}\right)}}
\def\zq#1{{{\g}_{\hbar}\left( #1 ; q\right)}}

\def\ndt{{\noindent}}

\def\mdp{\bigskip\noindent$\bullet$\bigskip\noindent}
\def\remark#1{\bigskip\ndt{\bf Remark:}{\ninepoint{\sl \ #1}}\bigskip\ndt}
\def\proof#1{\bigskip\ndt{\bf Proof:}{\medskip\ndt\ninepoint{\ #1}}\bigskip\ndt$\rm QED$}


\def\CA{{\cal A}}

\def\CC{{\cal C}}

\def\CE{{\cal E}}
\def\CF{{\cal F}}
\def\CG{{\cal G}}
\def\CH{{\cal H}}
\def\CI{{\cal I}}
\def\CJ{{\cal J}}
\def\CK{{\cal K}}

\def\CN{{\cal N}}
\def\CO{{\cal O}}

\def\CS{{\cal S}}

\def\CU{{\cal U}}
\def\CV{{\cal V}}
\def\CW{{\cal W}}

\def\CZ{{\cal Z}}

\def\p{\partial}
\def\pb{\bar{\partial}}

\def\intp#1{{\int\hskip -11pt - \hskip 5pt}_{\hskip -7pt #1}}

\def\zb{\bar{z}}

\def\Tr{{\rm Tr}}

\def\Det{{\rm Det}}

\def\dim{{\rm dim}}


\def\inbar{\,\vrule height1.5ex width.4pt depth0pt}

\font\cmss=cmss10 \font\cmsss=cmss10 at 7pt

\def\a{{\alpha}}

\def\b{{\beta}}
\def\d{{\delta}}

\def\g{{\gamma}}
\def\e{{\epsilon}}
\def\z{{\zeta}}
\def\ve{{\varepsilon}}
\def\vf{{\varphi}}
\def\m{{\mu}}
\def\n{{\nu}}
\def\u{{\Upsilon}}
\def\l{{\lambda}}
\def\s{{\sigma}}
\def\t{{\theta}}

\def\o{{\omega}}
\def\r{{\rho}}
\def\k{{\kappa}}

\def\boxit#1{\vbox{\hrule\hbox{\vrule\kern8pt
\vbox{\hbox{\kern8pt}\hbox{\vbox{#1}}\hbox{\kern8pt}}
\kern8pt\vrule}\hrule}}
\def\mathboxit#1{\vbox{\hrule\hbox{\vrule\kern8pt\vbox{\kern8pt
\hbox{$\displaystyle #1$}\kern8pt}\kern8pt\vrule}\hrule}}


\chardef\tempcat=\the\catcode`\@ \catcode`\@=11
\def\cyracc{\def\u##1{\if \i##1\accent"24 i%
    \else \accent"24 ##1\fi }}
\newfam\cyrfam
\font\tencyr=wncyr10
\def\cyr{\fam\cyrfam\tencyr\cyracc}


\def\darr#1{\raise1.5ex\hbox{$\leftrightarrow$}\mkern-16.5mu #1}

\def\half{{\textstyle{1\over2}}} 

\def\roughly#1{\raise.3ex\hbox{$#1$\kern-.75em\lower1ex\hbox{$\sim$}}}


\def\ndt{{\noindent}}
 

\def\ba{{\bf a}}
\def\bb{{\bf b}}

\def\fl{{\bf f}}
\def\bg{{\bf g}}

\def\bi{{\bf i}}
\def\bj{{\bf j}}
\def\bm{{\bf m}}
\def\bn{{\bf n}}
\def\bk{{\bf k}}
\def\bl{{\bf l}}

\def\bp{{\bf p}}

\def\bx{{\bf x}}

\def\bA{{\bf A}}

\def\bC{{\bf C}}

\def\bE{{\bf E}}
\def\bF{{\bf F}}

\def\bH{{\bf H}}
\def\bI{{\bf I}}

\def\bK{{\bf K}}

\def\bN{{\bf N}}
\def\bP{{\bf P}}
\def\bR{{\bf R}}
\def\bS{{\bf S}}
\def\bT{{\bf T}}

\def\bX{{\bf X}}

\def\bZ{{\bf Z}}



\def\lref{\begingroup\obeylines\lr@f}
\def\lr@f#1#2{\gdef#1{\ref#1{#2}}\endgroup\unskip}

\def\np#1#2#3{Nucl. Phys. {\bf B#1} (#2) #3}
\def\pl#1#2#3{Phys.Lett. {\bf #1B} (#2) #3}

\def\anp#1#2#3{Ann. Phys. {\bf #1} (#2) #3}
\def\pr#1#2#3{Phys. Rev. {\bf #1} (#2) #3}

\def\cmp#1#2#3{Comm.Math. Phys. {\bf #1} (#2) #3}
 
\def\jhep#1#2#3{JHEP {\bf#1}(#2) #3}
\def\jmp#1#2#3{J. Math Phys.{\bf #1} (#2) #3}

\def\jgp#1#2#3{J. Geom. Phys.{\bf #1} (#2) #3}

\def\atmp#1#2#3{Adv.~Theor.~Math.~Phys.{\bf #1} (#2) #3}


\lref\hiraku{H.~Nakajima, K.~Yoshioka, math.AG/0306198}

\lref\denis{D.~Bernard, \np{303}{1988}{77-93}, \np{309}{1988}{145-174}}

\lref\aslbernard{A.~S.~Losev, CERN-TH-6215-91, Aug. 1991}

\lref\knzam{V.~Knizhnik, A.~Zamolodchikov, \np{247}{1984}{83-103}}
\lref\shuryak{E.~Shuryak, Z.~Phys.{\bf C}38 (1988) 165-172, 141-145; \np{319}{1989}{521, 541}, \np{328}{1989}{85,102}}
\lref\coleman{S.~Coleman, ``The uses of instantons'', Erice Subnucl. (1977) 805 }
\lref\polyakovi{A.~Polyakov, ``Gauge fields and strings'', Harwood Academic Publishers, 1987 }
\lref\bpst{A.~Belavin, A.~Polyakov, A.~Schwarz, Yu. Tyupkin,
\pl{59}{1975}{85-87}}

\lref\samson{N.~Nekrasov, S.~Shatashvili, in progress}

\lref\tdym{D.~Gross, hep-th/9212149\semi D.~Gross, W.~Taylor,
hep-th/9301068, hep-th/9303046}

\lref\jm{M.~Jimbo, T.~Miwa, {\it Solitons and Infinite Dimensional Lie
Algebras}, Publ. RIMS Kyoto Univ. {\bf 19} ( 1983) 943-1001}

\lref\dougcft{M.~Douglas, hep-th/9311130, hep-th/9303159}
\lref\kawai{H.~Kawai, T.~Kuroki, T.~Morita, hep-th/0303210}
\lref\vafaavg{C.~Vafa, hep-th/0008142}
\lref\verker{ S.~V.~Kerov,
A.~M.~Vershik, ``Asymptotics of the Plancherel measure of the symmetric
group and the limit shape of the Young diagrams'', {\cyr DAN SSSR}, {\bf
233} (1977), no. 6,  1024-1027 (in Russian)}
\lref\verkeri{S.~V.~Kerov,
A.~M.~Vershik,
`` Asymptotic behavior of the maximum and
generic dimensions of irreducible representations of the symmetric
group,'' (Russian)  {\cyr Funkcional. Anal. i Prilozhen.}  {\bf 19}  (1985),  no. 1,
25--36}
\lref\kerovii{
S.V.~Kerov, ``Gaussian limit for the Plancherel measure of the
symmetric group'',  C. R. Acad. Sci. Paris S'r. I Math.  {\bf 316}  (1993),  no.
4, 303--308.}

\lref\dimer{R.~Kenyon, A.~Okounkov, S.~Sheffield, ``Dimers and amoebae'', to
appear
\semi
R. Kenyon, A.~Okounkov, in preparation}


\lref\logshep{B.F.~Logan, L.A.~Shepp, ``A variational problem for random Young
tableaux'',   Advances in Math.  26  (1977),  no. 2, 206--222}


\lref\kerovi{S.V.~Kerov, `` Interlacing measures'',  {\it  Kirillov's seminar on
representation theory},  35--83, Amer. Math. Soc. Transl. Ser. 2, 181,
Amer. Math. Soc., Providence, RI, 1998 \semi
S.V.~Kerov, ``Anisotropic Young diagrams and symmetric Jack
functions'',
(Russian)  {\cyr Funkcional. Anal. i Prilozhen.}  34  (2000),  no. 1, 51--64,
96;  translation in  Funct. Anal. Appl.  34  (2000),  no. 1, 41--51\semi
A.~M.~Vershik, ``Hook formulae and related identities'', {\cyr
Zapiski sem. LOMI}, {\bf 172} (1989), 3-20 (in Russian)\semi
S.~V.~Kerov, ``Random Young
tableaux'', {\cyr Teor. veroyat. i ee primeneniya}, {\bf 3} (1986), 627-628
(in Russian)}


\lref\gue{K.~Johansson, ``The longest increasing subsequence in a random
permutation and a unitary random matrix model'',   Math. Res. Lett.  5
(1998),  no. 1-2, 63--82\semi
J.~Baik, P.~Deift, K.~Johansson, `` On the distribution of the
length of the longest increasing subsequence of random permutations'',  J.
Amer. Math. Soc.  12  (1999),  no. 4, 1119--1178.\semi
A.~Okounkov, ``Random matrices and random permutations'',  Internat.
Math. Res. Notices  2000,  no. 20, 1043--1095\semi
A.~Borodin, A.~Okounkov, G.~Olshanski, ``Asymptotics of
Plancherel measures for symmetric groups'',  J. Amer. Math. Soc.  13
(2000),  no. 3, 481--515\semi
K.~Johansson, ``Discrete orthogonal polynomial ensembles and the
Plancherel measure'',  Ann. of Math. (2)  153  (2001),  no. 1, 259--296}


\lref\aoi{A.~Okounkov, ``Infinite wedge and random partitions'',   Selecta Math.
(N.S.)  7  (2001),  no. 1, 57--81}

\lref\vnm{P.~van Moerbeke, ``Integrable lattices: random matrices and random
permutations'',   Random matrix models and their applications,  321--406,
Math. Sci. Res. Inst. Publ., 40, Cambridge Univ. Press, Cambridge, 2001.}



\lref\anyons{V.~Pasquier, hep-th/9405104\semi
R.~Caracciollo, A.~Lerda, G.R.~Zemba, hep-th/9503229\semi
J.~Minahan, A.P.~Polychronakos, hep-th/9404192, hep-th/9303153\semi
A.P.~Polychronakos, hep-th/9902157\semi
E.~Langmann, math-ph/0007036, math-ph/0102005}

\lref\girvin{S.~Girvin, cond-mat/9907002}

\lref\phtran{M.~Douglas, V.~Kazakov, hep-th/9305047}

\lref\dv{R.~Dijkgraaf, C.~Vafa, hep-th/0206255, hep-th/0207106,
hep-th/0208048\semi R.~Dijkgraaf, S.~Gukov, V.~Kazakov, C.~Vafa,
hep-th/0210238\semi R.~Dijkgraaf, M.~Grisaru, C.~Lam, C.~Vafa, D.~Zanon,
hep-th/0211017\semi M.~Aganagic, M.~Marino, A.~Klemm, C.~Vafa,
hep-th/0211098\semi R.~Dijkgraaf, A.~Neitzke, C.~Vafa, hep-th/0211194}
\lref\cdws{F.~Cachazo, M.~Douglas, N.~Seiberg, E.~Witten, hep-th/0211170}

\lref\witdonagi{R.~Donagi, E.~Witten, hep-th/9510101}

\lref\donagirev{R.~Donagi, alg-geom/9705010}


\lref\sasha{A.~Orlov, nlin.SI/0207030, nlin.SI/0305001}

\lref\takashi{T.~Takebe, ``Representation theoretical meaning of the initial value
problem for the Toda lattice hierarchy I.'',   Lett. Math. Phys.  {\bf 21}
(1991),  no. 1, 77--84}


\lref\seibergminwala{S.~Minwala, M.~van Raamsdonk, N.~Seiberg,
hep-th/9912072}

\lref\estring{J.~A.~Minahan, D.~Nemeschansky, C.~Vafa, N.P.~Warner,
hep-th/9802168\semi T.~Eguchi, K.~Sakai, hep-th/0203025, hep-th/0211213}

\lref\bulkbndr{V.~Balasubramanian, P.~Kraus, A.~Lawrence, hep-th/9805171}

\lref\frbranes{D.-E.~Diaconescu, M.~Douglas, J.~Gomis, hep-th/9712230}

\lref\gorodentsevleenzon{A.~Klyachko, ``Moduli of vector bundles and numbers
of classes'', Funct. Anal. and Appl. {\bf 25} (1991) 67\semi G.~Ellingsrud,
L.~G\"ottsche, alg-geom/9506019\semi A.~Gorodentsev, M.~Leenson,
alg-geom/9604011} \lref\witdgt{E.~Witten, hep-th/9204083}

\lref\ikkt{N.~Ishibashi, H.~Kawai, Y.~Kitazawa, and A.~Tsuchiya,
\np{498}{1997}{467}, hep-th/9612115}

\lref\cds{A.~Connes, M.~Douglas, A.~Schwarz, \jhep{9802}{1998}{003}}
\lref\wtnc{E.~Witten, \np{268}{1986}{253}}

\lref\cstw{L.~Baulieu, A.~Losev, N.~Nekrasov, hep-th/9707174}

\lref\klemmzaslow{ A.~Klemm, E.~Zaslow, hep-th/9906046}

\lref\walg{A.~Gerasimov, A.~Levin, A.~Marshakov, \np{360}{1991}{537}\semi
A.~Bilal, I.~Kogan, V.~Fock, \np{359}{1991}{635}}

\lref\cftorb{A.~Lawrence, N.~Nekrasov, C.~Vafa, hep-th/9803015}

\lref\booksSW{ A.~Marshakov, ``Seiberg-Witten Theory and Integrable
Systems,'' {\it World Scientific}, Singapore 1999\semi H.~Braden and I.~Krichever, Eds. \ ``Integrability:
The Seiberg-Witten and Whitham Equations",
{\it Gordon and Breach} 2000.}

\lref\agmav{M.~Aganagic, M.~Mari\~no, C.~Vafa, hep-th/0206164}

\lref\moorewitten{G.~Moore, E.~Witten, hep-th/9709193}

\lref\gopakumarvafa{R.~Gopakumar, C.Vafa, hep-th/9809187, hep-th/9812127}
\lref\gopakumarvafaii{R.~Gopakumar, C.Vafa, hep-th/9802016, hep-th/9811131}

\lref\gopakumargross{D.~Gross, R.~Gopakumar, hep-th/9411021\semi R.~Gopakumar, hep-th/0211100}
\lref\mooreunpublished{G.~Moore, unpublished}

\lref\wittenone{E.~Witten, hep-th/9403195}

\lref\cg{E.~Corrigan, P.~Goddard, ``Construction of instanton and monopole
solutions and reciprocity'', \anp {154}{1984}{253}}
\lref\opennc{N.~Nekrasov, hep-th/0010017, hep-th/0203109}

\lref\kly{K.-Y.Kim, B.-H. Lee, H.S. Yang, hep-th/0205010}

\lref\donaldson{S.K.~Donaldson, ``Instantons and Geometric Invariant
Theory", \cmp{93}{1984}{453-460}}

\lref\nakajima{H.~Nakajima, ``Lectures on Hilbert Schemes of Points on
Surfaces''\semi AMS University Lecture Series, 1999, ISBN 0-8218-1956-9. }

\lref\neksch{N.~Nekrasov, A.~S.~Schwarz, hep-th/9802068,
\cmp{198}{1998}{689}}

\lref\freck{A.~Losev, N.~Nekrasov, S.~Shatashvili, hep-th/9908204,
hep-th/9911099} \lref\rkh{N.J.~Hitchin, A.~Karlhede, U.~Lindstrom, and
M.~Rocek, \cmp{108}{1987}{535}}

\lref\branek{H.~Braden, N.~Nekrasov, hep-th/9912019}

\lref\kazuyuki{K.~Furuuchi, hep-th/9912047}

\lref\wilson{G.~Wilson, ``Collisions of Calogero-Moser particles and adelic
Grassmannian", Invent. Math. 133 (1998) 1-41.}

\lref\abs{O.~Aharony, M.~Berkooz, N.~Seiberg,
hep-th/9712117,\atmp{2}{1998}{119-153}}

\lref\avatars{A.~Losev, G.~Moore, N.~Nekrasov, S.~Shatashvili,
hep-th/9509151}

\lref\abkss{O.~Aharony, M.~Berkooz, S.~Kachru, N.~Seiberg, E.~Silverstein,
hep-th/9707079, \atmp{1}{1998}{148-157}}

\lref\cecotti{S.~Cecotti, L.~Girardello, \pl{110}{1982}{39}}
\lref\smilga{A.~Smilga, Yad.Fiz. {\bf 43} (1986), 215-218}
\lref\sethi{S.~Sethi, M.~Stern, hep-th/9705046}

\lref\witsei{N.~Seiberg, E.~Witten, hep-th/9908142, \jhep{9909}{1999}{032}}

\lref\kkn{V.~Kazakov, I.~Kostov, N.~Nekrasov, ``D-particles, Matrix
Integrals and KP hierarchy'', \np{557}{1999}{413-442}, hep-th/9810035}

\lref\hoppe{J.~Hoppe, ``Quantum theory of a massless
relativistic surface ...''
Elementary Particle Research Journal (Kyoto) 80 (1989)}

\lref\DHf{J.~J.~Duistermaat, G.J.~Heckman, Invent. Math. {\bf 69} (1982)
259\semi M.~Atiyah, R.~Bott,  Topology {\bf 23} No 1 (1984) 1}
\lref\tdgt{M.~Atiyah, R.~Bott,  Phil. Trans. Roy. Soc. London {\bf A 308}
(1982), 524-615\semi E.~Witten, hep-th/9204083\semi S.~Cordes, G.~Moore,
S.~Rangoolam, hep-th/9411210}

\lref\atiyahsegal{M.~Atiyah, G.~Segal, Ann. of Math. {\bf 87} (1968) 531}

\lref\bott{R.~Bott, J.~Diff.~Geom. {\bf 4} (1967) 311}

\lref\torusaction{G.~Ellingsrud, S.A.Stromme, Invent. Math. {\bf 87} (1987)
343-352\semi L.~G\"ottche, Math. A.. {\bf 286} (1990) 193-207}

\lref\gravilit{M.~Bershadsky, S.~Cecotti, H.~Ooguri, C.~Vafa,
\cmp{165}{1994}{311}, \np{405}{1993}{279}\semi I.~Antoniadis, E.~Gava,
K.S.~Narain, T.~R.~Taylor, \np{413}{1994}{162}, \np{455}{1995}{109}}

\lref\calculus{N.~Dorey,T.~J.~Hollowood, V.~V.~Khoze, M.~P.~Mattis,
hep-th/0206063}

\lref\instmeasures{N.~Dorey, V.V.~Khoze, M.P.~Mattis, hep-th/9706007,
hep-th/9708036}

\lref\twoinst{N.~Dorey,V.V.~Khoze, M.P.~Mattis, hep-th/9607066}

\lref\vafaengine{S.~Katz, A.~Klemm, C.~Vafa, hep-th/9609239}

\lref\connes{A.~Connes,``Noncommutative geometry'', Academic Press (1994)}

\lref\macdonald{I.~Macdonald, ``Symmetric functions and Hall polynomials'',
Clarendon Press, Oxford, 1979}

\lref\nikfive{N.~Nekrasov, hep-th/9609219 \semi A.~Lawrence, N.~Nekrasov,
hep-th/9706025} \lref\nikgor{N.~Nekrasov, A.~Gorsky, hep-th/9401021 \semi
N.~Nekrasov, hep-th/9503157} \lref\nikdual{V.~Fock, A.~Gorsky, N.~Nekrasov,
V.~Roubtsov, hep-th/9906235}

\lref\dima{D.~Ivanov, hep-th/9610207}
\lref\giovanni{G.~Felder, hep-th/9609153}
\lref\olshanetsky{M.A.Olshanetsky, ``Painlev\'e type equations and Hitchin
systems'', in \booksSW}
\lref\kapustin{A.~Kapustin, hep-th/9804069} \lref\cm{F.~Calogero,
\jmp{12}{1971}{419}\semi J.~Moser, Adv. Math. {\bf 16}(1975), 197-220\semi
M.~Olshanetsky, A.~Perelomov, Inv. Math. {\bf 31} (1976), 93,
\pr{71}{1981}{313}\semi I.~Krichever, Funk. An. Appl. {\bf 14} (1980) 45}
\lref\hitchin{N.~Hitchin, Duke Math. J. {\bf 54}, vol. 1 (1987) }
\lref\fivedim{A.~Marshakov, A.~Mironov, hep-th/9711156\semi H.~Braden,
A.~Marshakov, A.~Mironov, A.~Morozov, hep-th/9812078, hep-th/9902205\semi
T.~Eguchi, H.~Kanno, hep-th/0005008\semi H.~Braden, A.~Marshakov,
hep-th/0009060\semi
 H. Braden, A. Gorsky, A. Odesskii, V. Rubtsov, hep-th/01111066\semi
C.Csaki, J.Erlich, V.V.Khoze, E.Poppitz, Y.Shadmi, Y.Shirman,
hep-th/0110188\semi T.~Hollowood, hep-th/0302165}

\lref\seibergfive{N.~Seiberg, hep-th/9608111}

\lref\ganor{O.~Ganor, hep-th/9607092, hep-th/9608108}

\lref\equivsheaf{A.~Knutsen, E.~Sharpe, hep-th/9804075}

\lref\bcov{ M.~Bershadsky, S.~Cecotti, H.~Ooguri, C.~Vafa, hep-th/9309140}

\lref\op{A.~Okounkov, R.~Pandharipande, math.AG/0207233, math.AG/0204305}

\lref\ob{S.~Bloch, A.~Okounkov, alg-geom/9712009}

\lref\robbert{H. Awata, M. Fukuma, S. Odake, Y.-H. Quano,
hep-th/9312208\semi H.Awata, M.Fukuma, Y.Matsuo, S.Odake,
hep-th/9408158\semi R.~Dijkgraaf, hep-th/9609022}

\lref\robberti{R.~Dijkgraaf, ``Mirror symmetry and elliptic curves'', in:
The moduli space of curves, Proc. of Texel Island Meeting, Birkhauser, 1994}

\lref\prtoda{T.~Eguchi, K.~Hori, C.-S.~Xiong, hep-th/9605225\semi T.~Eguchi,
S.~Yang, hep-th/9407134\semi T.~Eguchi, H.~Kanno, hep-th/9404056}

\lref\givental{A.~Givental, alg-geom/9603021}

\lref\maxim{M.~Kontsevich, hep-th/9405035}

\lref\whitham{A.~Gorsky,
A.~Marshakov, A.~Mironov, A.~Morozov, Nucl.Phys. {\bf B527} (1998) 690-716,
hep-th/9802007}

\lref\kricheverwhitham{I.~Krichever, hep-th/9205110, \cmp{143}{1992}{415}}

\lref\blowupwhitham{M.~Mari{\~n}o, G.~Moore, hep-th/9802185\semi
J.~Edelstein, M.~Mari{\~n}o, J.~Mas, hep-th/9805172}

\lref\takasaki{K.~Takasaki, hep-th/9901120}

\lref\zabrodin{P.~Wiegmann, A.~Zabrodin, hep-th/9909147\semi
I.K.~Kostov, I.~Krichever, M.~Mineev-Weinstein, P.~Wiegmann, A.~Zabrodin,
hep-th/0005259}

\lref\markman{J.~Hurtubise, E.~Markman, math.AG/9912161}

\lref\sw{N.~Seiberg, E.~Witten, hep-th/9407087, hep-th/9408099}

\lref\swsol{A.~Klemm, W.~Lerche, S.~Theisen, S.~Yankielowicz, hep-th/9411048
\semi P.~Argyres, A.~Faraggi, hep-th/9411057\semi A.~Hanany, Y.~Oz,
hep-th/9505074}

\lref\hollowood{T.~Hollowood, hep-th/0201075, hep-th/0202197}

\lref\nsvz{V.~Novikov, M.~Shifman, A.~Vainshtein, V.~Zakharov,
\np{229}{1983}{381}, \np{260}{1985}{157-181}, \pl{217}{1989}{103}}

\lref\seibergone{N.~Seiberg, \pl{206}{1988}{75}}

\lref\ihiggs{G.~Moore, N.~Nekrasov, S.~Shatashvili, hep-th/9712241,
hep-th/9803265 } \lref\potsdam{W.~Krauth, H.~Nicolai, M.~Staudacher,
hep-th/9803117} \lref\kirwan{F.~Kirwan, ``Cohomology of quotients in
symplectic and algebraic geometry'', Mathematical Notes, Princeton Univ.
Press, 1985}

\lref\wittfivebrane{E.~Witten, hep-th/9610234}

\lref\issues{A.~Losev, N.~Nekrasov, S.~Shatashvili, hep-th/9711108,
hep-th/9801061}

\lref\adhm{M.~Atiyah, V.~Drinfeld, N.~Hitchin, Yu.~Manin, Phys. Lett. {\bf
65A} (1978) 185}

\lref\seiberghol{N.~Seiberg, hep-th/9408013}

 \lref\warner{A.~Klemm, W.~Lerche, P.~Mayr, C.~Vafa, N.~Warner, hep-th/9604034}

\lref\wittensolution{E.~Witten, hep-th/9703166}

\lref\witbound{E.~Witten, hep-th/9510153}

\lref\twists{E.~Witten, hep-th/9304026 \semi O.~Ganor, hep-th/9903110 \semi
H.~Braden, A.~Marshakov, A.~Mironov, A.~Morozov, hep-th/9812078}

\lref\cmn{S.~Cherkis, G.~Moore, N.~Nekrasov, in progress}

\lref\dijkgraaf{R.~Dijkgraaf, hep-th/9609022}

\lref\iqbal{A.~Iqbal, hep-th/0212279}

\lref\wittenm{E.~Witten, hep-th/9503124}

\lref\vafawitten{C.~Vafa, E.~Witten, hep-th/9408074}

\lref\niklos{A.~Losev, N.~Nekrasov, in progress}

\lref\experiment{G.~Chan, E.~D'Hoker, hep-th/9906193 \semi E.~D'Hoker,
I.~Krichever, D.~Phong, hep-th/9609041}

\lref\flow{I.~Klebanov, N.~Nekrasov, hep-th/9911096\semi J.~Polchinski,
hep-th/0011193}

\lref\todalit{K.~Ueno, K.~Takasaki, Adv. Studies in Pure Math. {\bf 4}
(1984) 1}

\lref\kharchev{For an excellent review see, e.g. S.~Kharchev,
hep-th/9810091}

\lref\gkmmm{ A.Gorsky, I.Krichever, A.Marshakov, A.Mironov and A.Morozov,
Phys. Lett.  {\bf B355} (1995) 466; hep-th/9505035. }

\lref\witdonaldson{E.~Witten, \cmp{117}{1988}{353}} \lref\aj{M.~Atiyah,
L.~Jeffrey, \jgp{7}{1990}{119-136}}

\lref\swi{N.~Nekrasov, hep-th/0206161} \lref\lmn{A.~Losev, A.~Marshakov,
N.~Nekrasov, hep-th/0302191}

\lref\fucito{U.~Bruzzo, F.~Fucito, J.F.~Morales, A.~Tanzini,
hep-th/0211108\semi D.Bellisai, F.Fucito, A.Tanzini, G.Travaglini,
hep-th/0002110, hep-th/0003272, hep-th/0008225} \lref\flume{R.~Flume,
R.~Poghossian, hep-th/0208176\semi R.~Flume, R.~Poghossian, H.~Storch,
hep-th/0110240, hep-th/0112211} \lref\khoze{ N.~Dorey, T.J.~Hollowood,
V.~Khoze, M.~Mattis, hep-th/0206063, and references therein}

\lref\polyakov{A.~Polyakov, hep-th/9711002, hep-th/9809057}
\lref\ads{J.~Maldacena, hep-th/9711200\semi S.~Gubser, I.~Klebanov,
A.~Polyakov, hep-th/9802109\semi E.~Witten, hep-th/9802150}

\lref\loggaz{A.~Okounkov, hep-th/9702001}

\lref\kachru{S. Kachru, A. Klemm, W. Lerche, P. Mayr, C. Vafa,
hep-th/9508155\semi S.~Kachru, C.~Vafa, hep-th/9505105}

\lref\mmm{A.~Losev, G.~Moore, S.~Shatashvili, hep-th/9707250\semi
N.~Seiberg, hep-th/9705221}

\lref\cdef{H.~Ooguri, C.~Vafa, hep-th/0302109, hep-th/0303063\semi
N.~Seiberg, hep-th/0305248}

\lref\seibergback{N.~Seiberg, hep-th/0008013}

\lref\dbranes{J.~Polchinski, hep-th/9510017}

\chardef\tempcat=\the\catcode`\@ \catcode`\@=11
\def\cyracc{\def\u##1{\if \i##1\accent"24 i
\else \accent"24 ##1\fi }}
\newfam\cyrfam \font\tencyr=wncyr10
\def\cyr{\fam\cyrfam\tencyr\cyracc}


\Title{\vbox{\baselineskip 10pt \hbox{} \hbox{ITEP-TH-36/03}
\hbox{PUDM-2003}      \hbox{IHES-P/03/43} }  }{ \vbox{\vskip -30 true pt
\smallskip
   \centerline{SEIBERG-WITTEN THEORY}\bigskip \centerline{AND RANDOM PARTITIONS} \vskip2pt}}

\bigskip
\centerline{\authorfont{Nikita A. Nekrasov\footnote{$^{\dagger}$}{On leave of absence
from: ITEP, Moscow, 117259, Russia}$^{1}$, Andrei Okounkov$^{2}$}}
\bigskip\bigskip\centerline{\it $^{1}$ Institut des Hautes Etudes
Scientifiques, Bures-sur-Yvette F-91440 France} \centerline{\it $^2$
Department of Mathematics, Princeton University, Princeton NJ 08544 USA}
\bigskip \bigskip
\ndt

\abstractfont{We study ${\CN}=2$ supersymmetric four dimensional gauge theories, in a
certain ${\CN}=2$ supergravity background, called $\Omega$-background. The
partition function of the theory in the $\Omega$-background can be
calculated explicitly. We investigate various representations for this
partition function: a statistical sum over random partitions, a partition
function of the ensemble of random curves, a free fermion correlator.

These representations allow to derive rigorously the Seiberg-Witten
geometry, the curves, the differentials, and the prepotential.

We study pure ${\CN}=2$ theory, as well as the theory with matter
hypermultiplets in the fundamental or adjoint representations, and the five
dimensional theory compactified on a circle.}

\Date{June 2003}
\vfill\eject
\pageno=1
\centerline{\authorfont TABLE OF CONTENTS}\nobreak
{\bf     \medskip{\baselineskip=12pt\parskip=0pt\input NOI.tmp \bigbreak\bigskip}}


\newsec{INTRODUCTION}

\ndt Supersymmetric gauge theories are interesting theoretical laboratories.
They are rich enough to exhibit most of the quantum field theory phenomena,
yet they are rigid enough to contain a lot of exactly calculable information
\nsvz\seibergone\sw\swsol. They embed easily into string theory, and provide
an exciting arena for the search of string/gauge dualities
\warner\vafaengine\wittensolution\dv.

In the past year a lot of progress has been made in understanding some of
this rich structure using direct field theoretic techniques (\swi\lmn\ and
refs therein), in the case of the theories with extended supersymmetry, and
in \dv\ in the case of ${\CN}=2$ susy broken down to ${\CN}=1$ by the
superpotential.

In particular, a connection between four dimensional gauge theories and two
dimensional conformal field theories seemed to appear. Some earlier
indications for such a connection were observed in the study of ${\CN}=4$
super-Yang-Mills theory \vafawitten\estring.

In this paper we shall make this connection more transparent in the case of
${\CN}=2$ theories.

We shall also derive, by purely field theoretic means,
via direct instanton calculus, the solution for the low-energy effective
theories, proposed by Seiberg and Witten in 1994 \witsei\ and further
generalized in \swsol\nikfive.

\subsec{Notations}\ndt Throughout the paper we denote the colour indices by
the lower case Latin letters $l,m,n = 1, \ldots , N$,
 the vevs of the Higgs field by
$${\ba} = {\rm diag}(a_1, \ldots, a_N) \ , \quad a = \frac{1}{N} \sum_l a_l,
\quad {\tilde a}_l = a_l - a $$
 the dual vevs by $$ {\xi} = {\rm
diag}({\xi}_1, \ldots, {\xi}_N). $$
The vector
$$
{\r}_l = \frac{1}{N}\left( l - \frac{N+1}{2} \right)
$$
will occur often.

The gauge group $G=U(N)$, the group of
gauge transformations on ${\bR}^4$ which extend smoothly to ${\bS}^4$ will
be denoted by ${\CG}$. Its normal subgroup ${\CG}_{\infty}$, which consists
of the gauge transformations, trivial at ${\infty}$, will play a special
r{\^o}le.

\subsec{Organization of the paper}

The section $2$ is addressed at physicists, who want quickly jump
onto the subject. It contains
previously unpublished details on the noncommutative regularizations of the
theories we consider, as well as systematic introduction into
$\Omega$-backgrounds. In this paper we shall not touch upon the recently
revived \cdef\ $C$-backgrounds of \bcov\gopakumarvafa, postponing the
explanations of the relations between $\Omega$ and $C$ to some future work.

Section $3$ starts setting up the stage for the mathematical problem and the
reviews the ingredients needed for its solution. We present the formula for the partition
function of the pure ${\CN}=2$ gauge theory in the $\Omega$-background, as a
sum over random partitions. We also begin to formulate the same sum in terms
of random paths, which arise as the boundaries of the Young diagrams,
representing the random partitions. Mathematically oriented reader can skip
the section $2$ and proceed directly to $3$ (we also recommend \booksSW\donagirev\ for orientation).

Section $4$ attacks the problem of the calculation of the prepotential.
Physically it has to do with the limit, where the $\Omega$-background
approaches the flat space. In the random path representation, this limit is
the quasiclassical limit, which can be evaluated using the saddle point
method. This is exactly the idea of our derivation. We extensively discuss
the equations on the minimizing path, and their solution. We find that the
solution is most simply described in terms of the Seiberg-Witten curve.

Section $5$ explains the fermionic representation of the partition function
in the special $\Omega$-background, preserving twice as much supersymmetries
compared to the generic $\Omega$-background. Even though the formalism of
free fermions is well-known to mathematicians, under the name of the
infinite wedge representations of ${\bg}{\bl}({\infty})$ algebra, we supply
the necessary details. We find that a certain transform of the partition
function can be written as a matrix element (current conformal block) of the
exponentials in the $\widehat{U(N)}$ currents on the sphere ${\bS}^2$.

Section $6$ begins our quest for generalizations. We discuss the softly
broken ${\CN}=4$ theory, i.e. ${\CN}=2$ gauge theory with the matter
hypermultiplet in the adjoint representation. As most of the steps are
similar to the pure gauge theory case we move faster, and write down the
expression for the partition function in terms of partitions, paths, and
chiral fermions, which this time live on elliptic curve, determined by the
microscopic gauge coupling. Again we perform the saddle point evaluation,
and find that the prepotential is encoded, as conjectured by Donagi and
Witten \witdonagi, in the spectral curves of the elliptic Calogero-Moser integrable system.

Section $7$ considers gauge theories with matter in the fundamental
representation, and the gauge theories with the tower of Kaluza-Klein
states, coming from the compactification of the five dimensional theory on a
circle.

Section $8$ presents our conclusions and the discussion of the unsolved
problems.

\subsec{Acknowledgments}

We thank A.~Braverman, R.~Kenyon, A.~Losev, G.~Moore, A.~Morozov and S.~Smirnov for useful discussions.
Research was
partly supported by NSF under the grant DMS-0100593 (AO), by the Packard Foundation (AO),
by {\cyr RFFI} under the grant
01-01-00549 (NN), by the grant 00-15-96557 (NN) for the scientific
schools, and by Clay Mathematical Institute (NN).

AO thanks Institut Henri Poincare for hospitality. NN thanks Institute for
Advanced Study, New High Energy Theory Center at Rutgers University, Kavli
Institute for Theoretical Physics at University of California Santa Barbara,
Forschungsinstitut f{\"u}r Mathematik, ETH Zurich, University of S{\~a}o Paulo,
and ICTP, Trieste,
for hospitality during preparation of the manuscript.

\ndt
{\bf Note added:}
H.~Nakajima informed us that he and K.~Yoshioka found another proof \hiraku\
of our main theorem of the section $\bf 4$, which relates the partition function, prepotential and the Seiberg-Witten
curves. They are
using the blowup techniques.

\newsec{${\CN}=2$ GAUGE THEORY, DEFORMATIONS AND BACKGROUNDS}

In this section we remind the construction of \swi\lmn\ and also provide
quite a few new details. We consider ${\CN}=2$ supersymmetric gauge theory
with the gauge group $U(N)$. We study Euclidean path integral with the fixed
vev of the adjoint Higgs field, on ${\bR}^4$ with certain nonminimal
couplings. Most of the interesting dynamics in such theories is associated
with instanton \bpst\ effects \coleman.

\subsec{Lagrangian, fields, couplings}

The simplest way to write down the action of the four dimensional
super-Yang-Mills theory with extended supersymmetry is to use dimensional
reduction of the higher dimensional minimal supersymmetric theory.

In particular, starting with the six dimensional ${\CN}=1$ SYM we arrive at:

\eqn\phsym{\eqalign{ L_{flat} = & \frac1{4g_0^2}\int \ \sqrt{G} d^4 x \ {\Tr}
\biggl\{ -F_{IJ}F^{IJ} - 2 D_{I} {\phi} D^I {\bar \phi} - [{\phi},
{\bar\phi}]^2 \cr &\qquad\qquad - i {\bar\l}^{\dot\a}_{\bi}
{\s}_{\a\dot\a}^{I}D_{I} {\l}^{\a\bi} + \frac{i}{2} \left( {\phi}
{\e}^{\bi \bj} [{\bar\l}_{{\dot \a}\bi}, {\bar\l}^{\dot \a}_{\bj}] - {\bar
\phi} {\e}_{\bi\bj} [{\l}^{{\a}\bi}, {\l}_{\a}^{\bj}] \right) \biggr\} + \cr &
{{\vartheta}_0\over 2\pi} \int {\Tr} F \wedge F \cr}} Here, $A_I,  {\phi},
{\bar\phi}$ are the components of the six dimensional gauge field, which decompose as
the four dimensional gauge field and an adjoint complex Higgs field (or two
real Higgs fields); ${\l}_{\a \bi}$ are ${\CN}=2$ gluions
--
 the pair of
the four dimensional Weyl spinors, transforming in the adjoint
representation of the gauge group.

The indices ${\a}, {\b}=1,2$ correspond to the doublet of the $SU(2)_L$,
$\dot\a, \dot\b = 1,2$ is that for $SU(2)_R$, while ${\bi}, {\bj}=1,2$ are
the internal indices, which reflect the $SU(2)_I$ R-symmetry of the theory.
They are raised and lowered using the $SU(2)$ invariant tensor
${\ve}_{12} = - {\ve}_{21} = 1 = {\ve}^{21} = - {\ve}^{12}$.
Space-time Lorentz indices will be denoted throughout the paper by the upper
case Latin letters $I,J, .. = 1, 2, 3, 4$. The Pauli tensors, relating the
spinor and vector indices, are ${\s}_{I\a\dot \a}$.

In \phsym\ we have put the bare coupling constant $g_0$ and the bare theta
angle ${\vartheta}_0$. The bare coupling corresponds to some high energy
cut-off scale $\m$.

The action \phsym\ is the limit of the six dimensional action of the theory
put on a six manifold ${\bT}^2 \times {\bR}^4$ with the standard flat
product metric.

\subsec{$\Omega$-background}

However, in going from six to four dimensions one may have started with a
nontrivial six dimensional metric. In particular, by reducing on the two
torus one may have considered ${\bR}^4$ bundles with nontrivial flat $SO(4)$ connections,
such as the space ${\CN}_6$ with the metric: \eqn\twmtrc{ds^2  = A dzd{\zb} + g_{IJ} \left( dx^I + V^I dz +
{\bar V}^I d{\zb} \right)\left( dx^J + V^J dz + {\bar V}^J d{\zb} \right)}
with $V^I = {\Omega}^{I}_{J} x^J, \ {\bar V}^I = {\bar\Omega}^{I}_{J} x^J$, and the
area
$A$ of the torus to be sent to zero.
For $[\Omega, \bar\Omega ] =0$ the metric \twmtrc\ is flat.

However, for $\Omega \neq 0$ the background \twmtrc\ will break all
supersymmetries. Indeed, the spinors ${\e}_{\a}^{\bi}, {\bar\e}_{{\dot \a}\bi}$
generating supersymmetries of the \phsym\ are the components of the six
dimensional Weyl spinor. In order to generate the symmetry of the theory on
a curved background the spinor must be covariantly constant. In our case
this means that the spinor, viewed as four dimensional Weyl spinor,
 should be invariant under the holonomies around the two cycles of the
two-torus, which is possible only for discrete choices of
${\Omega},{\bar\Omega}$.

\subsubsec{$\Omega$-background in the physical formalism} Fortunately there exists a
continuous deformation of the ${\CN}=2$ theory preserving some fermionic symmetry (which also deforms along the way).
The trick is to use the R-symmetry, which is manifest in the four
dimensional theory --- the $SU(2)$ group, which acts on the internal index
${\bi}=1,2$ of the gluinos ${\l}$.

Namely, in addition to the nontrivial metric \twmtrc\ we turn on a Wilson
loop in the R-symmetry group, which should compensate some part of the
metric induced holonomy on the spinors.

As a result, in the limit $A \to 0$ with $\Omega, \bar\Omega$ fixed the
action \phsym\ gets extra terms: \eqn\omgt{{\Delta} L = {\Omega}^{I}_{J}
L_{I}^{(1) \ J} + {\bar\Omega}^{I}_{J}  {\bar L}_{I}^{(1) \ J} +
{\Omega}^{I}_{J} {\bar \Omega}^{K}_{L} L_{IK}^{(2) \ JL}} where
\eqn\dfmrs{L_{I}^{(1) \ J} = \int d^4 x \ \sqrt{G} \left(  x^J \left(  {\Tr}
F_{IK} D^{K} {\bar\Phi} + {\e}^{\bi\bj} {\Tr} {\bar\l}_{{\dot \a}\bi} D_{I}
{\bar\l}^{\dot \a}_{\bj} \right) + G^{JK} {\bar\s}_{IK}^{\bi\bj} {\Tr}
[{\bar\l}_{{\dot \a}\bi}, {\bar\l}^{\dot \a}_{\bj}] \right)} where the
tensor ${\bar\s}^{\bi\bj}_{IK} = - {\bar\s}^{\bi\bj}_{KI} =
{\bar\s}^{\bj\bi}_{IK} = {\sfrac{1}{2}}{\ve}_{IKJL} {\bar\s}^{\bi\bj\ JL}$
is the 't Hooft projector: ${\bf so}(4) = {\bf su}_L (2)\oplus {\bf su}_R
(2) \to {\bf su}_R (2) = \left( ({\half})\otimes({\half})\right)_{sym}$:
$${\s}_{IJ\dot \a}^{\dot \b} = \frac1{4} \left( {\s}^{\b\dot \b}_{I}
{\s}_{J\b \dot a} -  {\s}^{\b\dot \b}_{J} {\s}_{I\b \dot a} \right)$$ and,
for future use: $${\s}_{IJ\a}^{\b} = \frac1{4} \left( {\s}_{I\a\dot
\a}{\s}_{J}^{\b\dot \a} - {\s}_{J\a\dot \a}{\s}_{I}^{\b\dot \a} \right)$$
Finally, \eqn\dfmrt{L^{(2) JL}_{IK} = \int d^4 x \ \sqrt{G} \ x^J x^L \
G^{MN} {\Tr} F_{IM} F_{KN} } \subsubsec{${\Omega}$-background in the twisted
formalism}

Supersymmetric gauge theories with extended supersymmetry can be formulated
in a way which guarantees the existence of a nilpotent symmetry in any
curved background. This formulation, sometimes called twisted, or
topological, or cohomological, makes use of the R-symmetry of the theory.
The coupling to the curved metric is accompanied by the coupling to the
R-symmetry gauge field, which is taken to be equal to the corresponding
projection of the spin connection\witdonaldson\wittenone. In this way, the
fermions of the pure gauge theory become a one-form $\psi$, a self-dual
two-form $\chi$, and a scalar $\eta$. The bosons $A_I$ and $\phi$ are not
sensitive to the twisting.

The advantage of such a formulation of the theory is the clear geometric
meaning of all the terms in the action. It is well-known, that ${\CN}=2$
super-Yang-Mills in the twisted formulation provides an integral
representation for the Donaldson invariants of four-manifolds (for
$G=SO(3)$). In our study the gauge theory lives on ${\bR}^4$ which is boring
topologically. However, we study something different from the Donaldson
theory, due to $\Omega$-background. It corresponds to the $K$-equivariant
version of Donaldson invariants, $K = Spin(4)$ being the group of rotations.

In general, if the metric $G  = G_{IJ}dx^I dx^J$ of the Euclidean space-time
manifold has some isometries, one can deform the standard Donaldson-Witten
\witdonaldson\aj\ action, by coupling it to the isometry vector fields $V$
and $\bar V$, which should commute $[V, {\bar V}] =0$. Explicitly, the
action of the theory is given by:
\eqn\symac{\eqalign{L = & {1\over{2g_0^2}} \left( -  {\half}{\Tr} F \star F
+ {\Tr} \left( D_{A} {\phi} - \iota_{V} F \right) \star \left( D_{A}
{\bar\phi} - \iota_{\bar V} F \right) + {\half}{\Tr} [ {\phi}, {\bar\phi} ]^2 \
{\rm vol}_{g} \right) \cr + & {\Tr} \left( {\chi} \left( D_{A}
{\psi}\right)^{+} + {\eta} D^{*}_{A} {\psi} + {\chi} \star L_{V} {\chi} +
{\eta} \wedge\star L_{V} {\eta} + {\psi} \star L_{\bar V} {\psi} \right) \cr
+ & {\Tr} \left( {\chi} \star [ {\phi}, {\chi} ] + {\eta} \star [ {\phi},
{\eta} ] + {\psi}\star [{\bar\phi}, {\psi} ] \right) \cr + &
{{\vartheta}_{0} \over {2\pi}} {\Tr} F \wedge F \cr}} On ${\bR}^4$, we take,
as above, \eqn\vfld{V^{I} = {\Omega}^{I}_{J} x^{J}, \qquad {\bar V}^I =
{\bar\Omega}^{I}_{J} x^{J} } with \eqn\omgs{ {\Omega}^{IJ} = \pmatrix{ 0 &
{\e}_1 & 0 & 0 \cr  -{\e}_1 & 0 &  0 & 0 \cr 0 & 0 & 0 & {\e}_2 \cr  0 & 0 &
-{\e}_2 & 0 \cr}, \quad {\bar\Omega}^{IJ} = \pmatrix{ 0 & {\bar\e}_1 & 0 & 0
\cr  -{\bar\e}_1 & 0 &  0 & 0 \cr 0 & 0 & 0 & {\bar\e}_2 \cr  0 & 0 &
-{\bar\e}_2 & 0 \cr}} where ${\Omega}^{IJ} = G^{JK}{\Omega}^{I}_{K}$ etc.

\subsec{On supersymmetry}

For completeness, we list here the formulae for the supersymmetries of the
actions \phsym\omgt\symac. \subsubsec{Supersymmetry of the physical theory}

The supercharges transform in the representation $({\bf 2}, {\bf 1}, {\bf
2}) \oplus ({\bf 1}, {\bf 2}, {\bf 2})$ of the Lorentz $\times$ R-symmetry
group $SU(2)_L \times SU(2)_R \times SU(2)_I$. Introducing the corresponding
infinitesimal parameters: ${\z}_{\a}^{\bi} $ and ${\bar\z}_{\dot \a}^{\bi}$ the
supersymmetry transformations can be written \eqn\symphsym{\eqalign{ &
{\d}A_{I}  = - i {\bar\l}^{\dot \a}_{\bi} {\s}_{I\a\dot \a} {\z}^{\a \bi} -
i {\l}^{\a \bi} {\s}_{I\a\dot \a} {\bar\z}^{\dot \a}_{\bi}\cr & {\d}
{\l}_{\a \bi } = {\s}^{IJ}_{\a\b} {\z}^{\b \bi }F_{IJ} + i {\z}^{\bi}_{\a} D
+ i \sqrt{2} {\s}^{I}_{\a\dot \a} D_{I}B
{\e}^{\bi\bj}{\bar\z}^{\dot\a}_{\bj}\cr & {\d} {\bar\l}_{\dot \a}^{\bi}  =
{\s}^{IJ}_{\dot\a\dot\b} {\bar\z}^{\dot \b}_{\bi}F_{IJ} - i {\bar\z}_{\dot
\a \bi} D + i \sqrt{2} {\s}^{I}_{\a\dot \a} D_{I}{\bar B}
{\e}_{\bi\bj}{\z}^{\a \bj}\cr & {\d} B = \sqrt{2}{\z}^{\a\bi}{\l}_{\a\bi}\cr
&{\d}{\bar B} = \sqrt{2}{\bar\z}^{\dot\a}_{\bi}{\bar\l}^{\bi}_{\dot\a} \,
\cr} } with the auxiliary field $D = [B, {\bar B}]$.

\subsubsec{Twisted superalgebra}

The twisted formulation of the theory is achieved by replacing the Lorentz
group $K = SU(2)_L \times SU(2)_R \in  K \times SU(2)_I$ by another subgroup
of $K \times SU(2)_I$, namely $SU(2)_L \times SU(2)_d$, with $SU(2)_d$ being
diagonally embedded into $SU(2)_R \times SU(2)_I$. In other words, the
internal index $\bi$ is identified with another $SU(2)_R$ index,  $\dot\a$.
The fields are redefined, according to \eqn\rfdn{\psi_I = {\l}_{\a \dot
\b}{\s}_I^{\a\dot \b}, \quad \chi_{IJ} = {\s}_{IJ \dot\a
\dot\b}{\bar\l}^{\dot\a \dot\b}, \quad \eta =
{\e}_{\dot\a\dot\b}{\bar\l}^{\dot\a\dot\b}} Similarly, the supersymmetry
parameters ${\z}_{\a}^{\bi}$ become ${\z}^{I}$, ${\bar\z}_{\dot\a \bi}$
become ${\z}^{IJ}$ and ${\z}$. Of course, ${\z}^{IJ}$ is self-dual.

The supersymmetry algebra becomes: \eqn\symtwsym{\eqalign{ & {\d}A_{I}  = -
i {\z}{\psi}_{I} - i{\z}_{IJ} {\psi}^J - i  {\z}_I {\eta}\cr & {\d}
{\psi}_{I} = + {\z}^J F_{IJ}^{-} + i {\z}_I D + i {\z} D_I{\phi}\cr & {\d}
{\chi}_{IJ}  = {\z} F_{IJ}^{+}  - i {\z}_{IJ} D + i \left( {\z}_{[I}
D_{J]}{\bar \phi} \right)^{+}\cr & {\d}{\eta} = {\z}^{IJ} F_{IJ} - i {\z}D +
{\z}^I D_I {\bar \phi} \cr & {\d} \phi = {\z}^I {\psi}_I \cr &{\d}{\bar
\phi} = {\z}{\eta} + {\z}^{IJ}{\chi}_{IJ} \ . \cr} }

The geometric meaning of the transformations ${\d}$ is the following. The
space of the fields $A_I, {\phi}, {\psi}_I$, represents the
${\CG}$-equivariant de Rham complex of the space ${\CA}$ of gauge fields on
${\bR}^4$, together with the ingredients needed to construct a
Mathai-Quillen representative of the Euler class of a certain infinite
dimensional bundle over ${\CA}$, and the projection form\tdgt\ associated
with the projection ${\CA} \to {\CA}/{\CG}$.

The space ${\bR}^4$ is hyperk\"ahler, and possesses an action of the group
${\bR}^4$ of translations.

The transformation generated by ${\z}$ is the ${\CG}$-equivariant de Rham
differential. The transformations, generated by ${\z}^{IJ}$ correspond to
the ${\CG}$-equivariant ${\pb}_{{\CI},{\CJ},{\CK}}$ differentials,
corresponding to the three complex structures on ${\CA}$ induced from the
complex structures on ${\bR}^4$. Finally, ${\z}^I$ correspond to the
operators ${\iota}_{{\p}_I}$ of contraction with the vector fields on
${\CA}$, induced by the vector fields $$\frac{\p}{\p x^I}$$ generating
translations of ${\bR}^4$.

\subsubsec{Supersymmetry of the $\Omega$-background}

The transformations, which generate the symmetry of the $\Omega$-background
utilize the rotational symmetries of ${\bR}^4$. To the vector field $$ V =
{\Omega}^I_J x^J \frac{\p}{\p x^I} $$ there is an associated vector field on
${\CA}$ and the associated operation of contraction on the
${\CG}$-equivariant de Rham complex. In terms of the transformations ${\d}$
of twisted theory these are simply the transformation \symtwsym\ with the space-dependent transformation parameter $
{\z}^I = V^I (x) = {\Omega}^I_J x^J $.

The theory \omgt\ in invariant under the transformation, generated by: $$ (
{\z}, {\z}^I, {\z}^{IJ} ) = ( {\z}, V^{I}(x){\z}, 0 ) $$ The supercharge,
generating this transformation, will be denoted as ${\tilde Q}$.
\subsec{Noncommutative deformation}

\ndt The theory \phsym\ allows yet another deformation, which we shall
implicitly use to simplify our calculations. Let ${\Theta}^{IJ} = -
{\Theta}^{JI}$ be a constant Poisson tensor on ${\bR}^4$. The theory on
${\bR}^4$ can be deformed to that on the noncommutative space,
${\bR}^4_{\Theta}$. A naive way to define this deformation is to replace all
the products of functions (or components of various tensor fields) in
\phsym\ by the so-called Moyal product: \eqn\myl{f \star g \ (x)= {\exp}
\frac{i}{2} {\Theta}^{IJ} \frac{\p}{\p \xi^I} \frac{\p}{\p \eta^J}
\Biggl\vert_{\eta=\xi=0} \ f( x+\xi) g( x +\eta)} More conceptual definition
goes as follows (cf. \ikkt\seibergback). Consider the theory \phsym\
dimensionally reduced to zero space-time dimensions. We get some sort of
supersymmetric matrix model. Replace the matrices by the operators in the
Hilbert space ${\CH}$. Explicitly, we get the theory of six bosonic
operators \eqn\bsnc{X^I,  \quad {\Phi}, {\bar\Phi}; \qquad I = 1, \ldots, 4} and
eight fermionic ones (we use twisted formulation): \eqn\frmns{{\Psi}^I, \quad {\eta}, \quad  {\chi}^{IJ} = \half {\varepsilon}^{IJKL}
{\chi}_{KL}; \qquad\qquad I, J
= 1, \ldots, 4} where we lower indices using some Euclidean metric $g_{IJ}$:
$$\chi_{KL} = g_{KI}g_{LJ} {\chi}^{IJ} , $$ The action reads:
\eqn\act{\eqalign{L = -{1\over 4{\bg}^2} & {\Tr}_{\CH} \Biggl( g_{IK} g_{JL}
\left(  [ X^I, X^J] [X^K, X^L] +  {\chi}^{IJ} [X^K, {\Psi}^{L}] +
{\chi}^{IJ} [{\Phi}, {\chi}^{KL}] \right)\cr & \qquad + 2 g_{IJ} \left(
{\eta} [X^I, {\Psi}^J] + [{\bar\Phi}, X^I] [{\Phi}, X^J] + {\Psi}^I [
{\bar\Phi}, {\Psi}^J] \right) \cr & \qquad + \left( [\Phi, {\bar\Phi}]^2 +
{\eta} [ {\Phi}, {\eta}] \right)\Biggr) \cr & \quad + {\Tr}_{\CH} \Biggl( i
B_{IJ} [X^I, X^J] + \frac{{\vartheta}}{8\pi} \ {\varepsilon}_{IJKL} [X^I,
X^J][X^K,X^L] + L_0 \Biggr)\cr}} The action \act\ leads to the equations of
motion which do not depend on the parameters entering the last line in \act,
i.e. $B_{IJ}, {\vartheta}_0, L_0$. \mdp A special class of extrema of the
action \act\ with all the fermionic fields set to zero is achieved on the
operators $X^I, \Phi, \bar\Phi$ obeying: \eqn\absmn{\eqalign{& [X^I, X^J] =
i {\Theta}^{IJ} \ {\bI}, \qquad {\Theta}^{IJ} = 2g_0^2 \ g^{IK} g^{JL}
B_{KL} \cr & [X^I, {\Phi} ] = 0 = [X^I, {\bar \Phi}] = [{\Phi}, {\bar\Phi}]
\cr}} Let us fix a standard set of operators ${\bx}^I$ in ${\CH}$  obeying
\eqn\hsnbrg{[ {\bx}^I, {\bx}^J ] = i {\Theta}^{IJ} \ {\bf 1}} The algebra
\hsnbrg\ may be represented reducibly in $\CH$. In general, for
nondegenerate ${\Theta}^{IJ}$, \eqn\wspc{{\CH} = {\CH}_{0} \otimes W} where
${\CH}_0$ is irreducible representation of \hsnbrg, and $W$ is a
multiplicity space, which we shall assume to be a finite dimensional
Hermitian vector space, of complex dimension $N$. From now on, ${\bx}^I$
will denote the operators on ${\CH}_0$. The corresponding operators on $\CH$
will be denoted as ${\bx}^I \otimes {\bf 1}_{N}$. Expanding: \eqn\ggf{X^I =
{\bx}^I \otimes {\bf 1}_{N} + i {\Theta}^{IJ} A_{J}({\bx}), \ {\Phi}  =
{\phi}({\bx}) = {\bf 1} \otimes {\ba} + {\phi}_{\infty}({\bx}) \ , {\Psi}^I
= {\Theta}^{IJ} {\psi}_J ({\bx})} where ${\phi}_{\infty} ( x ) \to 0 , x
\to \infty$, we arrive at the naive formulation, with the metric on
${\bR}^4$, given by: \eqn\nwmtrc{\eqalign{& G^{IJ} = g_{KL} {\Theta}^{IK}
{\Theta}^{JL} \cr & {\vartheta}_0 = {\vartheta} \ {\rm Pf}({\Theta}) \cr &
\frac1{g_0^2} = \frac1{{\bg}^2} \ {\rm Pf}({\Theta}) \sqrt{{\det}g} \cr}} In
\ggf, both $A_J = A_{J|ln} \otimes E^{ln}$ and $\phi$ are valued in the $N
\times N$ matrices, anti-Hermitian operators in $W$. To arrive at the
ordinary gauge theory, we should take the limit ${\Theta} \to 0$, while
keeping $G_{IJ}, g_0^2, {\vartheta}_0$ finite. The curvature $F$ of the
gauge field $A$ is  given by: \eqn\crvt{[X^I, X^J] = i {\Theta}^{IJ} +
{\Theta}^{IK} {\Theta}^{JL} F_{KL}} \subsubsec{Combining $\Omega$ and $\Theta$}
The universal gauge theory \act\ can also be subject to the nontrivial
$\Omega$-background: \eqn\acti{\eqalign{ L = -{1\over 4{\bg}^2} \
{\Tr}_{\CH} & \Biggl( \ g_{IK} g_{JL} \Biggl(  \ [ X^I, X^J] [X^K, X^L] +
\cr & \qquad\qquad\qquad \qquad+ {\chi}^{IJ} [X^K, {\Psi}^{L}] + {\chi}^{IJ}
\left( [{\Phi}, {\chi}^{KL}] + 2 {\Omega}^{K}_{M} \chi^{ML} \right)
\Biggr)\cr & \qquad + 2 g_{IJ} \Biggl(
 \biggl([{\bar\Phi}, X^I] + {\bar\Omega}^{I}_{L} X^{L} \biggr)
\biggl( [{\Phi}, X^J] + {\Omega}^J_{K} X^K \biggr)+ \cr & \qquad\qquad\qquad
\qquad + {\eta} [X^I, {\Psi}^J ] + {\Psi}^I \left( [ {\bar\Phi}, {\Psi}^J] +
{\bar\Omega}^{J}_{L} {\Psi}^L \right) \Biggr) \cr & \qquad\qquad\qquad +
\left( [\Phi, {\bar\Phi}]^2 + {\eta}  [ {\Phi}, {\eta}] \right)\Biggr) \cr +
{\Tr}_{\CH} & \Biggl( i B_{IJ} [X^I, X^J] + \frac{{\vartheta}}{8\pi} \
{\varepsilon}_{IJKL} [X^I, X^J][X^K,X^L] + L_0 \Biggr)\cr}} The vacua of the
theory \acti\ are given by the operators which solve slightly different
equations then \absmn. Namely, the condition on $\Phi$ is now:
\eqn\absmnn{[X^I , {\Phi} ] = {\Omega}^I_J X^J, \quad [X^I , {\bar\Phi} ] =
{\bar\Omega}^I_J X^J} which is consistent with $[X^I, X^J] = i
{\Theta}^{IJ}$ only under certain conditions on ${\Omega}, {\Theta},
{\bar\Omega}$. Namely, assuming nondegeneracy of $\Theta$, with ${\o} =
{\Theta}^{-1}$, the matrices \eqn\hmls{{\CE}_{IJ} = {\Omega}^K_J {\o}_{KI},
\quad {\bar\CE}_{IJ} = {\bar\Omega}^K_J {\o}_{KI}, \qquad } must be
symmetric. Then the solution to \absmn\ is given by: \eqn\vca{X^I = {\bx}^I,
\quad {\Phi} = {\half} {\CE}_{IJ} {\bx}^I {\bx}^J, \quad {\bar\Phi} =
{\half} {\bar \CE}_{IJ} {\bx}^I {\bx}^J } Now, the gauge field $A_I$ and the
Higgs field $\phi$ are introduced via: \eqn\ggfi{X^I = {\bx}^I \otimes {\bf
1}_{N} + i {\Theta}^{IJ} A_{J}({\bx}), \ {\Phi}  = {\half} {\CE}_{IJ} X^I
X^J \otimes {\bf 1}_N + {\phi}({\bx}) }

\subsubsec{Supersymmetry of the noncommutative theory in $\Omega$ background}

\def\ddt{{\tilde Q}}
We shall only write down the supercharge of our interest:
\eqn\ssyiii{\eqalign{ {\ddt}X^I = {\Psi}^I \ , & \ {\ddt} {\Psi}^I = [
{\Phi}, X^I] + {\Omega}^I_{J} X^J \cr
 {\ddt} {\Phi}  = 0 \ , & \quad {\ddt} {\chi}^{IJ} = H^{IJ} \cr
{\ddt} H^{IJ} = & \ [ {\Phi}, {\chi}^{IJ} ]  - \left( {\Omega}^{I}_K
{\chi}^{JK} - {\Omega}^{J}_{K} {\chi}^{IK} \right)^{+}\cr
 {\ddt} {\bar\Phi} = \eta ,
\qquad & {\ddt} \eta = [ {\Phi}, {\bar\Phi}] \cr}} Here $H^{IJ}$ is an
auxiliary field, which is equal to $[X^I, X^J]^{+}$ on-shell.

\subsubsec{Observables in the supersymmetric gauge theory: $\Omega = 0$}

In the ordinary ${\CN}=2$ supersymmetric gauge  theory a special class of
observables play an important role. They are distinguished by the property,
 that
a certain supercharge annihilates them. Of particular interest for
application to Donaldson theory are the observables which are constructed
out of the invariant polynomials in the Higgs field $\phi$ by means of the
descend procedure. In terms of the twisted fields, these observables look as
follows. Let $P({\phi})$ be any $G$-invariant polynomial on the Lie algebra
${\bg} = Lie(G)$ of the gauge group $G$. Then: \eqn\obsrv{\eqalign{&
{\CO}_{P}^{(0)}(x) = P({\phi}(x)) \cr & {\CO}^{(1)}_{P}(C) = \oint_{C}
\frac{{\p}P}{{\p}{\phi}^a} {\psi}^a \cr & {\CO}^{(2)}_{P}({\Sigma}) =
\int_{\Sigma}  \frac{{\p}P}{{\p}{\phi}^a} F^a + \half
 \frac{{\p}^2 P}{{\p}{\phi}^a{\p}{\phi}^b} {\psi}^a \wedge {\psi}^b \cr
& \vdots \cr & {\CO}^{(4)}_{P} (X) = \int_{X} \half
 \frac{{\p}^2 P}{{\p}{\phi}^a{\p}{\phi}^b} F^a \wedge F^b + \ldots +
\frac{1}{24}
 \frac{{\p}^4 P}{{\p}{\phi}^a{\p}{\phi}^b{\p}{\phi}^c {\p}{\phi}^d}
{\psi}^a \wedge {\psi}^b \wedge {\psi}^c \wedge {\psi}^d \cr}} Here, $x, C,
\Sigma, \ldots, X$ represent a $0$, $1$, $2$, ..,$4$-cycles in the
space-time manifold, respectively. The main idea behind \obsrv\ is to use
the fact that the supercharge $Q$ acts on $P({\Phi}(x) + \ldots )$ as  the
de Rham differential. To compare the observables of the ordinary gauge
theory to those of the noncommutative gauge theory we shall utilize the
generating function(form): \eqn\gnobs{{\CO}_{P}  = P\left({\phi} (x) +
{\psi}_{I}(x) dx^I + \half F_{IJ}(x) dx^I \wedge dx^J \right) \in
{\Omega}^{*}( {\rm spacetime})} Its main property is: $$ d_x \langle
{\CO}_{P}(x) \ldots \rangle = 0 $$ \mdp The noncommutative gauge symmetry
does not allow, naively, for the local observables like in the first line of
\obsrv. However, it has as many gauge invariant observables as does the
ordinary theory. For the gauge group $U(N)$ the invariant polynomials can be
expressed as polynomials in the single trace operators \eqn\snglt{P_{n}
({\phi}) =  {\Tr} {\phi}^n} It is convenient to introduce the character:
\eqn\rslvntr{P_{\b}({\phi}) = {\Tr} e^{i{\b}{\phi}} =\sum_{n=0}^{\infty}
\frac{(i\b)^n}{n!} P_{n}} The analogue of \obsrv\gnobs\ is the following
closed form on ${\bR}^{4}$: \eqn\obsrvi{{\CO}_{P_{\b}} (x, dx) = {\b}^2
e^{-{\b\o}}\int d^4{\vartheta} d^4{\k} \ e^{- i{\b} ( {\k}_I x^I +
{\vartheta}_I dx^I)} \ {\Tr}_{\CH}  e^{i {\b} {\Phi} ({\vartheta},
{\kappa})}} where \eqn\sfld{{\Phi}({\vartheta}, {\kappa}) = {\Phi} +
{\kappa}_I X^I + {\vartheta}_I {\Psi}^I + {\half}{\vartheta}_I {\vartheta}_J
[X^I, X^J], \qquad {\o} = \half{\o}_{IJ} dx^I \wedge dx^J} The closedness of
${\CO}_{P_{\b}}$ is proved with the help of the symmetry acting on
${\vartheta}, {\kappa}$: \eqn\ssdr{ {\d} {\vartheta} = {\kappa}, \qquad
{\d}{\kappa} = 0 } \subsubsec{Observables: ${\Omega} \neq 0$}

The chiral observables get deformed when $\Omega \neq 0$.  First of all, in
the ordinary gauge theory, the generating form ${\CO}_{P}$ becomes
equivariantly closed under correlator: \eqn\eqvcl{\left( d + \iota_{V}
\right) {\CO}_{P} = 0} In the noncommutative gauge theory we get the same
statement, but now we need to modify the symmetry ${\d}$ to its equivariant
analogue: \eqn\eqvss{ {\d} {\vartheta}_I = {\kappa}_I, \qquad {\d}{\kappa}_I
= - {\Omega}_{I}^{J} {\vartheta}_{J} } and change \obsrvi\ to:
\eqn\obsrvii{\eqalign{& {\CO}_{P_{\b}; {\Omega}} =
\prod_{{\a}=1,2}\frac{e^{{\b}{\e}_{\a}}-1}{{\e}_{\a}} \
e^{{\b}{\o}_{\Omega}} \int d^4{\vartheta} d^4{\k} \ e^{- i{\b} ( {\k}_I x^I
+ {\vartheta}_I dx^I)} \ {\Tr}_{\CH}  e^{i {\b} {\Phi} ({\vartheta},
{\kappa})}\cr &\cr & \qquad {\o}_{\Omega} = \half \left({\o}_{IJ} dx^I
\wedge dx^J - {\CE}_{IJ} x^I x^J \right)\cr}} In what follows we consider
either $0$-observable, or integrated $4$-observable. In the first case, we
set $x=0$ (to be at the fixed point), and get the integral over the
$\vartheta, \kappa$ of ${\Tr}_{\CH} e^{i {\b}{\Phi}({\vartheta},
{\kappa})}$. In the second case, we get the integral over the $\vartheta,
\kappa$ of ${\Tr}_{\CH} e^{i {\b}{\Phi}({\vartheta}, {\kappa})}$ with an
extra ${\d}$-invariant Gaussian factor. Actually, \eqvss\ implies that the
integral over $\vartheta, \kappa$ is localized at $\kappa = \vartheta = 0$,
so that both $0$- and $4$-observables are equal, up to a
${\e}_{\a}$-dependent factor: \eqn\obsrviii{\eqalign{& {\CO}_{P_{\b};
{\Omega}}^{(0)} = \prod_{{\a}=1,2} \left( e^{{\b}{\e}_{\a}}-1\right) \
{\Tr}_{\CH} e^{i {\b}{\Phi}}\cr & \int_{{\bR}^4} {\CO}_{P_{\b};
{\Omega}}^{(4)} = \prod_{{\a}=1,2}\frac{e^{{\b}{\e}_{\a}}-1}{{\e}_{\a}} \
{\Tr}_{\CH} e^{i {\b}{\Phi}}\cr}}
The observables \obsrviii\ are most natural in the five dimensional gauge
theory compactified on the circle of the circumference ${\b}$. The
periodicity of ${\CO}_{P_{\b}}$ in ${\Phi}$ has there a simple origin --
large gauge transformations \nikfive. We shall return to this theory in the
section $\bf 7$.

\newsec{GAUGE THEORY PARTITION FUNCTION}

\ndt In the ordinary gauge theory the natural object of study is the
Euclidean path integral over the field configurations, such that
\eqn\hgscnd{{\phi}(x) \to {\ba}, \qquad x \to \infty} We are aiming to
calculate the path integral in the $\Omega$-background, with the boundary
conditions \hgscnd\ (and, in fact, using noncommutative deformation):
\eqn\prtnf{Z ( {\ba}, {\e}_1, {\e}_2; {\Lambda} ) = \int DA D{\psi} D{\eta}
D{\chi} D{\Phi} D{\bar\Phi} \ e^{- \int_{{\bR}^4} \ L \ \sqrt{g} d^4 x}} In
the noncommutative gauge theory the condition \hgscnd\ is phrased
differently. Think of ${\Phi}$ as of the element of the Lie algebra of the
group ${\CG}_{\Theta} \approx U({\CH})$ of gauge transformations (note that
normally one takes ${\CG}_{\Theta} = PU({\CH})$). The identification ${\CH}
= {\bC}^{N} \otimes {\CH}_0$ allows to view ${\Phi}$ as an $N \times N$
matrix of operators in ${\CH}_0$. The condition \hgscnd\ is now stated as
follows: \eqn\hgsii{{\Phi} = {\ba} \otimes {\bf 1}_{{\CH}_0} + {\half} {\bf
1}_{N} \otimes {\CE}_{IJ}{\bx}^I {\bx}^J + {\vf}({\bx})} where ${\vf}(x) \to
0$ as $x \to \infty$, faster then any power of $x$. \mdp The integral
\prtnf\ needs an UV cutoff ${\m}$. The bare couplings $g_0$ and
${\vartheta}_0$ are renormalized, thus generating an effective scale
$\Lambda$, \eqn\rgf{ {\Lambda}^{2N} \sim {\m}^{2N} e^{-{8{\pi}^2
\over{g_0^2}} + 2\pi i {\vartheta}_0} }

\remark{On $U(1)$ factor. The gauge theory with the gauge group $U(N)$ has
an interesting pattern of the renormalization group flow. The ordinary gauge
theory has a decoupled $U(1)$ factor, which does not exhibit any
renormalization of its coupling, and interacting $SU(N)$ part, with the
famous phenomenon of asymptotic freedom, reflected in \rgf. When the theory
is deformed by ${\Theta}$, the
perturbative loop calculations are altered, in a $\Theta$-dependent way, as
it introduces, among other things, an energy scale. In particular, the
$U(1)$ factor coupling constant starts running, in the energy range $$ \mu
\gg {\Lambda} \gg \frac{1}{\mu \Theta} $$ where the formula \rgf\ with $N=1$
holds (this result is a simple generalization of \seibergminwala, which can
also be justified using \nsvz\neksch). The $U(1)^{N-1} \subset SU(N)$ gauge
couplings experience different renormalization group flow, as they are
affected by the loops of the charged W-bosons. In principle, we should
introduce two distinct low-energy scales, one for $SU(N)$ part of the gauge
group, another for $U(1)$: ${\Lambda}, {\Lambda}_0$. As we shall eventually
remove noncommutativity, it makes sense to keep $\Lambda_0 \approx \mu$. We
shall encounter the ambiguity related to the $U(1)$ factor in the analysis
of the Seiberg-Witten curves. In  the paper we  concentrate
on the $SU(N)$ dynamics, and mostly set $a=0$.} \remark{ The theory \phsym\ has an anomalous
$U(1)_{gh}$ symmetry ({\it gh} for ghost). Under this symmetry, the adjoint
Higgs field ${\Phi}$ has charge $+2$, ${\bar\Phi}$ has charge $-2$, the
fermions ${\psi}$ have charge $+1$, $\chi, \eta$ have charge $-1$. In the
background of the instanton of charge $k$ the path integral measure
transforms under the ghost $U(1)$ with the charge $- 4 k N$. The $\Omega$
deformation breaks $U(1)_{gh}$. It can be restored by assigning ${\Omega},
{\bar\Omega}$ the charges $+2, -2$, respectively.}

\subsec{Partition function as a sum over partitions}

The partition function $Z({\ba}, {\e}_1, {\e}_2; {\Lambda})$ can be
explicitly evaluated as a sum over instantons. Moreover, the
$\Omega$-background lifts instanton moduli, leaving only a finite number of
isolated
points on the appropriately compactified instanton moduli space as a full set of
supersymmetric minima of the action. The evaluation of \prtnf\ is then
reduced to the calculation of the ratios of the bosonic and fermionic
determinants near each critical point. These points are labeled by the
colored partitions.

Consequently, \prtnf\  is given by the sum over $N$-tuples of partitions
\swi\lmn (see appendix $\bf B$ for the notations and definitions related to
partitions). Explicitly: \eqn\gnpri{Z ({\ba}; {\e}_1, {\e}_2, {\Lambda})  =
Z^{pert}({\ba}; {\Lambda},{\e}_1, {\e}_2) \sum_{{\vec\bk}} {\Lambda}^{2N
|{\vec\bk}|} Z_{{\vec\bk}} ({\ba}; {\e}_1, {\e}_2)}where:
\eqn\gnpr{\eqalign{
  Z_{{\vec\bk}}  ({\ba}; {\e}_1, {\e}_2) & \qquad = \prod_{l,n; i, j}
\frac{a_{l}-a_{n} + {\e}_1 ( i - 1) + {\e}_2 ( - j)}{a_{l}-a_{n} + {\e}_1 (
i - {\tilde k}_{nj} - 1) + {\e}_2 (k_{li} -j)} \cr &\qquad =   \prod_{l,n;
i, j} \frac{a_{l}-a_{n} + {\e}_1 ( -i ) + {\e}_2 ( j-1)}{a_{l}-a_{n} +
{\e}_1 ( {\tilde k}_{lj} -i) + {\e}_2 (j - k_{ni}-1 )} \cr
 \cr
& \qquad = \frac1{{\e}_2^{2N|{\vec\bk}|}} \prod_{(l,i) \neq (n,j)}
\frac{{\Gamma}\left(k_{li}- k_{nj}+ {\n} (j-i+1 ) + b_{ln}\right)
{\Gamma}\left( {\n}(j-i)+b_{ln}\right)}{ {\Gamma}\left( k_{li}- k_{nj} +
{\n}(j-i) + b_{ln}\right) {\Gamma}\left( {\n}(j-i+1) + b_{ln}\right)} \ ,
\cr &\cr}} \eqn\bn{b_{ln} = \frac{a_l - a_n}{{\e}_2}, \qquad {\n} =
-\frac{{\e}_1}{{\e}_2}} and\eqn\gnprii{Z^{pert}({\ba}; {\e}_1, {\e}_2,
{\Lambda}) =
\exp \left(\sum_{l,n} {\g}_{{\e}_1, {\e}_2} (a_l
- a_n ; {\Lambda}) \right)} where the function ${\g}_{{\e}_1, {\e}_2}$ is defined
in the appendix $\bf A$
\remark{ The product over $i,j$ in \gnpr\ in
$Z_{\vec\bk}$ is infinite and needs precise definition. Here it is: Fix a
pair $(l,n)$. Consider all the factors in $Z_{\vec\bk}$ corresponding to
$l,n$. Split the set of indices $(i,j)$ into four groups:
\eqn\frgrps{\eqalign{& {\bZ}_{+}^2 = S_{++} \cup S_{+-} \cup S_{-+} \cup
S_{--} \cr & S_{++} = \{ (i,j) \vert \ 1 \leq i \leq {\ell}({\bk}_l), \ 1
\leq j \leq k_{n1}\} \cr
 & S_{+-} = \{ (i,j) \vert \ 1 \leq i \leq {\ell}({\bk}_l), \ k_{n1} < j \} \cr
& S_{-+} = \{ (i,j) \vert  \ {\ell}({\bk}_l) < i , \ 1 \leq j \leq k_{n1}\}
\cr & S_{--} = \{ (i,j) \vert  \ {\ell}({\bk}_l) < i , \ k_{n1} < j \} \cr}}
The set $S_{++}$ is finite, the set $S_{--}$ contributes $1$. The set
$S_{+-}$ contributes: \eqn\plmncntr{Z_{+-; \vec\bk}^{ln} =
\prod_{i=1}^{{\ell}({\bk}_l)} \prod_{j=k_{n1}+1}^{\infty} \frac{a_{ln}
+{\e}_1 ( i -1) + {\e}_2 (-j)} {a_{ln} +{\e}_1 ( i -1) + {\e}_2 (k_{li}-j)}
 := }$$\prod_{i=1}^{{\ell}({\bk}_l)} \prod_{j=1}^{k_{li}} \frac1{a_{ln} +
{\e}_1 ( i -1 ) + {\e}_2 ( k_{li} - k_{n1} - j )}$$ and similarly for
$S_{-+}$. See also \loggaz.} \mdp
The special case ${\e}_2 = - {\e}_1 = {\hbar}$, i.e. ${\n}=1$
deserves a special attention. In this case, the expression
for the
partition function simplifies to: \eqn\prtn{\eqalign{ Z ({\ba}; {\hbar},
{\Lambda} )  & = \sum_{{\vec\bk}} {\Lambda}^{2N\vert {\vec\bk} \vert }
Z_{{\vec\bk}} ({\ba}; {\hbar}) \cr Z_{\vec\bk} ({\ba}; {\hbar}) & =
Z^{pert} ({\ba};{\hbar}) \m_{\vec\bk}^2({\ba}, {\hbar})\cr
\m_{\vec\bk}^2({\ba}, {\hbar}) &= \prod_{(l,i) \neq (n,j)} \left( {{a_{l} -
a_{n} + {\hbar} ( k_{l,i}- k_{n,j} + j - i)}\over{a_{l} - a_{n} + {\hbar} (
j-i)}} \right) \cr Z^{pert} ( {\ba}; {\hbar})  &=  \exp \left(\sum_{l, n}
\zu{a_{l} - a_{n}} \right) \cr}}
where the function $\zu{x}$ is defined in the appendix
{\bf A}.

\subsec{Plancherel measure on partitions}

In the case when $N=1$ and $\n=1$, the sum in \gnpri\ is over a
single partition ${\bk}$ and
the weight ${\m}_{\vec\bk}({\ba}, {\hbar})$ reduces to
\eqn\planch{{\m}({\bk}) = \prod_{i < j} \left( \frac{k_i - k_j + j - i}{j-i}
\right) = \prod_{\squarel \in {\bk}} \frac1{h(\square)}\,,}
where second the product is over all squares $\square$ in the diagram of
the partition $\bk$ and $h(\square)$ denotes the corresponding
hook-length.

The weight $\m(\bk)^2$ is known as the Plancherel measure
on partitions because of the relation
 \eqn\dmr{{\m}({\bk})  = \frac{{\dim}R_{\bk}}{\vert {\bk}
\vert!}\,,}
where $R_{\bk}$ is the irreducible representation of the symmetric
group corresponding to the partition ${\bk}$. Just like in
the classical Plancherel theorem, the
Fourier transform on the symmetric group an isometry of the
$L^2$-spaces with respect to the Haar (i.e.\ counting) measure on the
group and Plancherel measure on the set of its irreducible
representations.
Observe that from \dmr\ it follows that
\eqn\nrmls{\sum_{| {\bk} | = n} \ {\m}^2({\bk}) = \frac1{n!}}

The Plancherel measure is the most fundamental and
natural measure on the set of partitions. In many
aspects, the set of partitions equipped with the
Plancherel measure is the proper discretization of
the Gaussian Unitary Ensemble (GUE) of random matrices \gue. In
particular, the integrable structures of random
matrices are preserved and, in fact, become more
natural and transparent for the Plancherel measure, see \aoi\vnm.
Similarly, Plancherel random partitions play a
central role in the Gromov-Witten theory of
target curves, extending the role played by
random matrices in the case of the point target, see \op.

A pedestrian explanation of the relation between
Plancherel measure and GUE can be obtained by
rewriting the weight \planch\ as the product of the
Vandermonde determinant in the variables
$k_i-i$ and a multinomial coefficients, which, of course, is
the discrete analog of the Gaussian weight.

As we will see below in Section $\bf 5$, for $N>1$ and $\n=1$
the partition function is again related to the
Plancherel measure, but now periodically
weighted with period $N$. Finally, the case
$\n\ne 1$ leads to the Jack polynomial
analog of the Plancherel measure, see \kerovi.

\newsec{PREPOTENTIAL}

In this section we study the limit ${\e}_1, {\e}_2 \to 0$ of the partition
function \prtnf. In this limit, according to the field theory arguments
\swi\ the partition function must behave as: \eqn\swipr{Z({\ba};
{\e}_1,{\e}_2, {\Lambda}) = {\exp} \left(- \frac1{{\e}_1{\e}_2} {\CF} ({\ba};
{\e}_1, {\e}_2, {\Lambda})\right)}
where ${\CF}$ is analytic in ${\e}_1, {\e}_2$
for ${\e}_{1,2} \to 0$. We prove the conjecture of \swi, which identifies
${\CF}_0 ({\ba}, {\Lambda}) \equiv {\CF}({\ba}; 0,0, {\Lambda})$ with the
Seiberg-Witten prepotential of the low-energy effective theory.

\subsec{Quasiclassical limit and SW curve}

\ndt By setting ${\n}=1$ we deduce from the field theory prediction \swipr\
that for ${\hbar} \to 0$ the partition function \prtn\ has an asymptotic
expansion:
\eqn\asmpt{Z ( {\ba}; {\hbar}, {\Lambda} ) = {\exp} \left(
\sum_{g=0}^{\infty} \ {\hbar}^{2g-2} \ {\CF}_{g}( {\ba}, {\Lambda} )\right)}
where
\eqn\prptn{{\CF}_0 ( {\ba}, {\Lambda}) = -
{\half}
\sum_{l , n} (a_l - a_n)^2 \left( {\lg{a_l - a_n \over \Lambda }} - \frac{3}{2} \right) + \
\sum_{k=1}^{\infty} {\Lambda}^{2 k N} f_{k} ( {\ba})} is the so-called
prepotential of the low-energy effective theory. The coefficients $f_{k}$ for $k  = 1, 2$ were computed directly from instanton
calculus in \khoze\twoinst, for $k \leq 5$ in \swi.

The prepotential was
identified in \sw\swsol\gkmmm\ with the prepotential of the periodic Toda
lattice. The arguments for this identification were indirect. In
\experiment\ the coefficients $f_{k}$ for $k \leq 5$ were computed for the
Toda prepotential, in agreement with the later results of \swi.

The purpose of this paper is to give a proof of this relation to all orders
in the instanton expansion.

More precisely, we shall show that:
\bigskip
\boxit{\centerline{}\medskip\noindent The set $\left( a_1, \ldots, a_N;
{1\over{2\pi i}} \frac{{\p}{\CF}_0}{{\p} a_1}, \ldots, {1\over {2\pi i}} \frac{{\p}{\CF}_0}{{\p} a_N}\right)$
coincides with the set of periods of the differential \eqn\swdf{dS =
\frac1{2\pi i} z {dw\over w}} on the curve ${\CC}_u$ defined by \eqn\tdcr{ {\Lambda}^{N} \left( w
+\frac1{w} \right) = P_{N} (z) , \qquad P_N(z) = z^N + u_1 z^{N-1} +
u_2 z^{N-2} +
\ldots + u_N}}
\remark{Normally the factor $2{\pi}i$ is included in the definition of the
prepotential \booksSW. We chose not to do this to get real expressions for
${\CF}_0$ for
real $a_l$'s and $\Lambda$.}

For generic $u = (u_1, u_2, \ldots, u_N)$ the curve \tdcr\ is a smooth hyperelliptic
curve of genus $N-1$. The differential $dS$
has poles at two points ${\infty}_{\pm}$ over $x = \infty$.
The homology group $$H_{1}({\CC}_{u} - \{
{\infty}_{+}, {\infty}_{-} \}; {\bZ})$$ is isomorphic to ${\bZ} \oplus
{\bZ}^{2(N-1)}$, the first summand being generated by the small loop
${\s}_{\infty}$ around ${\infty}_{+}$ (the loop around ${\infty}_{-}$ is in
the same homology class). The intersection pairing identifies the dual space
with $$H_{1}({\CC}_{u}, \{ {\infty}_{+}, {\infty}_{-} \} ; {\bZ} ) \ , $$
the relative cycle, dual to ${\s}_{\infty}$ being the path
${\ell}^{\infty}$, connecting ${\infty}_{-}, {\infty}_{+}$.

\mdp Note that in the $N$-parametric family of curves \tdcr\ there is a set
of singular curves. The identification between the periods and $(a,
\frac{{\p\CF}_0}{{\p}a})$ is up to the action of the monodromy group. We
need to fix this identification in some open domain of the parameter
space.

\mdp
Consider the region ${\CU}_{\infty}$ of the parameter space,
where  $u_s \gg {\Lambda}^s$. In this domain the curve \tdcr\ can be
approximately described  as follows. Let ${\a}_1, \ldots, {\a}_N$ be the
zeroes of the polynomial $P_N(x)$: \eqn\plnm{P_N(x) = \prod_{l=1}^{N} ( x -
{\a}_l )} For small ${\Lambda}$ for each $l$ we can unambiguously find
${\a}_l^{\pm} \approx {\a}_l$, such that \eqn\brptns{P_{N} ( {\a}_l^{\pm} )
= \pm 2{\Lambda}^{N}} These are the branch points of the two-fold covering
${\rho}: {\CC}_u \to {\bP}^1$, ${\bP}^1 \ni x$.

\mdp Let ${\ba}_l$ be the
1-cycles on ${\CC}_u$ which are the lifts of the cycles on the $x$-plane,
which surround the cuts going from ${\a}_l^{-}$ to ${\a}_l^{+}$. Let
${\bb}_l$ be the relative 1-cycle, represented by the path on ${\CC}_u$
which starts at ${\infty}_{-}$, goes to ${\a}_l^+$, and then goes on the
second sheet to ${\infty}_{+}$. The cycles ${\ba}_l$, and ${\bb}_l$
have canonical intersection pairing:
\eqn\intrc{{\ba}_l \cap {\bb}_m =
{\d}_{lm}\,.}
The position of the $\ba$-cycles and $\bb$-cycles
on the curve ${\CC}_u$ is illustrated on the Fig.1
\Figx{9}{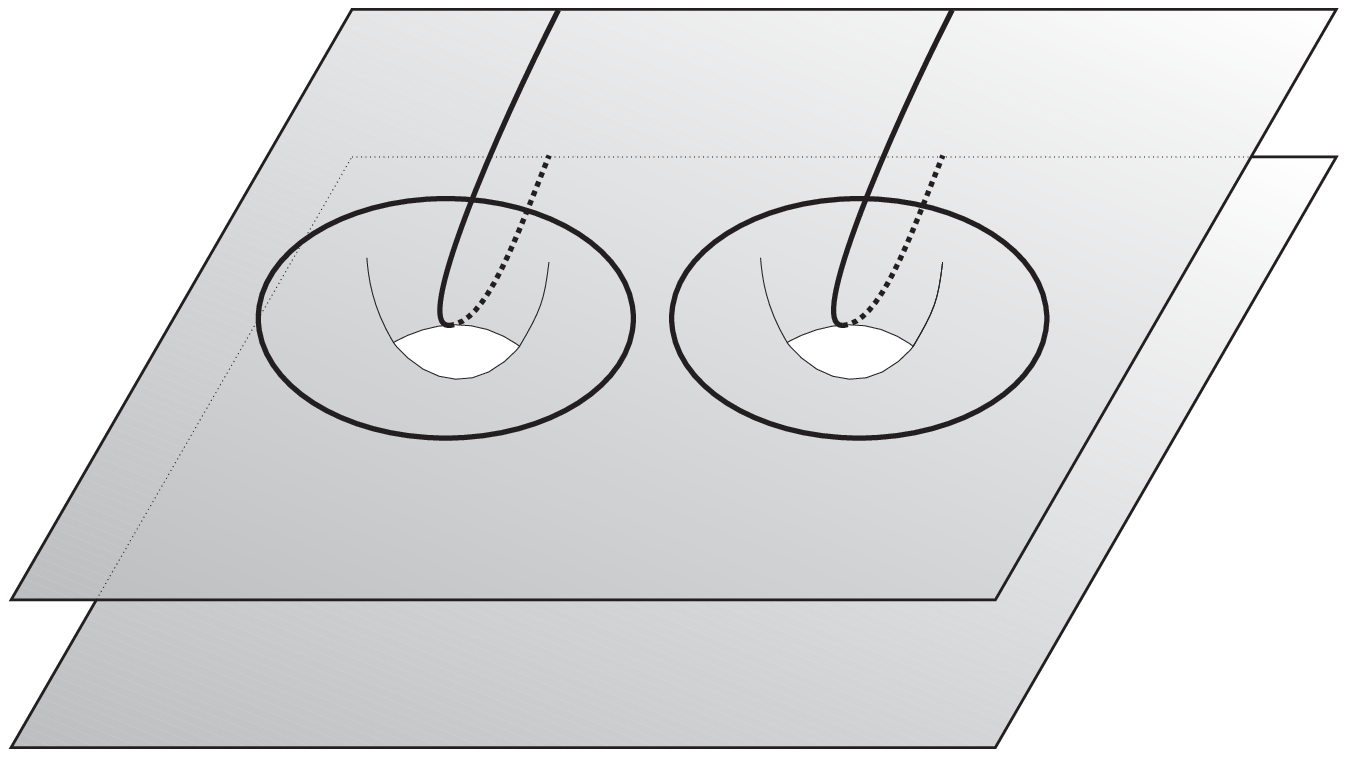}{The curve ${\CC}_u$ and the cycles on it, the closed
ones are ${\ba}_l$'s, the noncompact ones are ${\bb}_l$'s}
The cycles ${\s}_{\infty}$ and ${\ell}^{\infty}$ are related
to ${\ba}_l, {\bb}_m$ via: \eqn\ccls{{\s}_{\infty} = \sum_l {\ba}_l, \qquad
{\ell}^{\infty} = \sum_l {\bb}_l} The periods $a_l$ and
${{\p}{\CF}_0}\over{{\p}a_l}$ are defined via: \eqn\prdss{a_l =
\oint_{{\ba}_l} dS, \qquad {{\p}{\CF}_0 \over{{\p}a_l}} =
2\pi i \int_{{\bb}_l({\mu})} dS} where ${\bb}_l({\m})$ is the regularized
contour of integration, which connects $w\vert_{z = {\m}}$ and $w^{-1}\vert_{\m}$ instead of ${\infty}_{+}$
and ${\infty}_{-}$. \mdp The divergent (with $\m$) part of the periods
is easy to calculate: \eqn\dvrg{\frac1{2\pi i} \frac{\p{\CF}_0}{{\p}a_l} = 2 N {\m} - u_1
\lg{\m} + {\rm finite}\ {\rm part}} At the same time:
\eqn\uone{\oint_{{\s}_{\infty}} dS = a= \sum_l a_l = - u_1}
As we said above, in the sequel we set $a=0 = u_1$. Also, to avoid worrying
about the cut-off we work with the absolute cycles ${\bb}_l - {\bb}_{l+1}$.

\subsec{Partitions and their profiles}

The standard geometric object associated to a partition is
its diagram. In this paper, we will draw partition diagrams
in what is sometimes referred to as the Russian form (as opposed to the
traditionally competing French and English traditions of
drawing partitions). For example, the following
figure
\Figy{4.5}{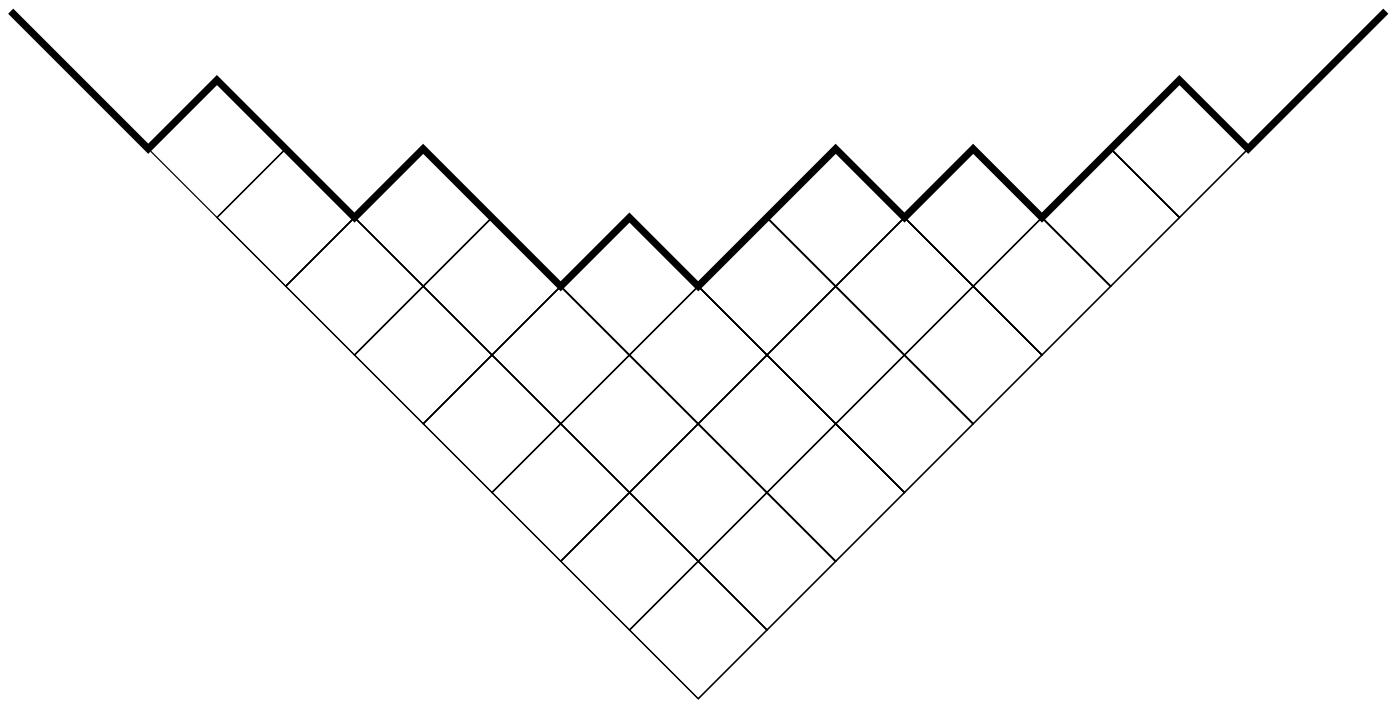}{\cyr Diagramma YUnga}
\ndt shows the diagram of the partition $(8,6,5,3,2,2,1,1)$. The upper
boundary of the diagram of $\bk$ is a graph of a piecewise-linear
function $f_\bk(x)$ which we will call the {\it profile} of the
partition $\bk$.
Explicitly,
\eqn\prof{
f_\bk(x) = |x| + \sum_{i=1}^{\infty} \biggl[ |x-k_i+i-1| - |x-k_i+i| + | x + i | - | x + i -1|
\biggr]}
The profile is plotted in bold in Fig.2 .

For general ${\e}_2 > 0 > \e_1$, it is convenient to extend
the definition of $f_\bk(x)$ by scaling the two axes by
$-\e_1$ and $\e_2$, respectively. For example, for
$(\e_1,\e_2)=(-1,\frac12)$, the scaled diagram of the
same partition $(8,6,5,3,2,2,1,1)$ and the
corresponding profile $f_\bk(x|\e_1,\e_2)$ will look as follows

\Figy{4.5}{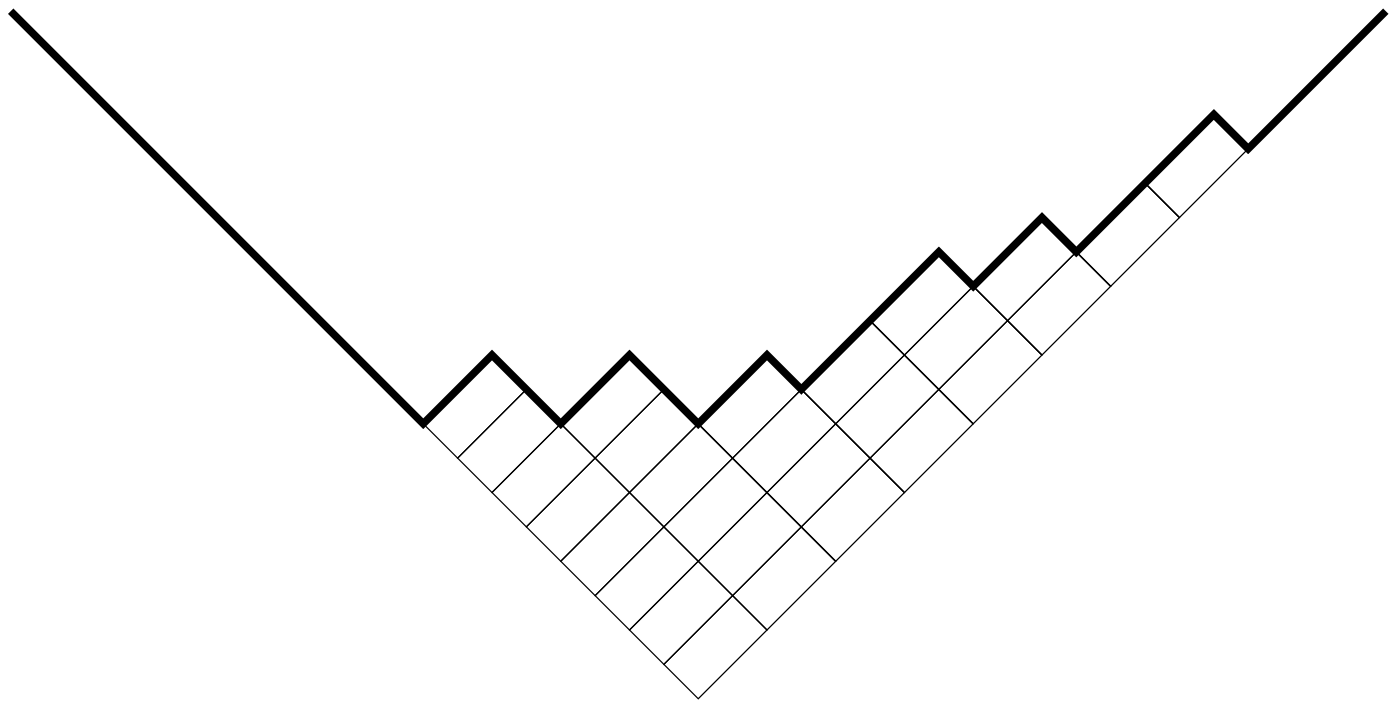}{Squeezed Young diagram}
We have
\eqn\gnepspr{\eqalign{
f_{\bk}(x|{\e}_1, {\e}_2) =& \ |  x |  +
\sum_{i=1}^{\infty} \biggl[ \ | x + {\e}_1 - {\e}_2 k_{i} -{\e}_1 i
| - | x - {\e}_2 k_{i} -{\e}_1 i | \cr &
\qquad\qquad\qquad\qquad\qquad - | x + {\e}_1 - {\e}_1 i
| + | x  - {\e}_1 i | \ \biggr]\cr =&\   |  x | +
\sum_{j=1}^{\infty} \biggl[ \ | x + {\e}_2 -
{\e}_1 {\tilde k}_{j} -{\e}_2 j | - | x - {\e}_1 {\tilde
k}_{j} -{\e}_2 j | \cr & \qquad\qquad\qquad\qquad\qquad - | x + {\e}_2 - {\e}_2 j | + | x  - {\e}_2 j | \ \biggr]\cr}}
By construction, the profile of a partition satisfies
\eqn\prfi{\eqalign{ f_{\bk}'(x|{\e}_1, {\e}_2)&=\pm 1\,,\cr
f_{\bk}(x|{\e}_1, {\e}_2)&\ge |x|\,, \cr f_{\bk}(x|{\e}_1, {\e}_2) &= |x|\,, \
{\rm for} \ |x| \gg 0\,.\cr}}
We also define the profile of a charged partition
$$
f_{a;{\bk}} (x|{\e}_1, {\e}_2)= f_{\bk} ( x - a|{\e}_1, {\e}_2)
$$
The
charge and the size are easily recovered from $f_{a;{\bk}}(x)$:
\eqn\chsz{\eqalign{ a =  \half \int_{\bR} \ dx \ x \ f''_{a;{\bk}}(x|{\e}_1,
{\e}_2) = & - \half \intp{\bR} \ dx \ f'_{a;{\bk}}(x|{\e}_1, {\e}_2), \cr
& \cr
 \vert \bk \vert =
\frac{a^2}{2{\e}_1{\e}_2}  - \frac1{4{\e}_1{\e}_2} \int \ dx \ x^2  \
f''_{a;\bk}(x|{\e}_1, {\e}_2) = & \frac1{2{\e}_1{\e}_2} \left( a^2 -
\int dx \ \left( f_{a; {\bk}}(x|{\e}_1, {\e}_2) - |x|
\right)\right)\cr} }
Here and in what follows, we denote by
$$
\intp{D} g(x) \, dx  = \lim_{L\to\infty,\delta\to 0}
\int_{D \cap [-L,L]\setminus {\rm sing}_\delta (g)} g(x) \,dx
$$
the principal value integral over a domain $D\subset \bR$,
where ${\rm sing}_\delta (g)$ denotes the $\delta$-neighborhood
of the singularities of $g(x)$.

\mdp

For a colored partition ${\vec\bk}$ and a vector ${\ba}$ we define:
\eqn\cmbnpr{f_{{\ba};\vec\bk}(x|{\e}_1, {\e}_2) = \sum_{l=1}^{N} \ f_{a_l;
{\bk}_l} (x|{\e}_1, {\e}_2)}
For example, for ${\e}_2 = - {\e}_1 = {\hbar}$, $a_1 = - a_2 =
11{\hbar}$ and the partition
$$
\{(7,4,3,3,2,1),(8,7,4,4,3,1)\}
$$
the corresponding profile looks as follows:
\Figx{6}{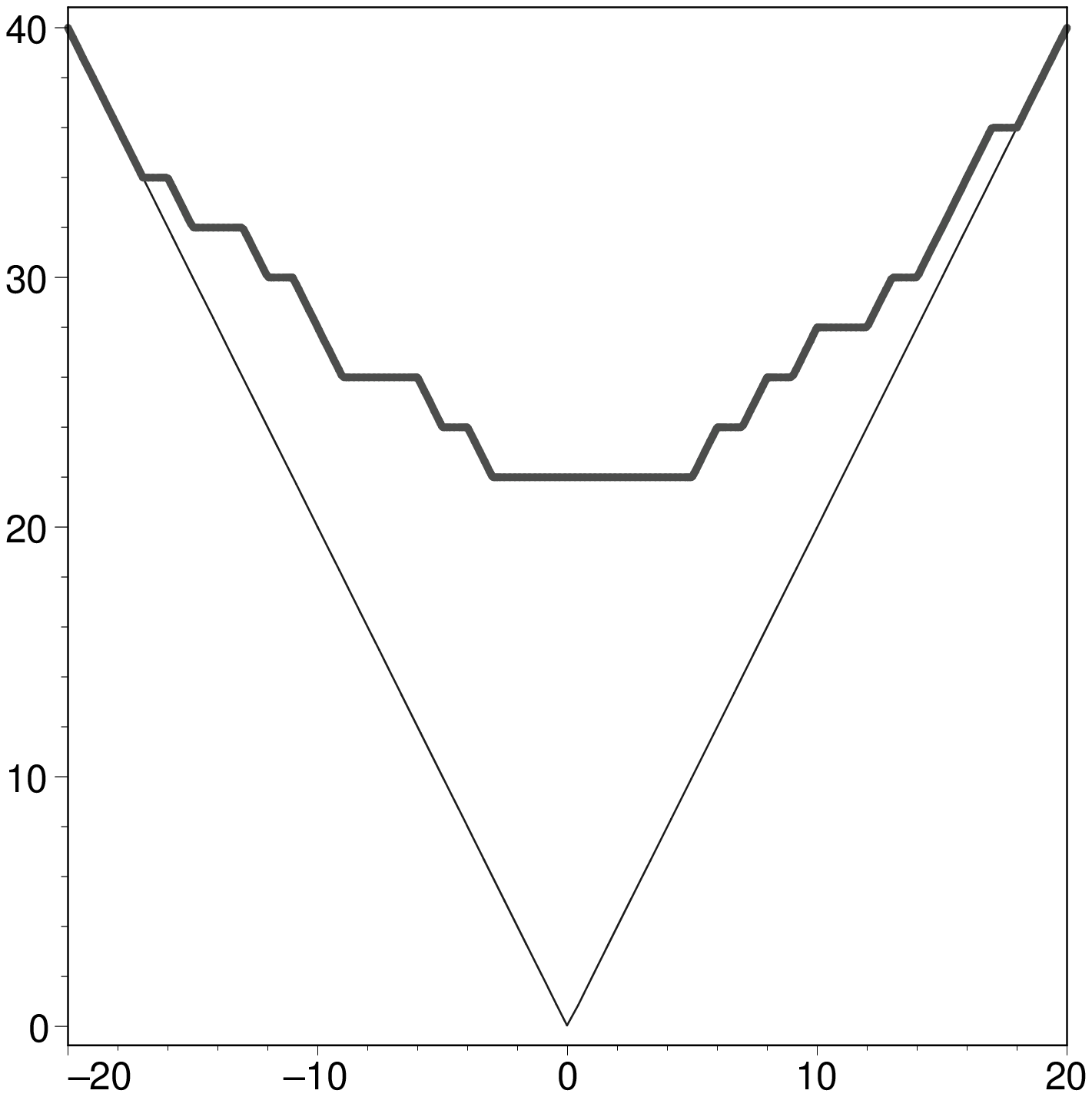}{The profile of the colored partition $
\{(7,4,3,3,2,1),(8,7,4,4,3,1)\}
$}

\subsec{Vacuum expectation values and resolvents}
The profile function is natural from the gauge theory point of
view: the vacuum expectation values of the single trace operators
${\Tr}{\phi}^n$ have a simple expression in terms of the profiles:
\eqn\vevs{\langle {\Tr} {\phi}^n \rangle_{\ba} = \frac{1}{Z ( {\ba}; {\e}_1,
{\e}_2,  {\Lambda})} \sum_{{\vec\bk}} \ {\Lambda}^{2N \vert {\vec\bk} \vert}
Z_{{\vec\bk}} ({\ba}; {\e}_1, {\e}_2, {\Lambda}) \ {\CO}_{n}[{\vec\bk}]}
where \eqn\hgs{{\CO}_{n}[{\vec\bk}] = \half\int_{\bR} dx \ x^n \
f''_{{\ba};\vec\bk}(x | {\e}_1, {\e}_2)}
The second derivative $f''$ of a partition profile is
a compactly supported distribution on $\bR$ which
for random partitions plays to the role similar
the role of the spectral measure for random
matrices. In particular, it is convenient to introduce
the resolvent $R(z |  {\e}_1, {\e}_2)$ of a colored partition $\vec\bk$ by
\eqn\rslvnt{R(z  |  {\e}_1, {\e}_2) = \half\int_{\bR} dx \
{{f''_{\ba;\vec \bk}(x |  {\e}_1, {\e}_2) \over{z-x}}\,.}
}
Note that:
\eqn\rslvntt{R(z | {\e}_1, {\e}_2 ) = \frac{N}{z} + \frac{a}{z^2} + O(\frac1{z^3}), \qquad z
\longrightarrow \infty}
In our context, the limit shapes $f$ of partitions
will be convex piece-wise analytic functions
for which the second derivative $f''(x)$ will be positive
and compactly supported.
\subsec{Real vs. complex}
In the gauge theory the
vacuum expectation values $a_l$ of the Higgs field are, in general,
 complex.
In this case the profile function $f_{{\ba}, {\vec\bk}}$ does not
make sense. However, the resolvent \rslvnt\ of a colored partition
 is well-defined and
can be effectively used in the analysis below. The formula \hgs\ is replaced
by: \eqn\hgss{{\CO}_{n}[{\vec\bk}] = \frac{1}{2\pi i}\oint z^n
R(z|{\e}_1, {\e}_2) dz} where the contour goes around $z = \infty$.

Complex  values of $a_l$ can be reached from the real values
by analytic continuation. By the same token, it is enough
to analyze the problem for values of $a_l$ in any open set of $\bR^N$.
In particular, it is enough to analyze the problem in the
asymptotic domain ${\CU}_{\infty}$, which is what we will do
next.

\subsec{The thermodynamic limit}

Our strategy in extracting the ${\e}_1, {\e}_2 \to 0$ limit
of the sum \gnpri\ is the following. The typical size of the partition
${\bk}$ contributing to the sum is of the order
$\vert {\bk} \vert \sim {1\over{{\e}_1
{\e}_2}}$, and is so large that the sum over the partitions can be approximated
by an integral over the space of continuous Young diagrams $f(x)$.
A continuous Young diagram, by definition, is a function
satisfying the following weakening of the condition \prfi\
\eqn\lpshtx{\eqalign{& f (x) =  \vert x \vert, \qquad \vert x \vert \gg 0\,,
\cr & \vert f(x) - f(x) \vert \leq \vert x - y \vert\,, \cr & \intp{\bR}
dx\ f'(x) = 0\,, \cr & \int_{\bR} dx\  \left( f(x) - |x| \right) < \infty\cr}}
Note that the condition $f'(x)=\pm1$ is replaced by the weaker
Lipschitz condition.

We will show that this path integral is dominated by a
unique saddle point (in fact, a
strict maximum). In a more mathematical language, this
means the following. By associating to every partition its profile,
we get from \gnpri\ a measure on the space of Lipschitz functions.
As ${\e}_1, {\e}_2 \to 0$, these measures concentrate around a
single point, that is, converge to the delta measure at a single
function. This function is the limit shape of our random partition.
We will construct this limit shape explicitly and show that
it has a very simple and direct relation to the Seiberg-Witten
geometry.

\mdp

We begin by observing that the Plancherel measure ${\m}({\bk})$ can be
written in terms of  profile  $f_{\bk}(x|\hbar)$ as follows
 \eqn\plchn{{\m}({\bk})
= {\exp}\  \left(-\frac1{8}\intp{x \neq y} \ f''_{\bk}(x|\hbar) f''_{\bk}(y|\hbar)
\ {\zu{x-y}} \ dx dy\right)\,.}
More generally, we have
\eqn\gnpr{
  Z_{\vec\bk}  ({\ba}; {\e}_1, {\e}_2, \Lambda) \ = \exp
\left(-\frac1{4} \intp{}\, \, \, dx dy \
f''_{\ba,\vec\bk}(x | {\e}_1, {\e}_2) \, f''_{\ba,\vec\bk}(y | {\e}_1, {\e}_2)\,  {\g}_{{\ve}_1, {\ve}_2} ( x - y, {\Lambda})\right)\,,}
as can be easily
checked with the help of the main difference equation.
Denoting the right-hand side of \gnpr\ by $Z_{f} ({\e}_1,
{\e}_2, \Lambda)$, we have
\eqn\gnpriv{Z ({\ba}; {\e}_1, {\e}_2, {\Lambda})  =
 \sum_{f \in {\Gamma}_{\ba}^{discrete}}  Z_{f} ( {\e}_1,
{\e}_2, \Lambda)\,,}
where the summation is over the set ${\Gamma}_{\ba}^{discrete}$ of paths  of
the form $f=f_{\ba,\vec\bk}$.

When ${\e}_1, {\e}_2 \to 0$ with $\n$ fixed,
the size of a typical partition ${\bk}_l$ in \gnpriv\ grows
like $\vert {\bk}_l \vert \sim \frac1{{\e}_1{\e}_2}$. In
this limit, the sum \gnpriv\ looks like an integral, over the space ${\Gamma}_{\ba}$ of
paths $f$ of the form
\eqn\pths{f(x) = \sum_{l=1}^{N} f_{l}(x-a_l)}
with each $f_l$ satisfying \lpshtx. For this integral,
${\e}_1{\e}_2$ plays the r{\^o}le of the Planck constant.
Our strategy, therefore, is to find a saddle point (in fact, the minimum)
 of the action
\eqn\actn{{\CE}_{\Lambda}(f) =  \frac1{4}\intp{y <  x} \ dx dy \ f''(x) f''(y)\  (x-y)^2
\left( \lg{\frac{x-y}{\Lambda}} - \frac{3}{2}\right) }
on the space ${\Gamma}_{\ba}$ of paths \pths.  The action \actn\ is
the leading term as ${\e}_1, {\e}_2 \to 0$ of the action in \gnpr:
\eqn\gnprv{ Z_f ({\ba}; {\e}_1, {\e}_2, {\Lambda}) \sim {\exp} \left(
\frac1{{\e}_1{\e}_2} {\CE}_{\Lambda}(f)\right)}
Integrating by parts we rewrite \actn\ as
\eqn\E{
    {\CE}_{\Lambda}(f)= - \frac1{2}
 \intp{x<y} (N+f'(x))(N-f'(y))\ \lg{\frac{y-x}{\Lambda}} \, dx \, dy}
which, for $N=1$, reproduces the result of Logan-Schepp-Kerov-Vershik\logshep\verker\verkeri\kerovii.
Thus, we have:
\eqn\crpnt{{\CF}_{0}({\ba}, {\Lambda}) = - {\rm Crit}_{f  \in
{\Gamma}_{\ba}} \quad {\CE}_{\Lambda}(f)}

\subsubsec{Profiles vs. eigenvalue densities}

In \swi\ an expression for $Z_{k}({\ba}, {\e}_1, {\e}_2) = \sum_{{\vec\bk}, \ | {\vec\bk}| = k} Z_{\vec\bk}$ was given
in terms of some contour integral over $k$ eigenvalues ${\phi}_I$, $I=1,
\ldots, k$. This expression follows straightforwardly from the ADHM
construction of the moduli space of instantons, and, therefore, easily
generalizes to the case of $SO, Sp$ gauge groups. It is, therefore, quite
remarkable that the formula \gnprv\ can be obtained directly from the
contour integral expression, avoiding the actual evaluation of the integral
(which is, of course, needed to get correclty the ${\CF}_g$'s with $g > 0$).

\eqn\epsint{Z_{k}({\ba}, {\e}_1, {\e}_2) = \oint \prod_{I=1}^{k} \left[ \frac{{\e}_1 + {\e}_2}{2\pi {\e}_1{\e}_2}
\frac{d{\phi}_I}{P({\phi}_I) P({\phi}_I + {\e}_1 + {\e}_2)}\right] \prod_{I \neq
J} \frac{{\phi}_{IJ} ( {\phi}_{IJ} + {\e}_1 + {\e}_2)}{({\phi}_{IJ}+
{\e}_1)({\phi}_{IJ} + {\e}_2)}}
where
\eqn\ntns{\eqalign{P(x) = \prod_{l=1}^{N} ( x - a_l) & \cr
{\phi}_{IJ} \ = & \ {\phi}_I - {\phi}_J \cr}}
We now multiply $Z_{k}$ by ${\Lambda}^{2kN}$ and sum over $k= 0, 1, \ldots,
$ to get $Z^{inst}$. Take the limit ${\e}_1, {\e}_2 \to 0$. The typical $k$
which will contribute most to the sum will be of the order $k \sim
\frac1{{\e}_1 {\e}_2}$. So we introduce the density of eigenvalues:
\eqn\dnst{{\rho}(x) = {\e}_1 {\e}_2 \ \sum_{I=1}^{k} {\d}( x - {\phi}_I)}
which is normalized in a $k$-independent way, which nevertheless guarantees
its finiteness in the limit we are taking.
Now, the difference with the ordinary 't Hooft-like limit of the ordinary matrix
integrals is the presence of the equal number of the
${\phi}_{IJ}$ terms in the numerator and the denominator of the measure
\epsint. This will change qualitatively the density dependence of the
effective potential on the eigenvalues, and the resulting equilibrium
distribution of the eigenvalues. In particular in our limit the former superymmetric matrix integral \epsint\ scales as ${\exp} k F$ as opposed to the
't Hooft's ${\exp} k^2 F$.

Nevertheless we have a sharp peak in the
measure, which justifies the application of the saddle point method. Indeed,
by expanding in ${\e}_1, {\e}_2$ we map the measure in \epsint\ onto:
\eqn\epsinnt{{\Lambda}^{2kN} Z_{k} ({\ba}, {\e}_1, {\e}_2) \sim {\exp}
\left( \frac1{{\e}_1{\e}_2} {\bE}_{\Lambda}[{\rho}] \right)}
where
\eqn\enrgi{{\bE}_{\Lambda}[{\rho}] = - \intp{x \neq y} \ dxdy\
\frac{{\rho}(x){\rho}(y)}{(x-y)^2} - 2 \int \ dx \ {\rho}(x)
\lg{P(x)\over{\Lambda}^{N}}}
Up to the perturbative piece $$\half\sum_{l,n} ( a_l - a_n)^2
\lg{{a_l - a_n}\over{\Lambda}}$$ the energy \enrgi\ coincides with the action \actn\ if we
identify:
\eqn\idntfc{\mathboxit{f(x) - \sum_{l=1}^{N} |  x - a_l | \ = \ {\rho}(x)}}
The same method applies to other theories considered in \swi\ihiggs.

Thus we have learned yet another interpretation of the limiting profile of the
colored partition.
In the rest of the paper we shall work with limit profiles, as the equations
on the extremum
are identical to those following from \enrgi.

\subsubsec{Surface tension}
Given a function $f(x)$ as in \pths,
how can we extract $a_l$'s?
This is easy to do in the region ${\CU}_{\infty}$ where
$$
a_l \ll a_{l+1}\,,
$$
in which case the supports of the functions $f_{a_l,{\bk}_l}$
do not intersect. Introduce parameters ${\xi}_1, \ldots, {\xi}_N$
which will play the role of dual variables to the charges $a_l$. From
\chsz\ we have the following
\pro{For $a_l \ll a_{l+1}$ we have
$$\sum_l {\xi}_l a_l = - \half\intp{\bR} \sigma(f'(x)) \, dx \,, $$
where $\sigma(y)$ is a concave, piecewise-linear function on $[-N,N]$ such
that
$$\sigma'(y) = \xi_l \,,
\quad y\in \left[-N + 2(l-1),-N + 2l \right]\,, $$
and $$ \sigma(-N)= - \sigma(N)= - \sum_l {\xi}_l \,. $$}
An example
of the graph of $\sigma$ is plotted on the Fig.5
\Figx{6}{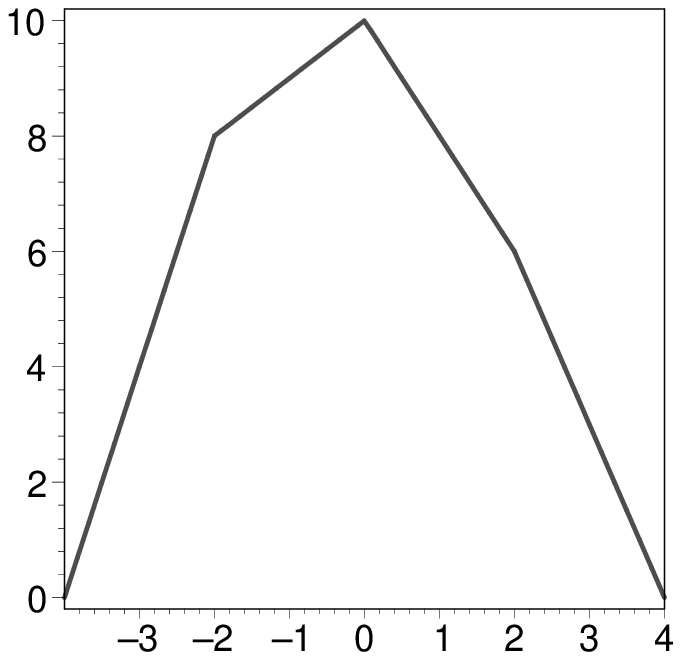}{Surface tension for $
\xi=(4,1,-2,-3)
$}
In Section $\bf 5$, we define a dual partition function $Z^D({\xi}, {\hbar}, {\Lambda})$.
We will see that the dual partition
function $Z^D$ can be interpreted in terms of a
periodically weighted (with period $N$)
Plancherel measure on partitions.
The periodic weights in this formalism are precisely $e^{\xi_l}$
and the function $\sigma$
becomes the corresponding surface tension function.

This story has many parallels and direct connections
to the theory of periodically weighted planar
dimers developed in \dimer.

It is a general principle that singularities of
surface tension (usually referred to as ``cusps'')
correspond to flat regions (``crystal facets'') of
the corresponding action minimizing shapes. In our case,
the singularities of the function $\sigma$
correspond to gaps between the supports of
the functions $f_{\star, l}(x-a_l)$ for the minimizer $f_{\star}(x)$.

\subsec{The equations for the limiting shape} In order to
minimize ${\CE}(f)$ with fixed $a_l$'s, we introduce Lagrange multipliers
${\xi}_l$, which we order as
\eqn\defxi{
{\xi}_1 >\dots > {\xi}_N \,, }
and look for the maximizer of the total action \eqn\tta{ {\CS}_{\Lambda}(f)
= - \CE_{\Lambda}(f) + \half\int \sigma(f'(x)) \, dx }
for fixed values of the $\xi_l$'s.
To get back the minimum of ${\CE}(f)$
with fixed $a_l$ we shall later perform the Legendre transform with respect
to ${\xi}_l$'s.
Note that the surface tension term $\half\int{\s}(f')$ in \tta\ under the conditions $a=0$ and \defxi\
can be made arbitrarily large by making the
individual profiles $f_l$ sufficiently separated from each other. The price
one pays for this is the decrease in the ``energy'' term
$-{\CE}_{\Lambda}(f)$, thus there is a competition between the two terms in
\tta\ which leads to the global maximum.

The action \tta\ is a concave functional. In fact, the
first term in \tta\ is strictly concave (with proper boundary
conditions), which can be seen by rewriting it as a certain
Sobolev norm, see \verkeri. Therefore, any
critical point of the action $\CS(f)$
is automatically a global minimizer. It is, therefore,
enough to look at the first variation of \tta.
Because of the singularities of $\sigma$, this
first variation will involve the one-sided derivatives.
\mdp
Taking the first
variation and integrating once by parts, we find the following equation:
\eqn\onev{ \intp{y\neq x} \ dy \  (y-x) \left( \lgm{\frac{y-x}{\Lambda}} - 1
\right) \, f''(y)  =
 \sigma'(f'(x)) \ .
}
This equation should be satisfied for any point $x$, for which $f'(x)$ is a
point of continuity of $\sigma'$. When $$ f'(x) \in\left\{ -N + 2l \ \vert \
l=1,\dots,N-1\, \right\} $$ then considering the left and
right derivatives separately we obtain the inequalities
\eqn\onevtwo{ \bX f(x) \in ( \sigma'(f'(x)-0),  \sigma'(f'(x)+0) ) \ , }
where, by definition,
\eqn\defbX{
\big[\bX f \big](x)=\intp{y\neq x} \ dy \  (y-x) \left( \lgm{\frac{y-x}{\Lambda}}
- 1 \right) \, f''(y)}
The transform $\bX f$ is closely related to the standard Hilbert transform
\eqn\hlbrt{ \left[\bH \, g \right] (x)  =
\frac{1}{\pi} \, \intp{y \neq x} \ dy \ \frac{g(y)}{y-x}\,.}
Indeed,
\eqn\bXbH{
[\bX f]''= {\pi}\bH(f'')  \,.
}
With ${\xi}_0 = + \infty,
{\xi}_{N+1} = - \infty$, the conditions \onev\ and \onevtwo\ can be recast in the
following form:
\pro{A function $f_{\star}(x)$ is a critical point of
${\CS}(f)$ iff
\eqn\prm{\eqalign{\bX\fs(x) = \xi_l \quad & {\ninepoint
whenever} \quad -N + 2l - 2 < {\fs}'(x) < -N+2l\cr
{\xi}_l > \bX\fs (x) >   {\xi}_{l+1}  \quad
 & {\ninepoint  whenever} \quad
{\fs}'(x) = -N+2l, \qquad l = 0 , \ldots, N \cr}}}
In other words, the function \eqn\bndcfm{{\vf}(x) = {\fs}'(x) + \frac1{\pi i}[{\bX\fs}]'(x)} defines a map from ${\bR}$ to the boundary of the domain ${\Delta}$
depicted on Fig. {\bf 6}. We now describe the solution to \prm\ in great
detail.

\subsec{Construction of the maximizer}

Let $\Phi(z)$ be the conformal map from the
upper half-plane to the domain $\Delta$ which
is the half-strip
\eqn\strp{{\Delta} =
\{{\varpi} \ \vert \ |\Re({\varpi})|<N, \Im({\varpi}) > 0\}}
with vertical slits along
\eqn\slts{
\{\Re({\varpi}) = -N+2l, \Im({\varpi})\in [0,\eta_l]\},\quad l=1,\dots, N-1\,.
}
\Figy{5}{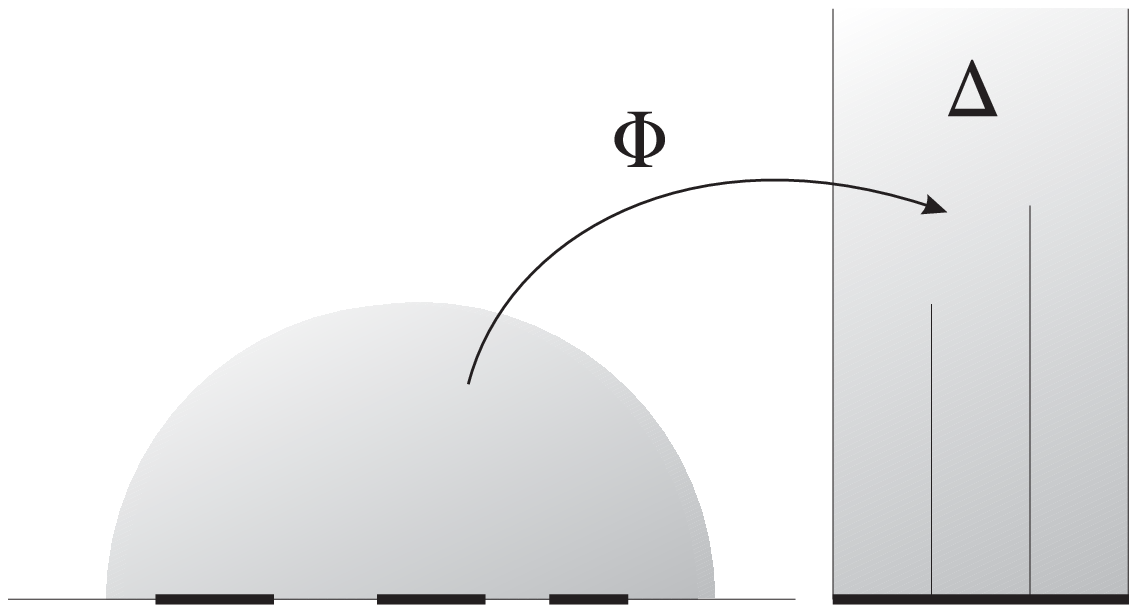}{Conformal map for $N=3$}
The positive reals $\eta_l$, $l=1\dots N-1$, are the parameters of
function $\Phi$.

We normalize $\Phi$ by the condition that it maps infinity
to infinity and
\eqn\asmpt{
\Phi(z) = N + \frac{2N}{\pi i} \, \log \frac{\Lambda}{z}\ + O(\frac1{z})\,,
\quad z \to \infty \,.
}
This fixes $\Phi$ up to an overall shift $x\mapsto x+{\rm const}$.
This ambiguity is related to the overall shift of the
limit shape (and hence, to the overall charge of our
colored partition) and is immaterial.
\subsubsec{Construction of the conformal map}
The map $\Phi$ can be, of course, found using the Schwarz-Christoffel
formula, but it is easier to construct is as the following sequence
of elementary conformal maps.

We take $P_{N}(z)$ to be a monic
real polynomial
$$
P_{N}(z)=z^N+\dots
$$
such that all roots $\alpha^\pm_l$ of the equation
\eqn\defalpm{
P_{N}(z)^2 - 4\Lambda^{2N} = \prod_{l=1}^N (z - \alpha_l^+)(z - \a_{l}^{-})
}
are real.
Let $w$ be the smaller root of the
equation
\eqn\zhu{
{\Lambda}^{N} \left( w + \frac1w \right) = P_{N}(z) \,.
}
The function
$$
w \mapsto {\Lambda}^{N} \left( w + \frac1w \right)
$$
is known as the
Zhukowski function and it maps the open
disk $|w|<1$ to the exterior of
the segment $[-2 \Lambda^N,2\Lambda^N]$. It also
maps reals to reals, and therefore the smaller root
of \zhu\ maps the upper half-plane to the disk
 $|w|<1$ with slits along the
real axis. It remains to take logarithm to
obtain the map $\Phi$. Concretely,
\eqn\vppi{
\Phi(z) = \frac2{\pi i} \lg{w} + N \,,
}
where $w$ is the smaller root of the equation \zhu\ and
we take the branch of the logarithm satisfying
$$
\Im \lg{w} \to 0 \,, \quad z \to +\infty \,.
$$
Since
$$
w \sim \frac{\Lambda^{N}}{z^N} \,, \quad z\to \infty
$$
we get the normalization condition \asmpt.

\subsubsec{Example: $N=2$}
As an example, consider the case $N=2$
\Figy{7}{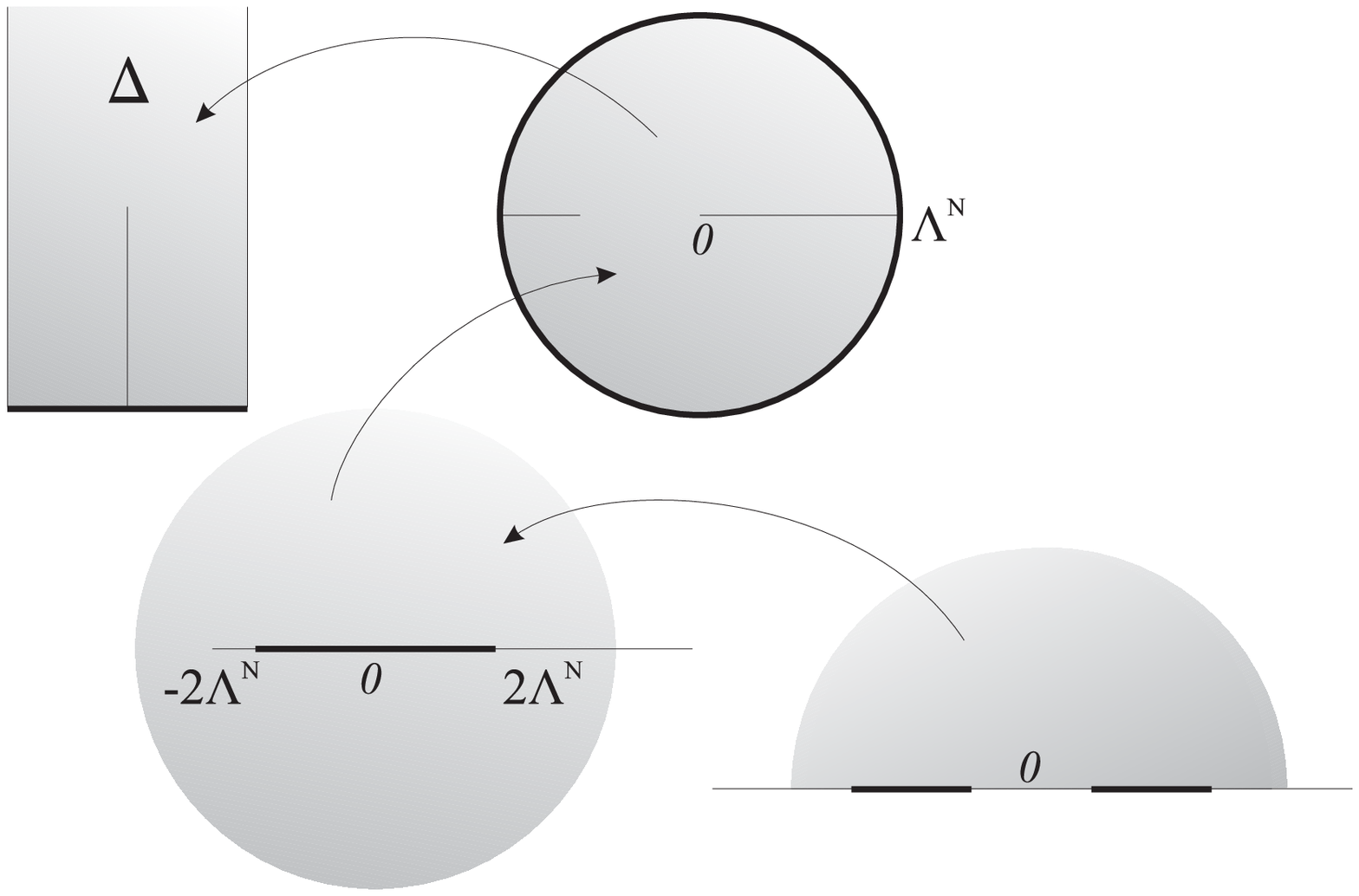}{A sequence of maps for $N=2$}
\ndt A real quadratic polynomial
$$
P_{2}(z)=z^2+\dots
$$
maps the upper half-plane to the entire
complex plane with a cut along the ray
$$
\left[\min_{x\in \bR} P_{2}(x), +\infty\right) \,.
$$
In particular, if the $\min_\bR P_{2}(x) < - 2\Lambda^2$,
that is, if both roots of
$$
P_{2}(x) = - 2\Lambda^2
$$
are real then the segment $[-2 \Lambda^2,2\Lambda^2]$
is contained in this cut.
\subsubsec{Gaps and bands}
For general $N$, the only difference with the $N=2$ case
is that the map $P_{N}$ winds the upper half-plane
$N/2$ times around the complex plane. Correspondingly,
the root of \zhu\ lives on a certain cover of
the disk $|w|<1$, which then gets
unfolded by the logarithm.

Observe that the conformal map $\Phi(z)$ has a well-defined
extension to the boundary $\bR$ of the upper half-plane
$$
{\vf}(x) = \Phi(x+i0)\,,
$$
where the notation is consistent with \bndcfm.

We label the roots of \defalpm\ so that the union of the
intervals $[{\a}^-_l,{\a}^+_l]$ is the
preimage of the base of the half-strip $\Delta$
under the map $\vf$, that is
\eqn\preiPh{
{\vf}^{-1}\left([-N,N]\right) = \bigcup_{l=1}^N
[\alpha^-_l,\alpha^+_l]\,.
}
These intervals are plotted in bold in the
above figures. We will call these intervals
the {\it bands} of $\Phi$ and the complementary intervals
--- the {\it gaps} of $\Phi$.
\pro{We claim that for a choice of ${\xi}_l$'s in \defxi\
we can find the corresponding values
of $\eta_l$, $l=1,\dots,N$, such that
the function $f_\star$ satisfying
\eqn\fstp{
\mathboxit{{\fs}'(x) = \Re\, {\vf}(x)}\,,}
is the maximizer of the action $\CS(f)$.
The bands and gaps of $\Phi$ will
correspond to the curves and flat
parts of the limit shape $f_\star$,
respectively.
They will also correspond
to the bands and gaps in the spectrum of
the Lax operator for the periodic Toda
chain, as will be explained in the next section.}
\mdp
Let us apply the Schwarz reflection principle
to any vertical part of the boundary of
$\Delta$ and the corresponding piece of
the real axis in the domain of $\Phi$.
Taking the resulting function modulo 2,
we obtain an $N$-fold covering map
\eqn\Phimo{
\Phi \mod 2 : \bC \setminus \bigcup_{l=1}^N
[\alpha^-_l,\alpha^+_l] \to
\{\Im {\varpi} >0\}/\mod 2
}
from the complex plane minus the bands to
the half-infinite cylinder, shown schematically on Fig.8.
\Figx{8}{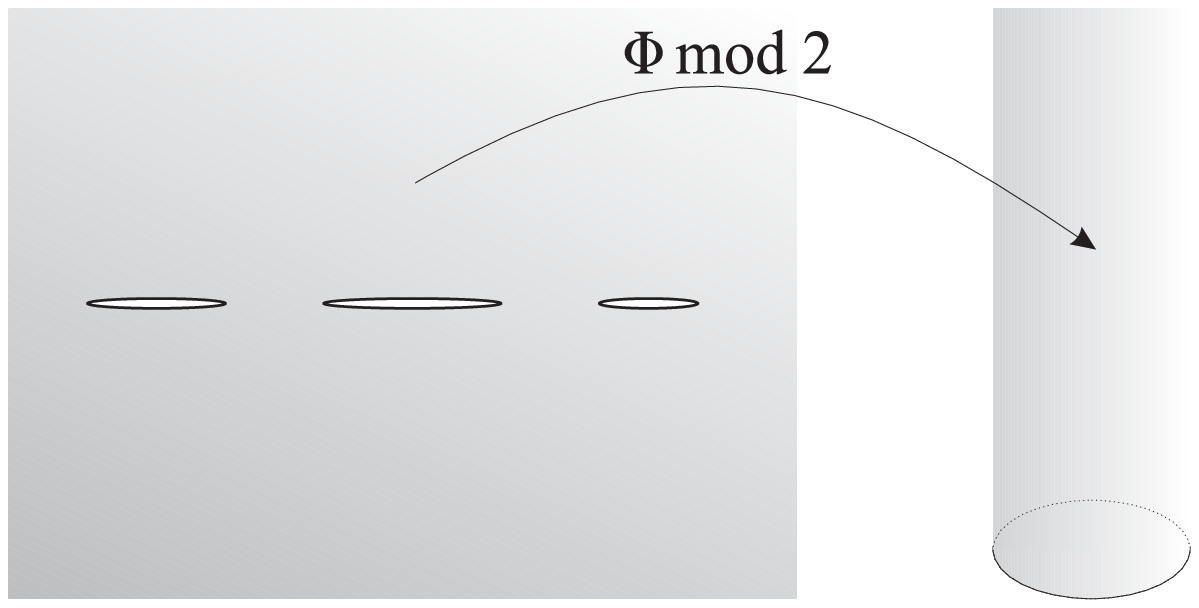}{Covering half-cylinder}
\ndt Note that this map has square-root branching precisely over
the points $i\eta_l$, $l=1,\dots,N-1$.
\ndt
It follows that $\Phi'(z)$ extends to an analytic
function in the complement of the bands. By
the reflection principle, the value of $\Phi'(z)$
on the other side of the cut is precisely $-\Phi'(z)$.
Also note that $\Phi'(z)$ is real (resp.\ purely
imaginary) on bands (resp.\ gaps).
We conclude that
\eqn\CPhi{
\Phi'(z) = \frac1{\pi i} \int_\bR \frac{\Re {\vf}'(x)}{x-z} \, dx =
- \frac{2}{\pi i} \, R_{\fs} (z)\,,
}
where $R_{\fs} (z)$ is (${\e}_1, {\e}_2 \to 0$ limit of) the resolvent of the limit profile $\fs$, defined
by
\fstp.
We can, therefore, identify:
\eqn\sprts{
l'{\rm th \ band} \equiv [{\a}_l^{-}, {\a}_l^{+}] = {\rm supp}f_{l}^{''}(x-a_l), \qquad l = 1, \ldots, N}
\subsubsec{Periods}
It follows from \CPhi\ that
\eqn\aper{a_l = \frac1{2\pi i} \oint_{{\ba}_l} z R_{\fs}(z) dz  = \oint_{{\ba}_l} dS}
where we used \vppi, and, in agreement with \swdf, we have defined
Seiberg-Witten differential:
\eqn\swdff{\mathboxit{dS = \frac1{2\pi i} z \frac{dw}{w}}}
\mdp
Taking the average of $\Phi'(z)$ on two sides of the
cut in \CPhi, we get
\eqn\bHRePhi
{
\bH \, \Re \vf'(x) = - \Im \vf'(x) \,.
}
Using \bXbH, we conclude from \bHRePhi\ that
$$
[\bX\fs]' (x) = - \pi \Im \vf(x) + {\rm const} \,.
$$
We claim that this constant is, in fact, zero. Indeed,
we have
\eqn\AsXp
{
[\bX\fs]' (x) = - \int \lg{\frac{|y-x|}{\Lambda}} \, f''(y) \,dy =
2N \lg{\Lambda\over |x|} + O\left(\frac1x\right)\  ,  \quad x\to\infty
}
Comparing this to \asmpt, we conclude that
\eqn\bXfI
{
[\bX\fs]' (x) = - \pi \Im \vf(x) \,.
}
Since, clearly, $\Im \vf(x) \ge 0$ and $\Im \vf(x)$
vanishes on the bands, it follows from \bXfI\ that $\bX\fs$ is
monotone decreasing and constant on the bands.
It is also clear that the function \fstp\ is
monotone increasing and
constant on the gaps, where it takes
the values
$$
-N+2l\,, \quad l=0,\dots, N \,.
$$
This
verifies the conditions \prm\ with $\xi_l$'s being
the values of $\bX\fs$ on the bands.
\mdp
To recover the $\xi_l$'s we just integrate $[\bX\fs]'$
along the gap, where
$$
i \Im\vf(x) = \vf(x) + {\rm const}\,.
$$
We find
$$
\xi_{l+1}-\xi_{l} = -\pi \int_{\alpha^+_{l}}^{\alpha^-_{l+1}}
 \Im \vf(x) \, dx =  -\pi i
\int_{\alpha^+_{l}}^{\alpha^+_{l+1}} x \,d\vf(x) \,,
$$
integrating by parts and using the vanishing of $\Im \vf(x)$
at the endpoints of a gap. In terms of the Seiberg-Witten differential and the cycles ${\bb}_l$
the last relation reads as:
\eqn\perdS
{
\xi_l-\xi_{l+1} = 2{\pi}i \oint_{\bb_l-\bb_{l+1}} dS
}
This together with the overall constraint $\sum \xi_l =0$
fixes the values of the $\xi_l$'s.
\subsubsec{Completeness}
We now show that by a suitable choice of
the slit-lengths $\eta_l$ we can achieve any
value of the parameters \defxi. The period
map
\eqn\etatoxi
{
\bR_{>0}^{N-1}\owns (\eta_1,\dots,\eta_{N-1}) \mapsto
(\xi_1 > \dots > \xi_N)\,, \quad 
}
is an continuous (in fact, analytic) map of
open sets of $\bR^{N-1}$. Because of the
uniqueness of the maximizer $f_\star$, this
map is one-to-one. Hence, if we can
additionally show that it maps boundary to
boundary, it
will follow that this map is onto.

It is clear that if $\eta_l \to 0$ for some
$l$ then $\xi_l - \xi_{l+1} \to 0$. Suppose
that for some $n=1,\ldots, N$,
$$
\eta_{n} =  \max_{l} \eta_l \to + \infty
$$
It is clear from the electrostatic
interpretation of the conformal map $\Phi$ that
the tip of the $n$-th slit will not be screened
by other slits, that is, for some constants ${\d}_{1},{\d}_{2} >0$
we have
$$
\int_{x\in [\alpha^+_l,\alpha^-_l],\, \Im \Phi(x) > (\eta_n-{\d}_1)} dx > {\d}_{2}\,,
$$
whence
$$
\xi_l-\xi_{l+1}=\int_{[\alpha^+_l,\alpha^-_l]}
\Im \Phi(x) \, dx > {\d}_2 ( \eta_{n}-{\d}_1) \to \infty.
$$
\subsubsec{Periods and the prepotential}
We now perform a check. We consider the partials
\eqn\lgndr{\frac{\p{\CS}(f_\star)}{\p {\xi}_l}}
and relate them to the $A$-periods $a_l$ \aper\ of the differential $dS$.

We have
\eqn\Sfp{
  \frac{\partial}{\partial \xi_l} \, \CS({\fs}) =
\left[\frac{\partial}{\partial \xi_l} \, \CS\right] (\fs) }
because any infinitesimal change in $\fs$ can only
decrease the value of $\CS({\fs})$ which forces  the variation of $\CS$ due to the
change in $\fs$ to vanish (this is essentially
a usual argument about the variation of the critical value of the action with respect
to the parameters of action).
The rest is trivial:
\eqn\vars{\eqalign{ {\d}{\CS} ({\fs})  = &
\half\int [{\d}{\s}]({\fs}') \ dx = - \half\int [{\d}{\s}]'({\fs}') \ x {\fs}''(x) dx =  \cr
& -\half \sum_{l} {\d}{\xi}_l \int_{{\a}_{l}^{-}}^{{\a}_{l}^{+}} x {\fs}''(x) dx =\cr
& -\frac1{4} \sum_l {\d}{\xi}_l \oint_{{\ba}_l} z d{\Phi}(z) \cr}}
which is exactly what we wanted, given \vppi\swdff.
\mdp
Thus,
\eqn\otvet{ \mathboxit{\eqalign{&\quad {\CF}_{0} ( {\ba}, {\Lambda} ) =
-{\CE}_{\Lambda} ( \fs) \cr &\quad R_{\fs} (z)dz  = d\lg{w}
\cr & \quad a_l = \oint_{{\ba}_l} dS \cr
& \quad {\xi}_l = 2\pi i \oint_{{\bb}_l} dS\cr}
}}
The integral in the last line is to be understood with the cut-off. Again,
in the $SU(N)$ theory, where only the differences $a_l - a_m$, and ${\xi}_l
- {\xi}_m$ make sense, the cut-off never show up. It is the Cheshir cat smile of
the noncommutative regularization.

\subsec{Lax operator}
The Seiberg-Witten curves arising from conformal maps
$\Phi$ can be parameterized as the spectral curves in
the periodic Toda chain corresponding to the {\it real}
initial conditions.

Consider the infinite Toda chain with
particle coordinates $q_i$ and momenta $p_i$.
Make it periodic by imposing the constraints
\eqn\percon
{
q_{i+N} = q_i - N \log \Lambda
}
for all $i$.
The Lax operator of this periodic Toda chain is a
discrete Schr\"odinger operator of the form
\eqn\LT{
L(w) =
\pmatrix{p_1 & e^{q_1 - q_2} & & & & w  \Lambda^N e^{q_N-q_1}\cr
e^{q_1 - q_2} & p_2 & e^{q_2-q_3} & \cr
& e^{q_2-q_3} & \ddots \cr
\cr
&&&&\ddots & e^{q_{N-1}-q_{N}}\cr
w^{-1}  \Lambda^N e^{q_N-q_1} &&&& e^{q_{N-1}-q_{N}} & p_N \cr
 }\,,
}
where $w$ is Bloch-Floquet multiplier.

The integrals of motions are summarized by the spectral curve,
which is the curve defined by the characteristic
polynomial
$$
\det(z-L(w)) = P(z) - \Lambda^N \left(w+\frac{1}{w}\right)\,.
$$
Here $P(z)$ is a monic polynomial of degree $N$.
Observe that  all roots
of the polynomials
\eqn\PofzLN
{
P(z)\pm 2\Lambda^N = \det(z-L(\mp 1))
}
are real because
$L(\pm 1)$ is a real symmetric matrix.
This fact plays an important role in the dynamics of the
periodic Toda lattice.

The band and gaps of the map $\Phi$\
are precisely the bands and gaps in the spectrum of the
associated periodic discrete Schr\"odinger operator $L$ on $\bZ$
$$
\left[L f\right](i) = e^{q_{i-1}-q_i} f(i-1) + p_i f(i) +
e^{q_i-q_{i+1}} f(i+1)\,, \quad i\in \bZ\,.
$$
Indeed, by \percon\
$L$ commutes with translation operator $T$
$$
\left[T f\right] (i) = f(i+N)
$$
and \LT\ is the restriction of $L$ onto the $N$-dimensional
$w$-eigenspace of $T$.
In a band, the Bloch-Floquet multiplier $w$ is complex
number of absolute value $1$ and hence $L$ has a
bounded eigenfunction.

It can be shown that all curves of the form \zhu\ with $N$ real
ovals arise in this way.  This is similar to the result of \dimer\
that the spectral curves of periodically weighted planar
dimers parameterize Harnack (also known as maximal) plane curves.
Incidentally, they are also M-curves, not just because they are used in the
M-theory construction of the gauge theory \wittensolution\ but also, and
mostly\foot{We should apologize to M-theorists for the fact,
that M-manifolds were introduced long time before the 11d sugra was
invented} because they have exactly real $N$ ovals.

For example, taking the lattice at rest leads to Chebyshev
polynomials. In this case all gaps shrink to points.

\subsec{An $SU(3)$ example}
Here is an example of a limit shape $f_\star$. Take
$$
P(z) = z^3 - 4 z \,.
$$
To visualize the curve $w+\frac1{w}=P(z)$, let's look
at the plots of $\Re(w)$ and $\Im(w)$ for $z\in[-3,3]$
plotted in the Fig.8 in bold and normal,
respectively
\Figy{8}{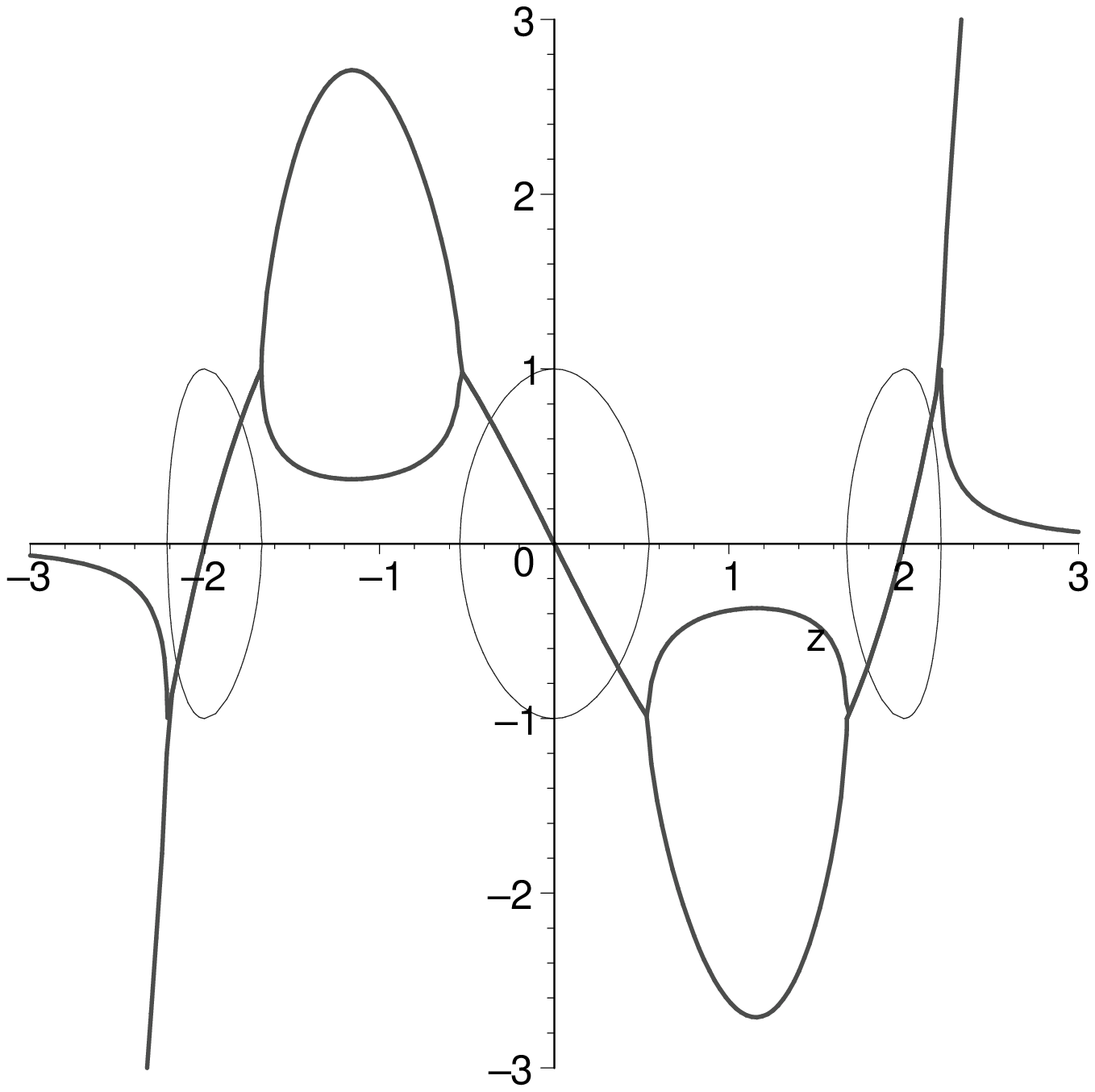}{}
The three parts of this curve correspond to the
three bands of the corresponding
limit shape $\fs$ plotted in the following
figure

\Figy{6}{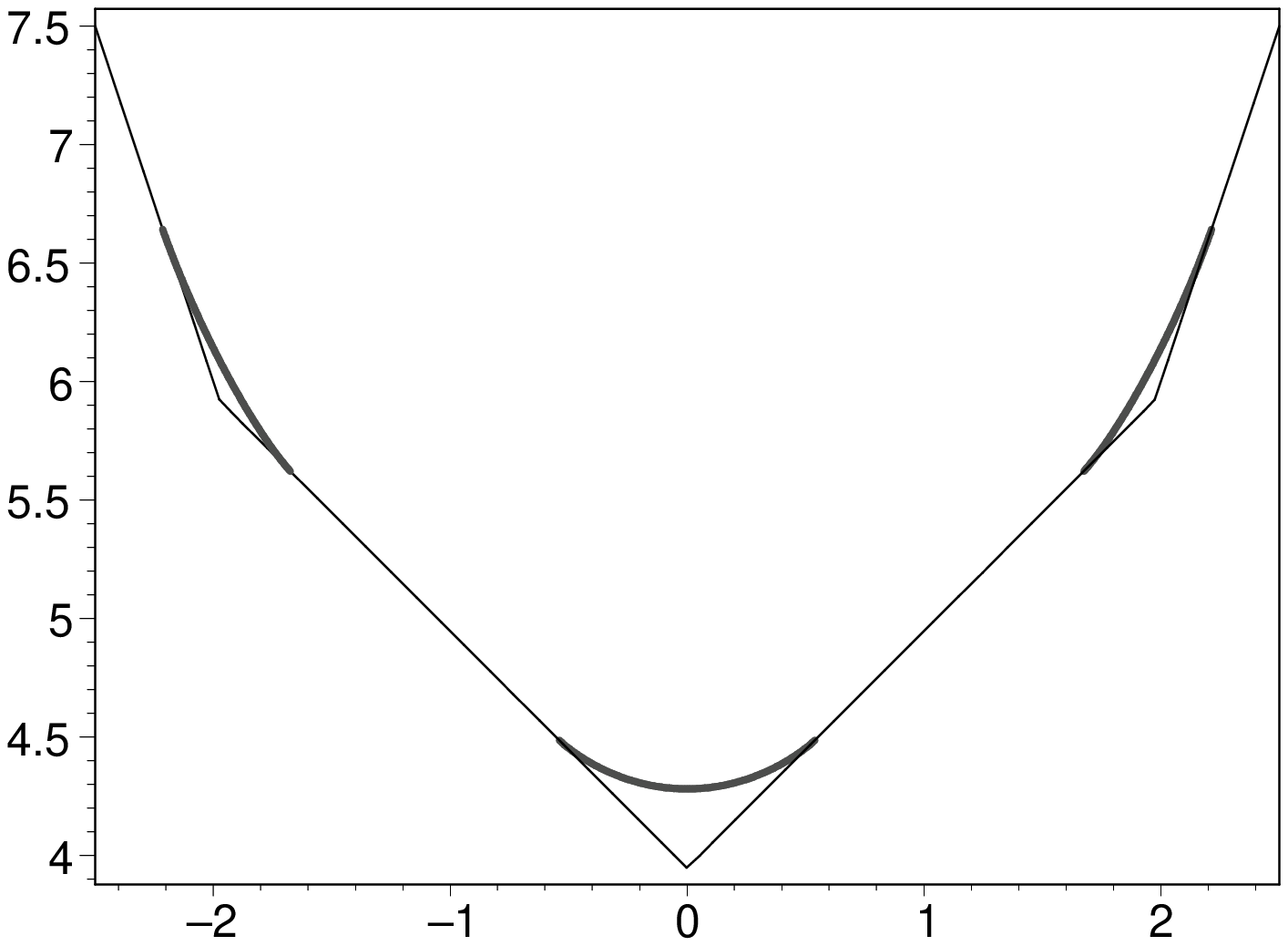}{}
\subsec{Higher Casimirs}

Just as in \lmn\ we could deform the theory by adding arbitrary higher
Casimirs to the microscopic prepotential: \eqn\mcrs{{\CF}_{UV} =
2{\pi}i \left[ \frac{{\tau}_0}{2} {\Tr}{\Phi}^2
 + \sum_{\vec n} {\tau}_{\vec n} \prod_{J=1}^{\infty} \left( \frac1{J}{\Tr}{\Phi}^J
\right)^{n_J} \right]}
The deformations by the single trace operators are especially simple, as
they would lead to the modification of the action ${\CS}(f)$ by the purely surface
term:
\eqn\hrcsm{
{\CS}(f; {\tau}_{\vec n}) = - {\CE}_{\Lambda}(f) + \frac1{2}\int \ dx \ {\s}(f') + \frac1{2} \int \ dx \
f''(x) \ \sum_{k=1}^{\infty} {\tau}_k \frac{x^{k+1}}{k+1}
}
Note that ${\tau}_1$ shifts $\lg{\Lambda}$.
Presumably the critical point of \hrcsm\ would be a solution of some generalization of Whitham
equations
\kricheverwhitham\gkmmm\booksSW\whitham\blowupwhitham\takasaki.

\newsec{DUAL PARTITION FUNCTION AND CHIRAL FERMIONS}

In this section we show that a certain transform of the partition function
\prtnf\ has a natural fermionic representation. The physical origin of these
fermions is still not completely clear. One way to understand them is to
invoke the Chern-Simons/closed string duality of Gopakumar-Vafa
\gopakumarvafaii, and then the fermionic representation of the large $N$
topological gauge theories \dougcft. Another possible origin is through the M-theory
fivebrane realization of the gauge theory. Mathematically this story is very
much related to the Heisenberg algebra representation in the cohomology of
the moduli space of the torsion free sheaves on ${\bC\bP}^2$, studied in
\nakajima.

\subsec{Dual partition function}

Let $\xi_1, \ldots, \xi_N$ be the complex parameters, $$ \sum_l \xi_l = 0 $$
Consider \eqn\dprtn{\eqalign{Z^{D} ( {\xi} ;p; {\hbar}, {\Lambda} ) = & \cr
& \sum_{\matrix{ &\scriptstyle{p_1, \ldots, p_N \in {\bZ},} \cr &
\scriptstyle{\sum_l p_l = p} }}  \ Z \left( {\hbar}  \left( p_l + {\r}_l
\right) ; {\hbar}, {\Lambda} \right) \ {\exp} \left( \frac{i}{\hbar} \sum_l
p_l {\xi}_l \right)\cr}} Clearly, \eqn\shft{Z^{D}({\xi};p+N; {\hbar},
{\Lambda}) =
 Z^{D}({\xi};p;{\hbar},
{\Lambda})}(shift all $p_l$'s by $1$). Thus, there are essentially $N$
partition functions one could consider. They are labeled by the level $1$
integrable highest weights of $\widehat{\bf sl}_N$. Moreover: \eqn\cls{Z^D
({\xi}; p+1; {\hbar}, {\Lambda}) = Z^D ({\xi}^{+}; p; {\hbar}, {\Lambda})}
where \eqn\clsn{{\xi}^{+}= ({\xi}_2, {\xi}_3, \ldots, {\xi}_{N}, {\xi}_1)}
To extract from $Z^D$ the partition function of interest we perform a
contour integral. In the search for prepotential we are actually interested
in the extremely high frequency Fourier modes of $Z^D$, as we want ${\CF}_0
( {\ba}; {\Lambda})$ as a function of finite $a_l = {\hbar}( p_l + {\r}_l
)$, with ${\hbar} \to 0$. This means that the inverse Fourier transform can
be evaluated using the saddle point, which we already analyzed.

Clearly, \eqn\dlpr{Z^D ( {\xi}; p; {\hbar}, {\Lambda}) = {\exp}
\sum_{g=0}^{\infty} {\hbar}^{2g-2} {\CF}^D_{g}({\xi}; p , {\Lambda})} where
${\CF}^D_{0}$ is in fact $p$-independent, and is given by the Legendre
transform of ${\CF}_0$: \eqn\lgdn{{\CF}^D_0 ({\xi}; {\Lambda}) = i \sum_l
{\xi}_l a_l + {\CF}_0 ( {\ba}; {\Lambda} ), \qquad {\xi}_l = i
\frac{{\p}{\CF}_0}{{\p}a_l} \ ,} i.e. ${\xi}_l$ must be given by the
${\bb}_l$ periods of the differential $dS$.
\remark{Note that in this section ${\xi}_l$ differ by the factor $i$ from
the ${\xi}_l$'s of the previous section.}

\subsec{Free fermions}

\ndt Introduce $N$ free chiral fermions ${\psi}^{(l)}$:
\eqn\frmns{\eqalign{& {\psi}^{(l)}(z) = \sum_{r  \in {\bZ}+{\half}}
{\psi}^{(l)}_r \ z^{-r} \left(\frac{dz}{z}\right)^{\half} \cr &
{\tilde\psi}^{(l)}(z) = \sum_{r  \in {\bZ}+{\half}} {\tilde\psi}^{(l)}_{r} \
z^{-r} \left(\frac{dz}{z}\right)^{\half} \cr &  \cr & \{ {\psi}^{(l)}_{r},
{\tilde\psi}^{(m)}_{s} \} = {\d}_{lm} {\d}_{r+s}\cr}} which can also be
packed into a single chiral fermion $\Psi$ \eqn\frmnsi{\eqalign{&{\Psi}_{r},
{\tilde\Psi}_{r} , \qquad r \in {\bZ} + {\half} \cr  & \{ {\Psi}_{r},
{\tilde\Psi}_{s} \} = {\d}_{r+s} \cr
 {\Psi}(z) = \sum_{r \in {\bZ}
+ {\half}} \ {\Psi}_r z^{-r} \left(\frac{dz}{z}\right)^{\half} & \ , \
{\tilde\Psi}(z) = \sum_{r \in {\bZ} + {\half}} \ {\tilde\Psi}_r z^{-r}
\left(\frac{dz}{z}\right)^{\half} \cr }} in the standard fashion \jm:
\eqn\pckg{{\Psi}_{ N (r + {\r}_l)} = {\psi}^{(l)}_{r}, \quad {\tilde\Psi}_{N
(r - {\r}_l)} = {\tilde\psi}^{(l)}_{r}}
The operators ${\Psi}_r$, ${\tilde\Psi}_s$ act in the standard fermionic
Fock space ${\CH}$ (sometimes called an infinite wedge representation).
It splits as a sum of Fock subspaces with fixed $U(1)$ charge, defined
below:
$$
{\CH} = \oplus_{p \in {\bZ}} {\CH}[p]
$$
\ndt Introduce affine $\widehat{U(N)}_{1}$ currents, which act within ${\CH}[p]$ for any $p$:
\eqn\crrnt{J^{ln}(z) =
: {\psi}^{(l)} {\tilde\psi}^{(n)} : = \sum_{k \in {\bZ}} \frac{dz}{z^{k+1}}
\sum_{r \in {\bZ}+{\half}} : {\psi}^{(l)}_{r} {\tilde\psi}^{(n)}_{k-r} :}
Here we normal order with respect to the vacuum $\vert 0 \rangle$, which is
annihilated by \eqn\vcm{\eqalign{& {\Psi}_{r} \vert 0  \rangle = 0, \quad r
> 0\cr & {\tilde\Psi}_s \vert 0 \rangle = 0, \quad s > 0 \cr}} which is
equivalent to \eqn\vcmm{\eqalign{& {\psi}^{(l)}_{r} \vert 0 \rangle = 0,
\quad r > 0, \cr & {\tilde\psi}^{(l)}_{s} \vert 0 \rangle = 0, \quad s >
0\cr}} The normal ordered product is simply: \eqn\nrmlo{\eqalign{ & :
{\Psi}_r {\tilde\Psi}_s : = \Biggl\{ \matrix{& {\Psi}_r {\tilde\Psi}_s,
\qquad s >0 \cr - & {\tilde\Psi}_s {\Psi}_r, \qquad r > 0 \cr}\cr & :
{\psi}_r^{(l)} {\tilde\psi}_s^{(m)} : = \Biggl\{ \matrix{& {\psi}_r^{(l)}
{\tilde\psi}_s^{(m)}, \qquad s >0 \cr - & {\tilde\psi}_s^{(m)}
{\psi}_r^{(l)}, \qquad r > 0 \cr}\cr}} One can also introduce vacua with
different overall $U(1)$ charges: \eqn\vcmi{\eqalign{& {\Psi}_r \vert p
\rangle = 0, \quad r> p, \cr & {\tilde\Psi}_s \vert p \rangle =0 , \quad s >
-p \cr &
{\vert p \rangle} \in {\CH}[p] \cr}} It is also useful to work with ${\Psi}$ and the corresponding
$\widehat{U(1)}_{1}$ currents: \eqn\crrps{{\CJ}(z) = : {\Psi}
{\tilde\Psi}:(z) = -\frac1{N} \sum_{l,m} J^{lm}(z^{-N}) z^{l-m}} and
Virasoro generators: \eqn\lnol{L_0 = \sum_{r \in {\bZ} + {\half} }  r  \ :
{\Psi}_r {\tilde \Psi}_{-r} : } Note: \eqn\zaryad{L_0 \vert p \rangle = {p^2
\over 2} \vert p \rangle} \subsubsec{Bosonization} It is sometimes convenient to
work with the chiral boson: \eqn\cbsn{\eqalign{{\phi}(z) = q - i {\CJ}_0 \
\lg{z} + & i \sum_{n \neq 0} \frac1{n} {\CJ}_n \ z^{-n} \cr
 {\p}{\phi} \equiv & z{\p}_z {\phi} = - i{\CJ} \cr
& [ q, J_0 ] = i \cr {\Psi}(z) = : e^{i {\phi}(z)} : \ , & \qquad
{\tilde\Psi}(z) = :e^{-i\phi(z)}:\cr}} We will also use a truncated boson,
with the zero mode $q$ removed: \eqn\trbsn{{\varphi}(z) = - i {\CJ}_0 \
\lg{z} + i \sum_{n \neq 0} \frac1{n} {\CJ}_n \ z^{-n}}
The space ${\CH}[p]$ is actually an irreducible representation of the
Heisenberg algebra generated by ${\CJ}$:
\eqn\hilp{{\CH}[p] = {\rm Span}_{n_1, \ldots, n_k >0 } \ {\CJ}_{-n_1} \ldots {\CJ}_{-n_k} \vert p
\rangle, \qquad
{\CJ}_0 \Biggr|_{{\CH}[p]} = p }
We also recall:
\eqn\frmshlp{{\Psi} : {\CH}[p] \to {\CH}[p+1], \qquad {\tilde\Psi}: {\CH}[p]
\to {\CH}[p-1]}
\subsec{Dual partition function as a current correlator} We claim:
\eqn\dprtnfrm{Z^{D} ({\xi}; p; {\hbar}, {\Lambda} ) = \langle p \vert
e^{{\sfrac1{\hbar}} \oint {\Tr} E_{+} (z) J (z)} e^{\oint
\sfrac1{\hbar}{\Tr} H(z) J (z) } {\Lambda}^{2 L_0} e^{-{\sfrac1{\hbar}}\oint
{\Tr} E_{-}(z) J(z)} \vert p \rangle}where the matrices $E_{\pm}, H$ are
given by: \eqn\mtrcs{\eqalign{& E_{+}(z) = z E^{N,1} + \sum_{l=2}^{N}
E^{l-1,l}, \cr & H ( z) = \sum_l {\xi}_l \ E^{l,l} \cr & E_{-}(z) = z^{-1}
E^{1,N} + \sum_{l=2}^{N} E^{l,l-1} \cr}}

\subsec{Affine algebras and arbitrary gauge groups} Note that \dprtnfrm\ can
be written using the Chevalley generators $e_i, f_i, h_i$ of the affine Lie
algebra $\widehat{\bf sl}_N$: \eqn\drptnfrmi{\eqalign{Z^{D}({\xi};p;{\hbar},
{\Lambda}) = & \left( u_{\hbar}, {\Lambda}^{2h_0} e^{{\bH}_{\xi}} u_{\hbar}
\right)_{V_{{\o}_{p}}}, \cr & \cr & u_{\hbar} = {\exp} \left(
\frac{N}{\hbar} \sum_{l=1}^{N} f_{l-1} \right) \ v_{0} \cr &
{\bH}_{\xi} =
\frac1{\hbar} \sum_{l=1}^{N-1} \left({\xi}_{l}-{\xi}_{l+1}\right) h_{l}
\cr}} where $V_{{\o}_{p}}$ is the integrable highest weight module with the
highest weight vector $v_0$ (annihilated by the simple roots $e_l$) and the
highest weight ${\o}_p$, $p=1, \ldots, N$.

\ndt The formula \drptnfrmi\ has an obvious generalization to any simple Lie
algebra $\widehat\bg$. Conjecturally, it gives the partition function in the
$\Omega$-background in the ${\CN}=2$ gauge theory with the gauge group,
$S$-dual to $G$ (Langlands dual).

\subsec{Dual partition function and ${\bf gl}({\infty})$} We now return to
the $A_{N-1}$ case.

\ndt In the language of the single fermion $\Psi$ the formula \dprtnfrm\
reads  as: \eqn\tda{Z^{D}({\xi}; p; {\hbar}, {\Lambda} ) = \langle p \vert
e^{{{\CJ}_1 \over {\hbar}}} e^{{\bH}_{\xi}} {\Lambda}^{2 L_0} e^{{\CJ}_{-1}
\over {\hbar}} \vert p \rangle} where ${\bH}_{\xi}$ is a diagonal matrix: $$
{\bH}_{\xi} =\frac1{\hbar} \sum_r {\xi}_{(r + {\half}){\rm mod}N} : {\Psi}_r
{\tilde\Psi}_{-r} : $$ Clearly, $[{\bH}_{\xi}, L_0 ] =0$. The formula
\tda\ expresses the dual partition function as an average with the
Plancherel measure of the $N$-periodic weight $e^{{\bH}_{\xi}}$.

\remark{On Toda equation} The function $Z^D$ obeys Toda equation (cf. \todalit):
\eqn\tdde{4 {\p}^2_{\lg{\Lambda}} \lg{Z^{D}(p)} = {{Z^{D}(p+1)
Z^{D}(p-1)}\over{(Z^{D}(p))^2}}}
In fact, the formula \tda\ identifies $Z^D$ as the tau-function of the Toda
lattice hierarchy, thanks to the results of \sasha\op\takashi, with the specific
parameterization of the times. We hope to return to this property of the
partition function in a future publication.

\subsec{The $U(1)$ case} \ndt In order to understand \tda\dprtnfrm\ we
consider first the case $N=1$. As the Plancherel measure is $a$-independent
in this case, we can study both the partition function and its dual at ease.
\ndt Let us represent the states in the fermionic Fock space as
semi-infinite functions of $\Psi_{r}$. The charge $p$ vacuum $\vert p
\rangle$ in this representation corresponds to the product: \eqn\vcmn{\vert
p \rangle = \ \prod_{r > -p}^{\longrightarrow} {\Psi}_r \equiv {\Psi}_{-p
+\sfrac1{2}} {\Psi}_{-p +\sfrac{3}{2}} \ldots } To every partition ${\bk}$
there corresponds the so-called {\it charge $p$ partition state}:
\eqn\chmprt{\eqalign{\vert p; {\bk} \rangle = & \prod_{i =1, 2,
\ldots}^{\longrightarrow} {\Psi}_{-p - \sfrac1{2} + i - k_i} \ = \prod_{1
\leq i \leq n}^{\longrightarrow} {\Psi}_{-p- \sfrac1{2} + i - k_i} \prod_{1
\leq i\leq n}^{\longleftarrow} {\tilde\Psi}_{p+\sfrac1{2}-i} \ \vert p
\rangle = \cr & \prod_{1 \leq i \leq n}^{\longrightarrow} {\Psi}_{-p-
\sfrac1{2} + i - k_i} \vert p-n\rangle\cr}} These states form a complete
basis in the space ${\CH}[p]$. The important fact, responsible
for \tda, is the variant of the boson-fermion correspondence:
\eqn\bsnf{\eqalign{{\exp} \frac1{\hbar} {\CJ}_{-1} \ \vert p \rangle = &
\sum_{{\bk}} {{\m}({\bk})\over{{\hbar}^{\vert \bk \vert}}} \ \vert p ; {\bk}
\rangle \cr &  \cr}} where factor ${\m}({\bk})$ is the Plancherel measure
\planch. It follows: \eqn\tdaabl{Z ( a, {\hbar}, {\Lambda}) =
{\Lambda}^{{}^{\sfrac{a^2}{{\hbar}^2}}} \ \langle 0 \vert e^{-\frac1{\hbar}
{\CJ}_1} {\Lambda}^{2 L_0} e^{\frac1{\hbar} {\CJ}_{-1}} \vert 0 \rangle}
which for $a = {\hbar}p$ can also be written as \eqn\tdaabli{Z ( a, {\hbar},
{\Lambda}) = \langle p \vert e^{-\frac1{\hbar} {\CJ}_1} {\Lambda}^{2 L_0}
e^{\frac1{\hbar} {\CJ}_{-1}} \vert p \rangle} \mdp To make use of \bsnf\ we
{\it blend} $N$ partitions ${\bk}_l$, $l = 1, \ldots , N$ into a single one,
$\bK$. Let $p_1, \ldots , p_N$ be some integers. Consider the following
countable set of distinct integers: \eqn\dsnt{\{ N \left( p_l + k_{li} - i
\right) + l-1  \vert \ l = 1, \ldots, N, \quad i \in {\bN} \} = \{ p+K_{I} -
I  \vert I \in {\bN}\} } where the sequence $K_{1} \geq K_{2} \geq \ldots $
is defined by \dsnt\ and the condition that it stabilizes to zero. Likewise,
$p$ is determined from \dsnt\ and depends only on $p_l$'s: \eqn\ttchrg{p =
\sum_l p_l} Let $$a_l = {\hbar} \left( p_l + {\r}_l \right)\ . $$ The size
of the blended partition $\bf K$ is expressed in terms of $a_l$'s (see
appendix $\bA$): \eqn\szcmb{\vert {\bf K} \vert = \frac{1-N^2}{24} -
\frac{p^2}{2} +  N \sum_{l} \frac{{\tilde a}_l^2}{2 \hbar^2} + N \sum_{l,i}
k_{li}} It is now straightforward to evaluate ${\m}_{\vec\bk} := {\m}({\bf
K})$: \eqn\vln{{\m}_{\bf K}^2 = Z_{\vec\bk}(a; {\hbar}), \qquad} and arrive
at \tda.

\newsec{${\CN}=2$ THEORY WITH ADJOINT HYPERMULTIPLET}

In a sense the most interesting ${\CN}=2$ theory is the theory with the
massive hypermultiplet in the adjoint representation. This theory is
ultraviolet finite, and is thus characterized by the microscopic coupling
${\tau}_0 = {{\vartheta}_0 \over 2\pi} + {{4\pi i }\over{g_0^2}}$, and by
the mass ${\bm}$ of the hypermultiplet. It is convenient to use the nodal
parameter $q = e^{2{\pi}i{\tau}_0}$ to count instantons.

In this case the prepotential of the low-energy effective theory is expected
to have the following expansion:
\eqn\prptnadj{\eqalign{{\CF}_0 ( {\ba}, {\bm}, q) = & {\pi}i {\tau}_0 \ \sum_l  \ a_l^2
- \qquad\cr
- {1\over 2}
\sum_{l \neq n} & \left[ (a_l - a_n)^2 {\lg{a_l - a_n}} -  (a_l - a_n+ {\bm})^2 {\lg{a_l - a_n+{\bm}}}
\right] +\cr & +
\sum_{k=1}^{\infty} q^k f_{k} ( {\ba}, {\bm})\cr}}
The previous calculations of the coefficients $f_k$ for low values of $k$
were done in \hollowood.
One of the remarkable features of the prepotential of the low-energy
effective theory of the theory with adjoint hypermultiplet is its relation
to the elliptic Calogero-Moser system\cm. Indeed, in \witdonagi\ an ansatz
for the family of curves encoding the prepotential was proposed, using a
version of Hitchin system \hitchin. This very construction of the elliptic
Calogero-Moser system was found earlier \nikgor. One of the advantages of
this realization is its simple extension to the Lie groups other then
$SU(N)$ (see, e.g.\markman). The coefficients $f_{k}$
of the prepotential of the Calogero-Moser system were calculated for $k=1,2$
in, e.g. \experiment.

\subsec{Partition function}

The partition function of the theory with adjoint hypermultiplet in the
$\Omega$-background is explicitly calculable (in \swi\ the countour integral
representation was given, the poles of the integral are exactly the same as
those of the contour integral for the pure theory) and the answer is (${\m} = {\bm}/{\hbar}$):
\eqn\prtfnms{\eqalign{  Z ({\ba}, {\bm} ; {\hbar}, q )  & = {\exp}
\sum_{g=0}^{\infty} {\hbar}^{2g-2} {\CF}_{g} ( {\ba}, {\bm} ;q) \cr = &
q^{\sinv{2{\hbar}^2} \sum_l a_l^2 - \sinv{24} N ({\m}^2-1)} \sum_{{\vec\bk}} q^{\vert {\vec\bk} \vert} \ Z_{{\vec\bk}}
({\ba}, {\bm} ; {\hbar},q )\cr}} where
\eqn\partchl{
 \eqalign{& Z_{\vec\bk} ({\ba},  {\bm}; {\hbar}, q) =
Z^{pert} ({\ba},  {\bm}; {\hbar}, q) \ {\m}^2_{\vec\bk}({\ba}, {\hbar}) \times \cr
& \qquad\qquad \prod_{(l,i) \neq (n,j)}
 {{\left( a_{l} - a_{n} + {\bm} +
{\hbar} \left( j-i\right)\right) }\over{ {\left( a_{l} - a_{n} + {\bm}+
{\hbar} \left( k_{l,i}- k_{n,j} + j - i\right)\right) }}}
 \cr & \qquad Z^{pert}({\ba},  {\bm}; {\hbar}, q) = {\exp} \sum_{l,n}  \zq{a_{l} - a_{n}} -
\zq{a_{l} - a_{n} + {\bm}} \cr}}

\subsec{Abelian theory}

\ndt Let us consider $N=1$ case first. Let ${\m} = \frac{\bm}{\hbar}$. In
the $N=1$ case we are to sum over all partitions ${\bk}$: \eqn\abl{Z (a,
{\bm} ; {\hbar}, q) = q^{\sfrac{a^2}{2{\hbar}^2} + \sinv{24}\left({\m}^2 -
1\right) } e^{-\zq{\bm}} \ \sum_{{\bk}} q^{\vert {\bk} \vert}
\prod_{\squarel
\in {\bk}} \left( h({\square})^2 - {\m}^2 \over h({\square})^2 \right)} The fact that
\prtfnms\ reduces to \abl\ for $N=1$ is a consequence of a simple identity
between the Chern characters: \eqn\idnt{\sum_{ i < j} \ e^{{\hbar}( k_i -
k_j + j - i)} - e^{{\hbar}(j-i)} - e^{{\bm} + {\hbar}(k_i - k_j + j - i)} +
e^{{\bm} + {\hbar}(j-i)} = \sum_{{\squarel} \in {\bk}} e^{{\bm} + {\hbar}
h({\square})} - e^{{\hbar} h({\square})}}
 Another expression for \abl\ will be useful immediately:
\eqn\abli{\eqalign{Z (a, {\bm} ; {\hbar}, q) = & q^{\sfrac{a^2}{2{\hbar}^2}}
e^{-\zq{\bm}} \ {\CZ}({\m}), \qquad {\CZ}({\m}) = q^{\sinv{24}\left({\m}^2 -
1\right) } \sum_{\bk} q^{\vert {\bk} \vert}{\CZ}_{\bk}({\m}) \cr
 & {\CZ}_{\bk}({\m}) = \prod_{i=1}^{{\ell}({\bk})}
\frac{({\ell}({\bk}) + k_i - i + {\m})! ({\ell}({\bk})+k_i - i -
{\m})!({\ell}({\bk})-i)!^2}{({\ell}({\bk}) - i + {\m})! ({\ell}({\bk}) - i -
{\m})!({\ell}({\bk})+k_i-i)!^2} \times \cr &
\qquad\qquad\qquad\qquad\qquad\qquad\qquad\qquad \times \frac{{\det}_{1\leq
i,j\leq {\ell}({\bk})}\Vert \frac1{k_i - k_j +j - i +{\m}}
\Vert}{{\det}_{1\leq i,j\leq {\ell}({\bk})}\Vert \frac1{j - i +{\m}}
\Vert}\cr}} \pro{The analogue of the formula \tdaabli\ for
the theory with adjoint matter is: \eqn\trc{{\CZ} ( {\m})   =
q^{\frac{{\m}^2}{24}}{\Tr}_{{\CH}_{0}} \ q^{L_{0}-\sfrac1{24}} \
{\CV}_{\m}(1) } where \eqn\bsn{{\CV}_{\m}(z) = \ : e^{i \m {\vf}(z)} : \ = \
{\exp}\left(- \m \sum_{n>0} {\CJ}_{-n}\frac{z^{n}}{n}\right) z^{{\m}{\CJ}_0}
\ {\exp}\left(\m \sum_{n> 0} {\CJ}_{n}\frac{z^{-n}}{n}\right) }
${\varphi}(z)$ is the bosonic field \trbsn, and ${\CH}_{\l}$ is the charge
${\l}$ subspace of the fermionic Fock space.}

\proof{We represent the trace \trc\ as a sum over partition states: $\vert
p;{\bk}\rangle$, where $p = \frac{a}{\hbar}$ is fixed. We start by
evaluating:
\eqn\mtrel{\langle {\bk} \vert {\CV}_{\m}(1) \vert {\bk} \rangle
= \frac{{\det}_{1 \leq i,j\leq {\ell}({\bk})}\Vert G_{k_i-i+{\ell}({\bk}),
k_j - j+{\ell}({\bk})} \Vert}{{\det}_{1 \leq i,j\leq {\ell}({\bk})}\Vert
G_{-i+{\ell}({\bk}),  - j+{\ell}({\bk})}\Vert} } where we employ Wick's
theorem, and where for $i,j \in {\bZ}_+$:
\eqn\goper{\eqalign{G_{ij} = & (-)^{i-j}
\frac{(i+{\m})!(j - {\m})!}{i!j!(i-j+{\m})}\ \frac{{\rm
sin}{\pi}{\m}}{{\pi}{\m}}\cr & =  {\rm Coeff}_{x^{i}y^{j}} \left(\frac{1-y}{1-x} \right)^{\m} \frac1{1-xy} = \cr&
\qquad\qquad = \langle 0 \vert {\tilde\Psi}_{j+\half} {\CV}_{\m}(1)
{\Psi}_{-\half-i} \vert 0\rangle \cr}}
The second line of \goper\
follows from: $$ \left[{\m} - y{\p}_y + x{\p}_x \right]
\left\{ \left(\frac{1-y}{1-x} \right)^{\m} \frac1{1-xy}\right\} =
{\m}
(1-y)^{{\m}-1}(1-x)^{-{\m}-1} $$
while the last line is most easily derived from the current action on the fundamental
fermions ${\Psi}, {\tilde\Psi}$: \eqn\bznz{\eqalign{ & \langle 0
|{\tilde\Psi}\left(y^{-1}\right) {\CV}_{\m}(1) {\Psi}(x) | 0  \rangle =\cr & \quad = {\exp} \left( {\m} \sum_{n
> 0} \frac1{n} \left( x^n -
y^n \right) \right) \times  \langle 0 |{\tilde\Psi}\left(y^{-1}\right) {\Psi}(x) | 0
\rangle = \cr & \qquad\qquad =\left(  \frac{ 1- y}{1-x}\right)^{\m}
\frac{(dx)^{\half}(dy)^{\half}}{1-xy} \cr}}from which \goper\ follows by employing the
expansion \frmnsi. } \remark{Note that an immediate
consequence of the equalities \abl\  and \trc\
is following identity
\eqn\trcc{
\eqalign{\prod_{n>0} (1-q^n)^{\mu^2-1}  &=
\sum_{\bk} q^{|{\bk}|}
\,\prod_{\square\in\bk}
\frac{h(\square)^2-\mu^2}{h(\square)^2} \cr
&\cr
& {\CZ}({\m}) = {\eta}(q)^{{\m}^2-1}\cr}
}
(see appendix $\bf C$ for notations)
which interpolates between the
Dyson-Macdonald formulas for the root systems $A_n$.
Indeed, when $\mu=n$, the exponent on the left in
\trcc\ equals
$$
\dim \, SU(\mu) = \mu^2-1\,,
$$
while on the right we have a sum over all partitions
with no hooks of length $\mu$, which one can
easily identify with the sum over the weight lattice
of $SU(n)$ entering the   Dyson-Macdonald formula.
}

\subsubsec{Higher Casimirs and Gromov-Witten theory of elliptic curve}

Just as in \lmn\ we can consider deforming the theory by the higher
Casimirs. Just as in \lmn\ these are represented in the free fermion
formalism by the ${\CW}$-generators, and the trace \trc\ becomes the
$W_{1+\infty}$ character, in the presence of the ``vertex'' operator
${\CV}_{\frac{\bm}{\hbar}}$.

For ${\bm}=0$ this trace becomes {\it exactly} the $W_{1+\infty}$ character,
as studied in \robbert\ob. As shown in \op\ this trace also coincides with
the all-genus partition function of the Gromov-Witten theory of an elliptic
curve (see also \robberti). It would be nice to find a Gromov-Witten dual of
the ``vertex'' operator ${\CV}_{\m}$, and establish precise
relation of \trcc\ to \op\ob\robbert.

Note that the ``flow to the pure ${\CN}=2$ theory'' which is represented by
the limit ${\bm} \to \infty$, $q \to 0$, s.t. ${\Lambda}^2 = {\bm}^2 q$
stays finite, leads to the dual Gromov-Witten theory of  ${\bC}{\bP}^1$ with
$\lg{\Lambda^2}$ playing the r\^ole of the K\"ahler class.
\subsec{Nonabelian case}

Using \abl\ it is simple to express the partition function \prtfnms\ in
terms of free fermions: \eqn\nnbl{Z({\ba} ; {\hbar}, {\bm} , q ) =
{\Tr}_{{\CH}_{{\ba}\over{\hbar}}^{(N)}} \left( q^{L_{0}} {\CV}_{\m}(1) \right)} where $${\CH}_{\bp}^{(N)} =
\otimes_{l=1}^{N} {\CH}[{\bp}_{l}]$$ stands for the Fock space of $N$ free
fermions, with specified charges $( {\bp}_1,\ldots, {\bp}_N)$  ($U(1)^{N}$
weight, in the language of affine Lie algebras).

We also have an expression for the dual partition function:
\eqn\nnbli{Z^{D}({\xi}, p ; {\hbar}, {\bm}, q) = {\Tr}_{{\CH}_{p}} \left(
q^{L_{0}} {\CV}_{\m}(1) \ e^{{\bH}_{\xi}}\right)}
Of course, the modular properties of the partition functions
\abli\trcc\nnbli\
reflect the S-duality of the ${\CN}=4$ theory \vafawitten.

\subsec{Path representation}

As in the pure gauge theory case it is extremely useful to represent the
partition function as a sum over the profiles -- paths $f(x)$. The result is:
\eqn\pramp{\eqalign{Z(& {\ba}, {\bm};   {\hbar}, q) = \sum_{f \in
{\Gamma}_{\ba}^{discrete}} Z_{f} ({\ba}, {\bm}; {\hbar}, q) \cr & Z_{f} ({\ba}, {\bm};
{\hbar}, q) = \exp \ \Biggl(  -\frac1{4} \int dx dy \ f''(x) f''(y) \zq{x-y}
+ \cr & \qquad\qquad\qquad\qquad\qquad\qquad\qquad\qquad +\frac1{4} \int dx
dy \ f''(x) f''(y) \zq{x - y + {\bm}} + \cr &
\qquad\qquad\qquad\qquad\qquad\qquad\qquad\qquad\qquad\qquad +\frac1{4{\hbar}^2} \lg{q} \int dx
\ x^2\ f''(x) \Biggr)\cr}}
Interchanging $x$ and $y$ one sees that this
is symmetric with respect to $\bm\mapsto - \bm$.

\subsec{Variational problem for the prepotential} \ndt Again, just as in the case of the pure ${\CN}=2$
gauge theory we are looking for the extremum of a certain action functional
on the space ${\Gamma}_{\ba}$ of profiles $f(x)$ with fixed partial charges. By adding the surface tension term
to the action we get a minimization problem on the space of all
profiles, i.e. functions that satisfy $$ f(x) - N|x|
=0 \,, \quad |x| \gg 0 \,, $$ and the Lipschitz condition $|f'(x)|\le N$.
The Lipschitz condition in our case will be satisfied automatically, since
all extrema will be convex. We also impose the total zero charge condition,
$a=0$, i.e.
\eqn\zrch{\int_{\bR} x \ f''(x) \ dx  = 0}
In this section we set
${\bm} = im$.

\subsubsec{Path energy and surface tension}
Let \eqn\yadro{ L(x) =
\frac{1}{2} \, x\ \lg{\frac{x^2}{x^2+m}}  - m \ {\rm arctan} \frac{x}{m}} We
have: \eqn\deryadra{\eqalign{& L'(x) = \frac{1}{2} \,
\lg{\frac{x^2}{x^2+m^2}} , \cr & L''(x) = \frac{1}{x} - \frac{1}{2}\,
\left(\frac{1}{x+im}+\frac{1}{x-im}\right)\,. \cr}} \mdp
We introduce a generalization of the energy functional \actn:
\eqn\actnadj{\eqalign{ {\CE}_{q,m} (f) = & - \frac1{2}\intp{x<y} ( N + f'(x))(N-f'(y))
L'(y-x) \ dx\ dy + \cr & - \frac{i{\pi\tau}}{2} \int_{\bR} \ x^2 \ f''(x) dx\cr}}
where the first term is the leading $\hbar \to 0$ asymptotics
of \pramp, and the
second term comes from the weight $q^{|\vec\bk|}$. In the limit $q \to 0, m
\to \infty$, so that
${\Lambda} = m q^{1\over 2N}$ stays finite, ${\CE}_{q,m} \to
{\CE}_{\Lambda}$ (this reflects the flow of the theory with adjoint
hypermultiplet to the pure ${\CN}=2$ super-Yang-Mills theory).
\mdp
The functional
which we want to maximize is:
\eqn\func{{\CS}_{q,m}(f) = - {\CE}_{q,m}(f) + \half\int \sigma(f'(x)) \, dx \,,}
where the last term
is the usual surface tension term. Again, the functional \func\ is convex,
therefore every local maximizer is also a global maximizer.
\subsubsec{Variational equations}
Varying $f'(x)$ in \func, and integrating by parts, we
obtain the following equation
\eqn\eextr{
  \intp{y\neq x} f''(y)\, L(y-x) \, dy  -  \sigma'(f'(x))+ 2\pi i \tau x  = \const
}
which should be satisfied at any point of continuity of $\sigma'(f'(x))$.
The constant in \eextr\ is the Lagrange multiplier corresponding to the
constraint \zrch.
At the points of discontinuity of $\sigma'(f'(x))$, the equality \eextr\
should be replaced by the corresponding two-sided inequalities.

The function $\sigma'$ is piecewise constant, therefore differentiating
\eextr\ with respect to $x$ we get the equation
\eqn\eextwo{
   \int f''(y)\, L'(y-x) \, dy  = 2\pi i \tau \,. }
\subsec{The spectral curve}
The facets of the limit shape correspond to the singularities of the
surface tension $\sigma$, therefore the maximizer $f$ will have $N-1$ of
them. This means that the support of $f''$ will be $N$ intervals
(bands): \eqn\sprtsi{\supp f'' = \bigcup_{l=1,\dots,N} \supp f''_l}
and, as in \sprts, we denote:
\eqn\sprtsii{\supp f''_l = [{\a}_l^{-}, {\a}_l^{+}]}
Consider
the resolvent $R(z) \equiv R_{f}(z; 0,0)$ introduced in \rslvnt. It has jump discontinuities along the bands with the jump
equal to ${\pi} i f''(x)$.
Set, by definition (cf. \hoppe\kkn), $$ G(z) = R\left(z-\frac{im}{2}\right) -
R\left(z+\frac{im}{2}\right) \,. $$ This function is analytic in the domain
$U$ which is the Riemann sphere with 2N symmetric cuts $$ U = \bar {\bC}
\backslash \bigcup_{l=1}^{N} \, \left[{\a}_l^{-}\pm \frac{im}{2}, {\a}_l^{+}\pm \frac{im}{2}\right]
\,. $$ We have $$ G(z) = O(z^{-2}) \,, \quad z\to \infty\,, $$ and hence we
can consider the multivalued function $$ \int_{\infty}^z G(y) \, dy \,. $$
The period of $\int G(y) \, dy$ around any cut is equal to the integral of
the jump of $G(z)$ along the cut. By construction, the jump of $G(z)$ is
$\pm {\pi} i f''(x)$ and, as $\int_{{\a}_l^{-}}^{{\a}_l^{+}} f''(y) \, dy = 2$
we have the well-defined function \eqn\ffun{\eqalign{ F(z) = & \frac1{2\pi i } \int_{\infty}^z G(y)
\, dy  \mod \, \bZ \cr & = \frac1{4\pi i} \int_{\bR} dy \ f''(y)\
\lg{z - \frac{im}{2}-y \over z + \frac{im}{2} -y} \ \mod \ \bZ
\cr}}
We will show that $F$ is  an $N$-fold branched
cover of the standard cylinder $$ C_{\tau}=\{{\varpi} \ \vert \ |\Im({\varpi})|< -i \tau/2 \} \mod  \,
\bZ $$ by the domain $U$. Clearly, $F$  maps the infinity of $U$ to the
origin ${\varpi} = 0$ of $C_{\tau}$.
\Figx{9}{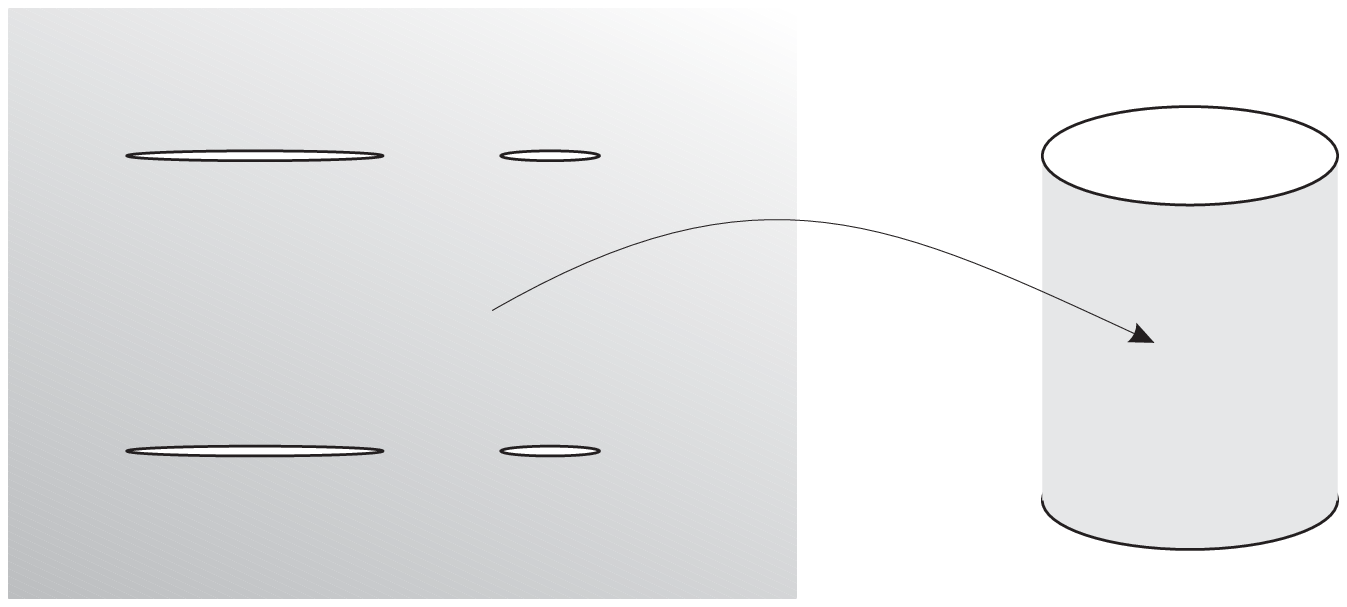}{Covering finite cylinder}
The following proposition describes the behavior of the function $F(z)$
on the both sides of the two cuts corresponding to any band $[{\a}_l^{-},{\a}_l^{+}]$
\pro{
 For any $x\in [{\a}_l^{-},{\a}_l^{+}]$, we have
\eqn\Fc{ F\left(x+\frac{im}{2}\pm i0\right) - F\left(x-\frac{im}{2}\mp i0
\right) = \tau \,. }
}
\proof{ This is, in fact, a reformulation of \eextwo. Indeed, let $x_\pm$
denote the two arguments of $F$ in \Fc. We need to show that
\eqn\intG{\eqalign{&
  \tau=\int_{x_-}^{x_+} G(y) \, dy = \cr
& \frac{1}{4\pi i} \int_{x_-}^{x_+}
  dz \, \int_{\bR} \left(\frac{1}{y-z+\frac{im}2} -
\frac{1}{y-z-\frac{im}2}\right) \,
  f''(y)\, dy \,. \cr}}
Since the integrand is nonsingular on the domain of integration, we can
change the order of integration, which gives
\eqn\intGtwo{
  \frac{1}{2\pi i} \int \, L'(y-x) \, f''(y) \, dy = \tau \,,
}
which is precisely \eextwo.}

\ndt
Since $f''(x)$ is real, we have $$ R(\bar z) =  \overline{R(z)}\,, $$
from which it follows that \eqn\rlts{ G(\bar z) = - \overline{G(z)}\,, \quad
F(\bar z) = \overline{F(z)} \,. } Proposition  \Fc\ now implies that $$ \Im
F\left(x\pm \frac{im}{2}\right) = \mp i \tau/2 \,, \quad x\in [{\a}_l^{-},{\a}_l^{+}] \,.
$$ which means that $F$ maps the boundary $\partial U$ to the boundary
$\partial C_{\tau}$. It follows that $F$ maps $U$ onto $C_{\tau}$ and, moreover, since $F$
has degree $N$ on the boundary, the map $$ F: U \to C_{\tau} $$ is an $N$-fold
branched cover.
\mdp Proposition \Fc\ also gives the analytic continuation of the function
$F(x)$ across the cuts of $U$. The Riemann surface ${\CC}$ of the function
$F$ is obtained by taking countably many copies of the domain $U$ and
identifying the point $$ x+\frac{im}{2}\pm i0\,, \quad x\in [{\a}_l^{-},{\a}_l^{+}] $$ on
one sheet with the point $$ x-\frac{im}{2}\mp i0 $$ on the next sheet.

By construction, $F$ extends to the map from $\CC$ to the infinite cylinder
$$ F: \CC \to \bC \mod \, \bZ $$ satisfying the property
\eqn\Feq{ F^{-1}({\varpi}+\tau) = F^{-1}({\varpi}) + im  \,. }
More geometrically this means that the curve ${\CC}$ is imbedded into the
total space ${\CA}$ of affine bundle over the elliptic curve $E_{\tau} = {\bC}/{\bZ}\oplus{\tau\bZ}$ which is
obtained from $C_{\tau}$ by identifying its boundaries:\eqn\embcr{{\CC} \subset {\CA} = \{ ( z, {\varpi} ) \vert z \in {\bC},
{\varpi} \in {\bC} \} / (z, {\varpi}) \sim ( z, {\varpi} + 1 ) \sim ( z +
{\bm}
, {\varpi} + {\tau} )}
The map $F$ is the
restriction on ${\CC}$ of the projection ${\CA} \to E_{\tau}$. The
coordinate $z$ is the coordinate on the fiber of ${\CA}$.

This implies, in turn,  that $\CC$ is the spectral curve of the elliptic Calogero-Moser
system \cm.
Indeed, the latter is the spectral curve of the Lax operator (see appendix $\bf C$):
\eqn\EcmLax{\eqalign{& {\CC}: \qquad {\Det}_{l,n}\left( L({\varpi}) - z \right) = 0 \cr \qquad & \qquad
L_{ln}({\varpi}) = {\d}_{ln} \left( p_n + {\bm} \frac1{2\pi i} \lg{{\t}_{11}({\varpi})}'\right) +
 \frac{\bm}{2\pi i} \left( 1 - {\d}_{ln} \right)
\frac{{\t}_{11}({\varpi} + q_l - q_n){\t}_{11}^{\prime}(0)}{{\t}_{11}({\varpi}){\t}_{11}(q_l - q_n)}  \cr}}
and is naturally viewed as the holomorphic curve in ${\CA}$
representing the homology class of $N[E_{\tau}]$ under the
identification ${\CA} \approx {\bC} \times E_{\tau}$: $(z, {\varpi}) = ( p + {\bm} \lg{{\t}_{11}({\varpi})}', {\varpi})$.
This condition, together with the behaviour near ${\varpi} = 0$ fixes
uniquely the $N$-parametric family of curves ${\CC}$. The coordinates
$u = (u_1, \ldots, u_N)$ on the base of the family of curves can be defined using the characteristic polynomial
of the Lax operator \EcmLax. An alternative set of (local) coordinates is given by the so-called action variables, which
can be defined as a set of periods of the complexified Liouville one-differential $z d{\varpi}$ on ${\CC}$.
In our construction these periods are linear combinations of $a_l$'s,
${\xi}_l$'s and $m$, see below.
\remark{ For $N=1$, the function $F^{-1}({\varpi})$ is, up to normalization, the
classical Weierstra\ss\ function $\zeta({\varpi})$.}The limit shape $f$ is
reconstructed from the fact that $f'(x) \mod 2\bZ$ is equal to the jump of
the function $F(x)$ across the cut of $U$.

\subsubsec{Periods}
We shall now study the periods of the differential \eqn\swdfadj{dS = z d{\varpi} \vert_{\CC} = zdF(z) = \frac1{2\pi i}
z \left( R \left( z - \frac{im}{2} \right) - R \left( z + \frac{im}{2} \right) \right) dz} on the
Riemann surface $\CC$.

\ndt
For $l=1, \ldots, N$ let ${\ba}_l^{\pm}$ denote the 1-cycle, which circles
around the cut $[ {\a}_{l}^{-} \pm \frac{im}{2}, {\a}_{l}^{+} \pm
\frac{im}{2} ]$. Clearly:
\eqn\aper{\oint_{{\ba}^{\pm}_l} dS = a_l \pm \frac{im}{2}}
For $l=1,\dots, N-1$, let ${\b}_l$ be the cycle on
$\CC$ joining some points $x + \frac{im}{2}$ and $y+\frac{im}{2}$, where $$
x \in \left({\a}_l^{-},{\a}_l^{+}\right)\,, \quad y \in \left({\a}_{l+1}^{-},{\a}_{l+1}^{+}\right)\,, $$
above both cuts on some sheet of $\CC$ and below both cuts on the next sheet
of $\CC$.

By construction of $\CC$, we have
$$
  \oint_{\b_l} z dF(z)  =
\int_{x+\frac{im}2+i0}^{y+\frac{im}2+i0} z dF(z)  +
\int_{y-\frac{im}2-i0}^{x -\frac{im}2-i0} z dF(z) = $$ $$
\int_{y-\frac{im}2-i0}^{y + \frac{im}2+i0} z dF(z)  -
 \int_{x-\frac{im}2-i0}^{x + \frac{im}2+i0} z dF(z) \,.
$$
By the same principle as in the proof of Proposition \Fc, we obtain
\eqn\obt{\eqalign{
  \int_{x-\frac{im}{2}-i0}^{x + \frac{im}{2}+i0} & zdF(z)  =
\frac1{4\pi i}
\int_{\bR} f''(t) dt \times \cr &\left[ \left( t + \frac{im}{2} \right) \lg{x-t + i 0 \over x
- t - im} + \left( t - \frac{im}{2} \right) \lg{x-t-i0 \over x
- t + im} \right] = \cr &
\frac{1}{2\pi i} \int_{\bR} \left(x L'(t-x) + L(t-x) \right) \,f''(t) \, dt \cr &=
\frac{1}{2\pi i}\, \xi_{l+1} + \const \cr}}
thanks to \eextr\intGtwo\ and the fact that
$\sigma'(f'(x))=\xi_{l}$ because $x$ lies on the $l$'th band.
We summarize: \eqn\otvetadj{\mathboxit{\eqalign{{\xi}_{l} - {\xi}_{l+1} & = 2\pi i \oint_{\beta_l} dS \cr
a_l \pm \frac{im}{2} &  = \oint_{{\ba}_{l}^{\pm}} dS\cr
{\CF}({\ba}, {\bm}, q) &  = - {\CE}_{q,{\bm}}({\fs})\cr}}}

\newsec{ON ANOTHER MATTER}

The results of the previous chapters can be generalized in various ways. One
can incorporate hypermultiplets in arbitrary representations. One can add a
tower of Kaluza-Klein states. One can study quiver gauge theories. We shall
only sketch the results for the fundamental matter, or for five dimensional
theories. The other cases will be treated in future publications.

\subsec{Hypermultiplets in the fundamental representation}

In this section  ${\bm}$ denotes the vector of masses: $$ {\bm} = {\rm
diag}(m_1, \ldots, m_{k}) $$
We will denote by ${\fl} =1, \ldots, k$ the flavour index.

The partition function of the theory with $k$ matter hypermultiplets in the
fundamental representation of $U(N)$ is given by
the following sum over colored partitions
\swi: \eqn\prtfm{\eqalign{ Z({\ba},
{\bm}; {\hbar}, {\Lambda})& = \sum_{\vec\bk} {\Lambda}^{(2N-k)|{\vec\bk}|}\ Z_{\vec\bk}({\ba}, {\bm};
{\hbar}, {\Lambda})\cr Z_{\vec\bk} ({\ba}, {\bm}; {\hbar}, {\Lambda}) &=
Z^{pert}({\ba}, {\bm}; {\hbar}, {\Lambda}) \prod_{l,i,\fl}
\frac{{\Gamma}\left( \frac{m_{\fl} + a_l}{\hbar} + k_{li} +1 - i \right)}{
{\Gamma}\left( \frac{m_{\fl} + a_l}{\hbar}+1 - i \right)} \
\m_{\vec\bk}^2({\ba}, {\hbar})\cr Z^{pert}({\ba}, {\bm}; {\hbar},
{\Lambda}) &= \exp\left( \sum_{l,n} \zu{a_l-a_n} + \sum_{l,{\fl}} \zu{a_l +
m_{\fl}}\right) \cr}}
As before, the partition function \prtfm\ can be written
as a sum over profiles $f=f_{\ba,\vec\bk}$ as follows
\eqn\prtfmp{\eqalign{Z({\ba}, {\bm};   {\hbar},
{\Lambda}) &= \sum_{f} Z_{f} ({\ba}, {\bm}; {\hbar},
{\Lambda}) \cr Z_{f} ({\ba}, {\bm}; {\hbar}, {\Lambda}) &= \exp \ \Biggl( -
\frac1{4} \int dx dy \ f''(x) f''(y)\, \zu{x-y} + \cr &
\qquad\qquad\qquad\qquad\qquad\qquad \frac1{2} \sum_{\fl} \int
dx \ f''(x) \,\zu{x + m_{\fl}}\Biggr)\cr}}
Observe that the action in \prtfmp\ is the old action \gnpr\
evaluated at a modified function $\tilde f$ such that
\eqn\tildef
{
\tilde f'' = f'' - \sum_\fl \delta(x+m_\fl)
}
Therefore, the variational problem for the limit shape $f$ is
essentially the variational problem solved in Section $\bf 4$,
the only modification being that the maximizer $\tilde f_\star$ is required
to have corners at the points $x=-m_\fl$.

We will solve this problem under the assumption that
$m_\fl$ are real and $|m_\fl|\gg 0$. Without loss of generality,
we can assume that the number $k$ of flavors is exactly $2N-1$
since letting some $m_\fl$ go to infinity reduces the number
of labeled.

Let $\tilde\Delta$ be the half-strip
$$
\{|\Re({\varpi})|<N, \Im({\varpi}) > 0\}\,,
$$
but now with two kinds of vertical slits. As before,
we have $N-1$ slits going from the points
$$
-N+2l+i\eta_l\,, \quad l=1,\dots, N-1
$$
down to the real axis. Additionally, we introduce
$2N-1$ slits from the points
$$
-N+l+i\tilde\eta_l\,, \quad l=1,\dots,2N-1
$$
going up to the infinity. Let $\tilde\Phi$ be the conformal
map from the upper-half plane to the domain $\tilde\Delta$
as in the following figure

\Figy{5}{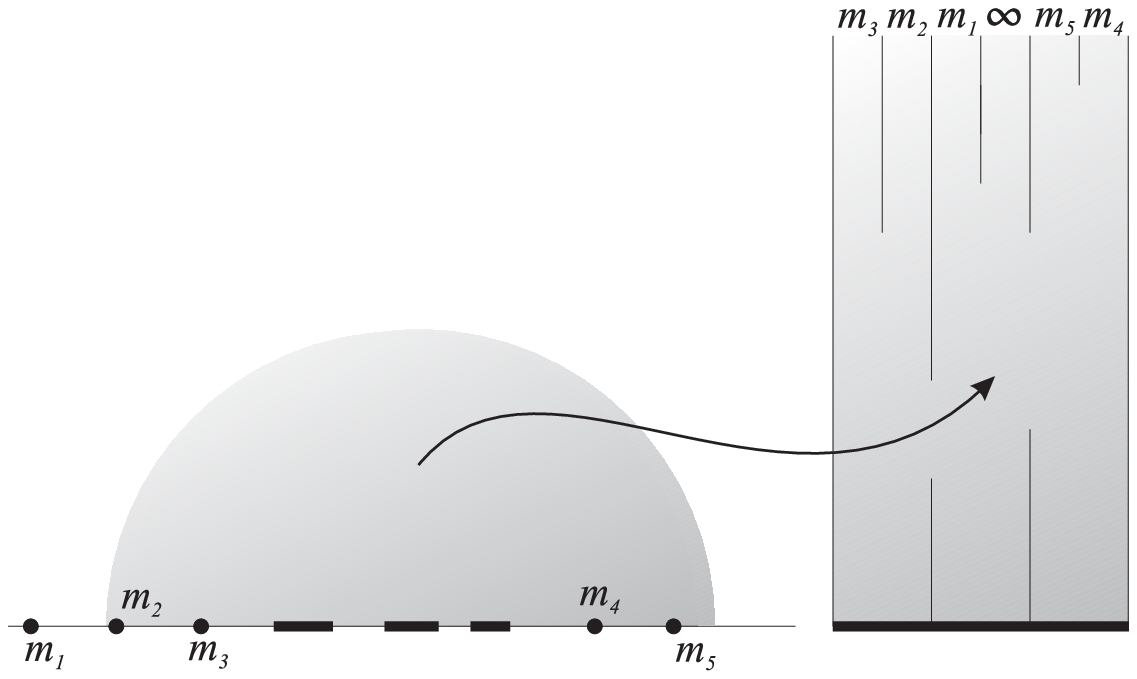}{The map $\tilde\Phi$ for $N=3$}

\noindent
The numbers $\eta_l$ are, as before, the parameters of this
map, as well as the number $\Lambda$.
The numbers $\tilde\eta_l$ are uniquely fixed by the
requirement that
$$
\Phi^{-1}(\infty) = \{m_1,\dots,m_{2N-1},\infty\} \,.
$$

Let $P(z)$ be a monic degree $N$ polynomial with all real roots.
Suppose that $\Lambda$ is sufficiently small and the $m_\fl$'s
are large enough. If
$$
m_1 < \dots < m_r \ll 0 \ll m_{r+1} < \dots < m_{2N-1}
$$
then, as before, one checks that the map
\eqn\tilPh{
\tilde\Phi(z) = \frac2{\pi i} \ln \frac{w}{\sqrt{\Lambda}} -N +r +1\,,
}
where $w$ is the smaller root of the
equation
\eqn\swmmtr
{
w+\frac{\Lambda}{w} = \frac{P(z)}{\sqrt{Q(z)}} \,, \quad Q(z)=
\prod_{\fl=1}^{2N-1} (z+ m_\fl)\,,}
is the required conformal map. Its construction is
illustrated in the following figure which is
a modification of Fig.\ 7:
\Figy{7}{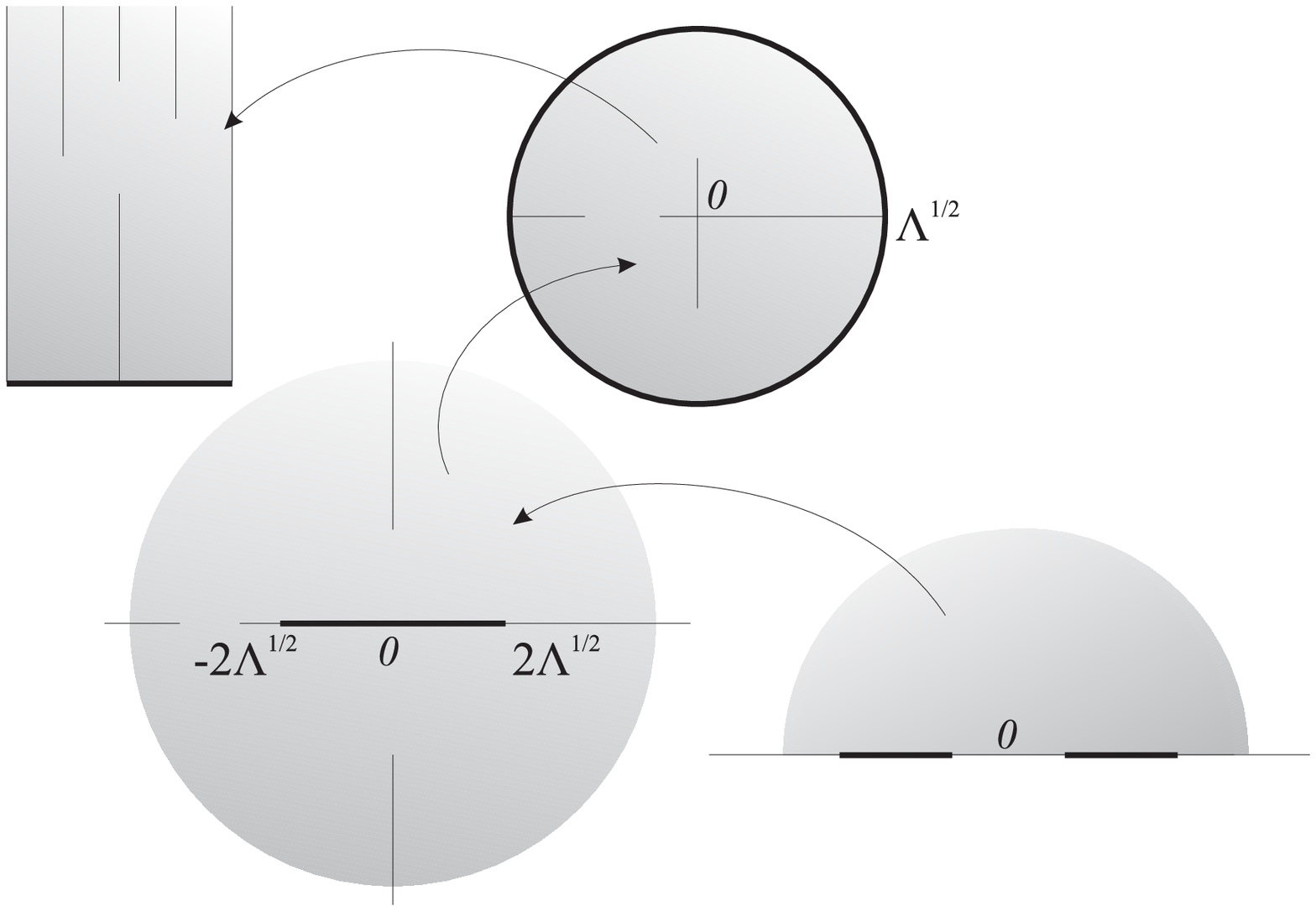}{The map $\tilde\Phi$ as a composition of maps for $N=2$}
The branch of the logarithm in
\tilPh\ is chosen so that
$$
\Im \lg{w} \to 0\,, \quad z\to \infty \,.
$$

We now claim that the maximizer $\tilde f_\star$ is given by
the same formula as before
\eqn\tfstar
{
\tilde f_\star(x)' = \Re \, \tilde\Phi(x) \,,
}
which simply means that the nontrivial part of
 the maximizer $f_\star$ is
determined by the formula
\eqn\tffstar
{
f_\star(x)'= \Re \, \tilde\Phi(x) \,, \quad |x|< \min \{|m_\fl|\}\,,
}
whereas $f_\star(x)=N|x|$ outside of this interval.

Indeed, $\tilde f_\star(x)'$ clearly has the required
jumps at the points $x=-m_\fl$. The
equation
$$
(\bX \tilde f_\star)' = - \pi \Im \Phi
$$
is checked in the exact same way as before.
In particular, the asymptotics
\eqn\AsPht
{\tilde\Phi(z) \sim \frac{1}{\pi i} \ln \frac{\Lambda}{z} +
{\rm real\ constant}\,, \quad z\to\infty
}
agrees with the fact that from \AsXp\ we have
\eqn\AsXpt
{
(\bX\tilde f_\star)'(z) = \ln \Lambda  - \ln |z| +
O\left(\frac1 z\right)\,, \quad z\to \infty\,.
}

This verifies the predictions of \sw\swsol. If $2N = k$ then
the curve will be covering elliptic curve, and again can be explicitly
constructed.

\subsec{Five dimensional theory on a circle}

Consider five dimensional pure supersymmetric gauge theory with eight
supercharges, compactified on the circle ${\bS}^1$ of circumference ${\b}$.
In addition, put the ${\Omega}$-twist in the noncompact four dimensions. The
resulting theory is a deformation of \phsym\ by ${\b}$-dependent terms.

The theory was analyzed in \nikfive\swi\ and the result of the calculation
of the partition function is the following: \eqn\prtfnfive{\eqalign{& Z
({\ba}; {\hbar}, {\b}, {\Lambda} )  = \sum_{{\vec\bk}}
({\b}{\Lambda})^{2N\vert {\vec\bk} \vert } Z_{{\vec\bk}} ({\ba}; {\hbar},
{\b}) \cr & \qquad Z_{\vec\bk} ({\ba}; {\hbar}, {\b}) = Z^{pert}
({\ba};{\hbar}, {\b}) \ \m_{\vec\bk}^2({\ba}, {\b}, {\hbar})\cr &
\qquad\qquad \m_{\vec\bk}^2({\ba}, {\b}, {\hbar}) = \prod_{(l,i) \neq (n,j)}
{{{\rm sinh}\frac{\b}{2} (a_{l} - a_{n} + {\hbar} ( k_{l,i}- k_{n,j} + j -
i))}\over{{\rm sinh}\frac{\b}{2} ( a_{l} - a_{n} + {\hbar} ( j-i))}}  \cr &
\qquad Z^{pert} ( {\ba}; {\hbar},{\b})  = {\Lambda}^{\sfrac{1-N^2}{12}} {\exp} \sum_{l, n} \zu{a_{l} -
a_{n} | {\b}} \cr}} Note that we included a power of ${\b}$ in $\Lambda$ in
\prtfnfive\ in order to have a simple four dimensional limit ${\b} \to 0$,
$\Lambda$ finite. This ``renormalization'' group relation was actually
derived in \vafaengine\nikfive.

\subsubsec{Path representation}

The partition function \prtfnfive\ is easily written as a path sum:
\eqn\prtfnfivep{Z ({\ba}; {\hbar}, {\b}, {\Lambda} ) = \sum_{f \in
{\Gamma}_{\ba}} \exp \left(- \frac1{4}\intp{x \neq y} f''(x) f''(y) \,\zu{x-y |
{\b}}\right)}

\subsubsec{Prepotential}

From \prtfnfivep\ we derive, as before, the equations on the critical path,
which, in turn, determine the prepotential. Assuming $\b>0$, we
have
\eqn\prpfive{
{\CF}_0({\ba}, {\b}, {\Lambda}) =  \frac1{4}\,{\CE}_\beta(f_{\star}) \,,
}
where $\fs$ is the maximizer and the functional $\CE_\beta$ is defined by:
\eqn\CEbeta{
{\CE}_\beta(f) = \intp{x < y} dxdy\ ( N + f'(x)) (N - f'(y)) \,\lg{\frac{2}{\b
\Lambda} {\rm sinh}\frac{{\b}| x - y |}{2}} \,.
}
Again, we modify the functional \prpfive\ by a surface energy term
to obtain the following total action functional:
\eqn\fvact{{\CS}(f) = \frac1{4}\,{\CE}_\beta(f) + \frac1{2}\int_{\bR} {\s}(f') \ dx}
We now introduce a five dimensional analogue of the transform ${\bX}f$ (see appendix ${\bf A}$ for the definition of the function $\g_0(x;\b)$):
\eqn\fdxt{
\left[{\bX}_{\b}{f}\right](x) =
\frac1{2}\intp{y \neq x} dy \,\, {\rm sgn}(y-x)\, {\g}_{0}'(|y-x|  ; \b ) \, f''(y)\,.
}
Note that the relation
\eqn\gampp
{
\gamma_0(x;\beta)'' =  \log \frac{2}{\beta\Lambda}
 \sinh \frac{\beta x}{2} \,,}
implies that
\eqn\fdxtt{\left[{\bX}_{\b}{f}\right]'(x) = -
\frac1{2}\intp{y \neq x} dy \, \log \left(\frac{2}{\beta\Lambda}
 \sinh \frac{\beta |x-y|}{2}\right)\, f''(y) \,.
}
The equations on the minimizer for \fvact\ are formulated exactly as in the Proposition \prm\ with the replacement of ${\bX}f$ by ${\bX}_{\b}f$.

The solutions to these equations can be constructed as follows.
Let $\Phi(z)$ be the conformal map from the
horizontal strip
\eqn\horstrip{
0< \Im z < \frac\pi\b
}
to the slit half-strip $\tilde\Delta$ with $N-1$ slits of the
form \slts\ and an additional
vertical semi-infinite slit
$$
{\Re}{\varpi} = 0\,,\quad {\Im}{\varpi} > {\tilde\eta}\,,
$$
as in Section ${\bf 7.1}$ above. We normalize $\Phi$ by the
requiring that it maps the two ends of the horizontal
strip to the two ends of $\tilde\Delta$. This
fixes it up to precomposing with an overall shift by a
real constant.

The exponential map
$$
z \mapsto X = e^{\b z}
$$
maps the  strip \horstrip\ to the upper half-plane,
mapping $\bR$ to $\bR_{>0}$. Therefore, we can use
the results of the previous section to conclude that
$\Phi$ has the form
$$
\Phi(z) = N + \frac{2}{\pi i}\, \log w \,,
$$
where $w$ is the smaller root of the equation
\eqn\rltd{X^{-\frac N2}P_N(X)=(\beta\Lambda)^N
\left(w+\frac1w \right)\,, \quad X = e^{\b z}\,.
}
Here $P_N$ is a monic polynomial of degree $N$ and
$\log w$ is normalized, as usual, by the
requirement that $\Im \log w \to 0$ as $z\to +\infty$
along the real axis.

In the form \rltd, the parameter $\tilde\eta$ of
the map $\Phi$ becomes a function of $\Lambda$ and
other parameters.
The equation \rltd\ defines the spectral curve of
the relativistic Toda chain, as predicted in \nikfive.

Applying the Schwarz reflection principle to any
vertical part of the boundary of $\tilde\Delta$ and
taking the resulting function modulo 1, we obtain
a $2N$-fold map from the horizontal cylinder of
circumference $2\pi/\beta$ with $N$ cuts along the real axis
to the unit half-cylinder, as shown in the
following figure
\Figx{8}{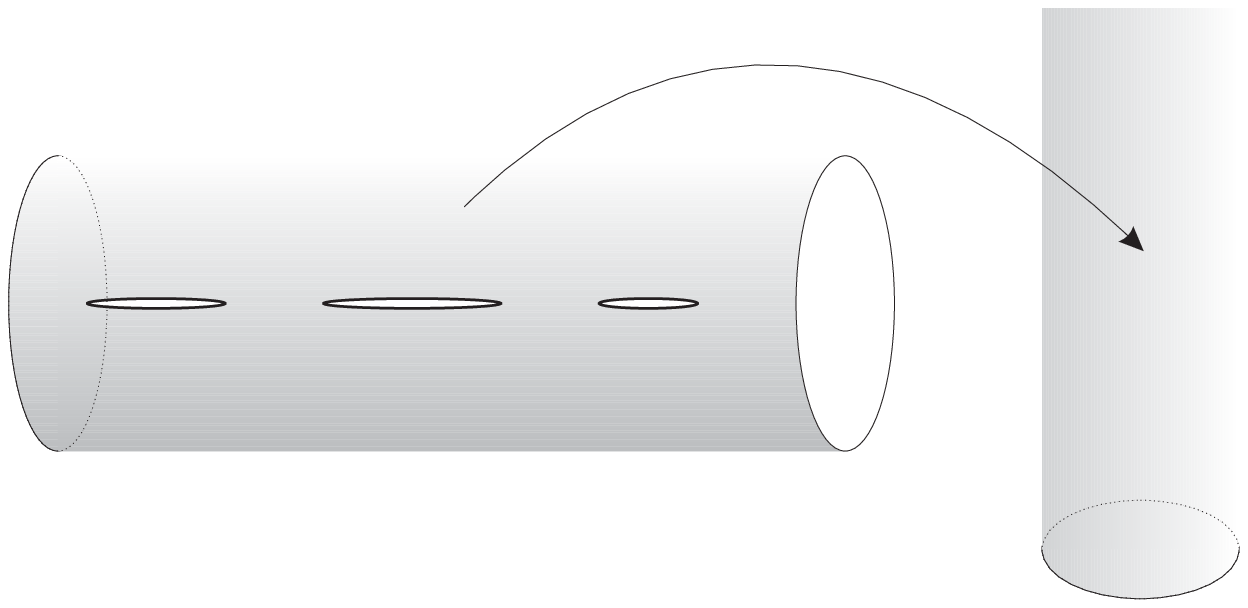}{The map $\Phi\mod 1$}
Here each cut covers 2-to-1 the finite boundary of the
half-cylinder, while both ends of the cylinder cover
$N$-to-1 each the infinite end of the half-cylinder.

As before, it follows that $\Phi'(z)$ extends to
a $2\pi i/\beta$-periodic function in the
entire complex plane with $N$ cuts along the
real axis, periodically repeated. The values
of $\Phi'(z)$ on the two sides of the cut are
real and opposite in sign. We have
\eqn\asmpbephi
{
\Phi(x+i0)= \pm \left(i \frac{N\beta x}{\pi} + N\right)
+ \frac{2N}{\pi i} \log \beta \Lambda\,,
\quad x\to\pm \infty\,,
}
whence
\eqn\asmpbephip
{
\Phi'(x+i0) = \mp \frac{N\beta}{\pi i}\,,
\quad x\to\pm \infty\,,
}
It follows that the function $\Phi'(z)$ satisfies the
following integral equation
\eqn\inteth{
\Phi'(z) = \frac{\beta}{2\pi i} \int_{\bR} \,\Re \Phi'(x)\,
\coth\left(\beta\,\frac{(x-z)}2\right) \, dx  \,.
}
Moreover, by comparing the asymptotics, we can integrate \inteth\ to
\eqn\intethh{
\Phi(z) = N+ \frac{i}{\pi } \int_{\bR} \,\Re \Phi'(x)\,
 \log \left(\frac{2}{\beta\Lambda}
 \sinh \frac{\beta (z-x)}{2} \right)\, dx  \,.
}

We claim that the maximizer $\fs$ is given in terms of
the map $\Phi$ by the usual formula
$$
\fs'(x) = \Re \Phi(x+i0) \,.
$$
Indeed, from \intethh\ and \fdxtt\ we conclude that
\eqn\intXbeta{
\left(\bX_\beta \fs\right)'(x) = - \frac{\pi}2 \Im \Phi(x) \,.
}
The difference with the previously considered cases lies
in the fact that to reconstruct the values of $\xi_l$'s (and $a_l$'s) we have to
integrate the multi-valued differential
$$
dS_5 =\frac1{\b}\, \lg{X} \frac{dw}{w}
$$
on the curve \rltd, as indeed proposed in \nikfive. Also note
that domain where the map $\Phi$ is defined is different.
This leads to the new form of the
Seiberg-Witten differential.

\subsubsec{Free field representation} Again we start with $N=1$ case. Let $Q =
e^{\b\hbar}$. In this case we are to sum over all partitions:
\eqn\fvsm{\eqalign{Z
(a, {\hbar}, {\b}, {\Lambda}) = \sum_{\bk}  (-1)^{|{\bk}|}
({\b}\Lambda)^{2|{\bk}|} & \ \prod_{{\squarel} \in {\bk}} \left(\frac{\b}{2{\rm
sinh}\left(\frac{{\b}{\hbar}h({\square})}{2}\right)}\right)^{2} = \cr &
{\exp}
\sum_{n=1}^{\infty} \frac{({\b}{\Lambda})^{2n}}{{4n \  {\rm sinh}^2 \left(
\frac{{\b}{\hbar}n}{2} \right)} }\cr}} The free field representation of the sum
\fvsm\ is most simply done using the vertex operators:
\eqn\vrtx{{\Gamma}_{\pm}( {\b}, {\hbar} )  = {\exp} \sum_{\pm n > 0}
\frac1{n(1-Q^n)} {\CJ}_{n}} Then \eqn\fvsmi{Z (a, {\hbar}, {\b}, {\Lambda})
= \langle 0 \vert {\Gamma}_{+}( {\b}, {\hbar} ) ({\b}{\Lambda})^{2L_0}
{\Gamma}_{-}( {\b}, {\hbar} ) \vert 0 \rangle} and, for $N>1$ we would get:
\eqn\fvsmii{Z^{D} ( {\xi} ;p; {\b}, {\hbar}, {\Lambda} ) =  \langle p \vert
{\Gamma}_{+}( {\b}, {\hbar} ) ({\b}{\Lambda})^{2L_0} e^{{\bH}_{\xi}}
{\Gamma}_{-}( {\b}, {\hbar} ) \vert p \rangle  }

\newsec{CONCLUSIONS AND FUTURE DIRECTIONS}

\subsec{Future directions and related work}

In this paper we extensively analyzed the partition functions of various
${\CN}=2$ supersymmetric gauge theories, subject to the
${\Omega}$-background of ${\CN}=2$ supergravity. In all cases the partition
function was identified with the statistical sum of an ensemble of random
partitions with various Boltzmann weights. The grand canonical ensemble
corresponds to the fixed theta angle in the gauge theory. In the
thermodynamic limit the statistical sums are dominated by a saddle point, a
master partition, from which one can extract easily various important
characteristics of the low-energy effective gauge theory, such as the
prepotential of the effective action, which is identified with the free
energy per unit volume in the dual statistical model, provided we relate the volume
to the parameters ${\e}_1, {\e}_2$ of the
${\Omega}$-background, via $$V = \frac1{{\e}_1{\e}_2} \ .$$
The particle number is clearly  the instanton charge. So the gauge theory partition
function corresponds to the grand canonical ensemble, with $\lg{\Lambda}$
being the chemical potential.
The average particle number per unit volume, the density, is easy to
determine:
\eqn\isntdens{\eqalign{{\rho} = \Biggl\langle \frac{k}{V} \Biggr\rangle & =
{{\e}_1{\e}_2 \over{2N \ Z^{inst} ({\ba}, {\e}_1,{\e}_2 , {\Lambda})}} \frac{\p}{\p\lg{\Lambda}} Z^{inst}({\ba}, {\e}_1,  {\e}_2, {\Lambda})
 \longrightarrow_{\kern -.2in \scriptscriptstyle{V \to \infty}} \cr & {1\over 2N}
{{\p}\over{{\p}\lg{\Lambda}}} {\CF}_{0}^{inst}({\ba}, {\Lambda}) = u_2 - \sum_l \frac{a_l^2}{2} \cr}}
(recall \tdcr).
In the region ${\CU}_{\infty}$ the right hand side of \isntdens\ is clearly
very small, it is of the order:
$$
{\rho} \sim {{\Lambda}^{2N} \over a^{2(N-1)}} \sim M_{W}^{2} e^{-{8\pi^2 \over g_{eff}^2(M_{W})}}
$$
where $M_{W}$ is the typical $W$-boson mass, and $g_{eff}(M_{W})$ is the
effective coupling at this scale. The dependence of the instanton density on
the coupling is the typical one in the dilute gas approximation (cf.
\coleman\polyakovi). In the semi-classical region there are very few
instantons per unit volume.

Of course, a cautious reader may wonder
about the dimensionality of the instanton density (which normally should be
(mass)$^4$). The point is that the noncommutative regularization which we
used in the calculations introduces a scale $\sim \sqrt{\Theta}$, which is
responsible for the dimensional transmutation we see in \isntdens. Indeed,
the physical volume, occupied by  $k$ instantons, sitting on top of each
other, is roughly $k {\Theta}^2$. Such instanton ``foam''\branek, or perhaps,
liquid, may well be related to more complicated phenomenological pictures, say instanton liquid
\shuryak. Moreover,
as we go deeper into the moduli space of vacua, the diluteness of the instanton gas also must cease to hold.
However, the
analytic properties of the partition function \prtnf\ are powerful enough to
uniquely fix it by the instanton gas expansion in ${\CU}_{\infty}$.

Note, that one of the interpretations of our result is the existence of the
master field in the ${\CN}=2$ gauge theories, which, however, is working
only for special (chiral) observables, unlike the master field of the large
$N$ gauge theories \gopakumargross. The latter, however, is much more elusive.

In our story, the master field is constructed as follows. First of all, the
labeled
partitions ${\vec\bk}$ we sum over are in one-to-one correspondence with
the (noncommutative) gauge fields, describing certain arrangements of the
nearly point-like instantons, sitting nearly on top of each other. The
``master'' partition with the profile $f_{\star}$ corresponds to
statistically most favorable configuration.

We extensively exploited in this paper the fact that ${\Omega}$-background
acts as a box, similarly, in some
respects, to the AdS space for the supergravity theory. One may wonder, why
couldn't we use the more traditional ways of regularizing
infrared divergences of the gauge theory. In fact, the number of options is
rather limited. One may study gauge theories on the compact four
manifolds. To preserve supersymmetry one needs to turn on certain
non-minimal couplings, which effectively twist the gauge theory.
However, in these approaches one cannot learn directly the properties of the
gauge theory in the infinite volume, as all vacua are averaged over, and
either most of them do not contribute to the correlation functions of the
chiral operators (which is the case for the simple type manifolds), or
(for the manifolds with $b_{2}^{+} \leq 1$) the contribution of the various
vacua is related to the prepotential in a complicated fashion
\moorewitten\issues. It is out of such attempts to regulate the ${\CN}=2$ theory
that one naturally arrives at the concept of the ${\Omega}$-background.

The topic which we completely neglected in our paper is the application of
our formalism to the theories with ${\CN}=1$ supersymmetry. Recently, there
was a lot of excitement related to the exact calculations of the effective
{\sl super}potentials of the theories, obtained from ${\CN}=2$ theories by
the superpotential deformations
\dv\kawai\cdws. It should be straightforward to apply our techniques in
these setups as well. One encouraging feature of our formalism in the case
of pure ${\CN}=2$ and softly broken ${\CN}=4$ theories, is the striking
similarity of the expression we have for the partition functions and the
matrix models calculating the effective superpotentials of the ${\CN}=1$
gauge theories with an additional adjoint chiral multiplet, or with three adjoint chiral
multiplets, with specific superpotential.

The similarity is the measure --- in the first case it is the regularized
Vandermonde determinant of the infinite matrix ${\hat\Phi}$ with the eigenvalues $a_l +
{\hbar}(k_{li} - i)$. In the second case it is the ratio of the
determinants: $$
{{\Det}^{\prime}(ad{\hat\Phi})\over{{\Det}(ad{\hat\Phi}+ {\bm})}}$$
The fact that the eigenvalues concentrate (for small ${\hbar}$) around
$a_l$'s also has a matrix model counterpart. Namely, to get the theory with
$U(1)^N$ gauge symmetry one adds to the $U(N)$ theory a tree-level
superpotential with extrema near $a_1, \ldots, a_N$. In the limit of
vanishing superpotential one is left essentially with the Haar measure on
matrices, together with the prescription to distribute eigenvalues near the
extrema of the phantom of the superpotential \dv.

The differences are also easy to see: on our case the matrix is infinite,
but the eigenvalues are discrete and are summed over. In the Dijkgraaf-Vafa case, the matrix is
finite, but the eigenvalues are continuous and integrated over. We have
${\hbar} \to 0$, they have ${\hat N} \to \infty$. The issue is tantalizing
and is under investigation \samson.
\vfill\eject
\hfill{\it 0h, East is East, and West is West, and never the twain shall
meet,} $\ldots$

\medskip\ndt
Another issue which we didn't touch much upon is the string theory dual of
our calculation. There are several points one may want to stress.

First of all, the summation over partitions that we encountered is very similar
to the summation over partitions one encounters in
the two dimensional $U({\hat N})$ Yang-Mills theory, in the large ${\hat N}$ limit. In
particular, when working on a two-sphere, one finds a master partition
\phtran, and moreover, for the values of the 't Hooft coupling constant
the profile of the master partition has the facets, just like ours.

Of course, technically speaking, the two dimensional YM theory corresponds to
the $U(1)$ theory in our case, and the 't Hooft coupling there corresponds
to the higher Casimir coupling ${\tau}_3$ in our game. But the phenomena
have similar origin. Now, the two dimensional Yang-Mills theory has a dual closed string
representation
\tdym, and so does our four dimensional partition function. And
again, just like $1/{\hat N}$ played the r\^ole of the string coupling
constant, ${\hbar}$  plays this r\^ole in the four dimensional
story.

In the $U(1)$ case all this is more or less well-known \op\lmn\ by now.
However, the full string dual of the $U(N)$ theory is yet to be discovered
(see \lmn\ for the discussion, see also \iqbal\agmav).

We should add that from the point of view of Gromov-Witten theorists
the dual type A topological string theory, whatever it is, must be rather simple. Most of them
should have one dimensional target spaces. The rough correspondence between
the gauge theory and the dual GW theory states that the $U(N)$ theory with $g$
adjoint hypermultiplets (which is not asymptotically free/conformal for $g > 1$)
is dual to the GW theory of $N$ copies of the Riemann surface of genus $g$.
The  words ``$N$ copies'' still do not quite have a formal meaning.

In the large ${\hat N}$ description of the two dimensional Yang-Mills theory
a prominent r\^ole was played by the formalism of free fermions \dougcft.
Of course, the chiral fermions of \dougcft\ are ours ${\Psi}, {\tilde\Psi}$
(somehow four dimensional gauge theories do not see the anti-chiral sector).

One of the exciting problems is to understand better
the nature of these chiral fields in the string realizations of ${\CN}=2$ gauge
theories. The conjecture of \swi\
was that they arise as the modes of the chiral two form propagating on the
worldvolume of the NS5/M5 brane, trapped by the $\Omega$-background. In the
bosonized form, these can also be mapped to the truncated version of the
Kodaira-Spencer field, propagating along the Seiberg-Witten curve. We are
planning to investigate these issues in the future.

Yet another exciting area of research, revolving around our partition
functions is their two dimensional (anyonic) interpretation in the ${\n}\neq 1$ case.
The connections to the Jack polynomials, which are the eigenfunctions of the
Sutherland many-body Hamiltonians with ${\n}({\n}-1)$ coupling constant,
suggest strongly some relations to the physics of the quantum Hall effect,
and also to the theory of analytic maps \zabrodin. The latter also seem to
be responsible for the higher Casimir deformation \hrcsm.

Seiberg-Witten curve itself seems to arise as some quasiclassical object in
the theory of some many-body (${\hat N} \to \infty$) system. Indeed, the
space $z,w$ where it is embedded, can be viewed as a one-particle phase
space. The particles are basically free for ${\n}=1$, and have some phantom
interaction for ${\n}\neq 0,1$, which only reveals itself in the generalized
Pauli exclusion principle (Haldane exclusion principle). They fill some sort
of Fermi sea, bounded by the Seiberg-Witten curve. Perhaps, the relation to
the random dimers \dimer\ will also prove useful in understanding these
issues. Another possibly useful feature of our energy functional is its
seimple relation to the two dimensional local theory of a free boson.
Indeed, the bi-local part of \enrgi\ is nothing but the induced boundary
action, with ${\rho}(x)$ being the boundary condition in the theory of the
free boson, living on the upper half-plane. Of course, when the subleading
in ${\hbar}$ corrections are taken into account the theory will cease to be
free. It is also interesting to point out that our extremizing configuration
${\fs}(x)$
corresponds to some sort of multiple D-brane boundary state, with $N$
D-branes located at $x \sim a_l$, $l = 1, \ldots, N$.

It is of course very tempting to develop these pictures further, connect
instanton gas/liquid/cristall to the (Luttinger?) liquid of the dual anyons,
and learn more about the properties of the fivebranes from all this.
Perhaps, the results of \anyons\girvin\ will prove useful along this route.

Finally, the theory with adjoint matter should be closely related to
conformal field theory on an elliptic curve. We have uncovered some of the
relation in the partition function being interpreted as a trace in the
representation of a current algebra. We should also note that the elliptic
Calogero-Moser system, whose spectral curve, as we showed above, encodes the
quasiclassical/thermodynamic limit of the partition function, allows a
nonstationary generalization, related to KZB equations, which, conjecturally,
governs our full partition function \knzam\denis\aslbernard\giovanni\dima\olshanetsky.

\subsec{Summary of the results}

In this paper we have advanced in the study of the vacuum
structure of gauge theories. We considered ${\CN}=2$ supersymmetric gauge
theories in four dimensions, and have subject them to the so-called
${\Omega}$-background. In this background one can calculate exactly the
partition function, in any instanton sector. The result has the form of a
statistical sum of the ensemble of random partitions.
The limit where the $\Omega$-background approaches flat space, is the most
interesting for the applications, as there one is supposed to learn about
the properties of the supersymmetric gauge theory in flat spacetime.
In terms of the random partitions this is a thermodynamic, or quasiclassical
limit. The flat space limit of the free energy coincides with the
prepotential of the low-energy effective theory.

We have evaluated the sum over instantons, by applying the saddle point
method, and thus have succeeded (to our knowledge, for the first time in the literature)
in producing the all-instanton direct calculation of the prepotential.

In all cases considered: pure gauge theory, theory with the matter
hypermultiplets, five dimensional theory compactified on a circle -- the
saddle point corresponds to some {\it master partition}, which is the
analogue of the {\it eigenvalue distribution} in the theory of
random matrices. And in all cases one can encode the solution in some family
of algebraic curves, endowed with a meromorphic (sometimes multivalued)
differential, whose periods contain the information about the prepotential.

We have also found interesting representation of the full partition function
(which, in addition to the prepotential, contains also certain higher
gravitational couplings ${\CF}_g$ of the gauge theory) as a partition
function of the theory of chiral fermions/bosons on a sphere (for the pure
gauge theory), on a torus (for the theory with adjoint matter), or some
$q$-analogue thereof (for the five dimensional theory).

We have also uncovered numerous puzzling relations between
various seemingly unrelated topics which leave a lot of work for the future.

\ndt{\bf Disclaimer.} Opinions presented in this paper do not necessarily
reflect author's point of view. There are about $\sim 2\pi i \kern -.051in \slash\kern .1in$  misprints in this
paper. We are however confident that the most of them cancel each other.

\appendix{A}{The function $\zu{x}$}

\subsubsec{Free case: ${\e}_2 = - {\e}_1 = {\hbar}$}
The function $\zu{x}$ is characterized by the following properties:
\item{1.} Asymptotic expansion for ${\hbar} \to 0$:
\eqn\asmt{\zu{x} = \sum_{g=0}^{\infty} {\hbar}^{2g-2} {\g}_{g}(x)}
\item{2.} Finite-difference equation:
\eqn\fntdf{\zu{x+{\hbar}}+\zu{x-{\hbar}}-2\zu{x} = \lg{x\over\Lambda}} The
conditions \asmt\fntdf\ specify $\zu{x}$ uniquely up to a linear function in
$x$. All the terms ${\g}_{g}(x)$, $g > 0$ are uniquely determined:
\eqn\gms{\eqalign{& {\g}_0 (x) = \frac1{2} x^2 \lg{x\over\Lambda} -
\frac{3}{4} x^2\cr & {\g}_1 (x) = - \frac1{12} \lg{x\over\Lambda} \cr & {\g}_2 (x) = -
\frac1{240} \frac1{x^2} \cr & \quad \vdots \quad \cr & {\g}_{g} (x) =
\frac{B_{2g}}{2g ( 2g - 2)} \frac1{x^{2g-2}} , \ g>1\cr}} where $B_n$'s are
the usual Bernoulli numbers: $$ \frac{t}{e^{t}-1} = \sum_{n=0}^{\infty}
\frac{B_{n}}{n!} t^{n} $$ The function $\zu{x}$ is closely related to the
gamma function: \eqn\gm{\zu{x+\frac{\hbar}{2}} - \zu{x-\frac{\hbar}{2}} =
\lg{\frac{1}{\sqrt{2\pi}} \ {\hbar}^{\frac{x}{\hbar}}\
{\Gamma}\left(\frac{1}{2} + \frac{x}{\hbar}\right)}}

Another definition of the function $\zu{x}$ is through the
zeta-regularization: \eqn\zdf{\zu{x} =
\frac{d}{ds}\Biggl\vert_{s=0}
\frac{{\Lambda}^s}{{\Gamma}(s)} \int_{0}^{\infty} \frac{dt}{t} \ t^s \ \frac{e^{-t
x}}{(e^{\hbar t}-1)(e^{-\hbar t}-1)} }

\remark{Note, that
$$ \zu{0} = - \frac1{12} $$}
The function $\zu{x}$ for $\Lambda = 1$ arises as the free energy of the $c=1$ string with
${\hbar}$ being the string coupling, and $x$ the cosmological constant. It
is also related to the free energy of the topological type A string on the
conifold (see also below). The adepts of the applications of matrix models in susy gauge theories
\dv\ praise yet another property of $\zu{x}$:
\eqn\vlun{\lg{{\rm Vol}U(N)} = {\g}_{1}(N;1)}

\subsubsec{Anyon case: general ${\e}_1 , {\e}_2$}
\def\zz#1{{{\g}_{{\e}_1, {\e}_2}\left( #1 ; {\Lambda}\right)}}
\eqn\zzdf{\zz{x} = 
\frac{d}{ds}\Biggl\vert_{s=0} \frac{\Lambda^s}{{\Gamma}(s)} \int_{0}^{\infty}
\frac{dt}{t} \ t^s \ \frac{e^{-t x}}{(e^{{\e}_1 t}-1)(e^{{\e}_2 t}-1)}}
In the case ${\e}_2 = - {\e}_1 = {\hbar}$ this reduces to $\zu{x}$:
$$
\zu{x} = {\g}_{-{\hbar}, {\hbar}} \left( x ; {\Lambda} \right)
$$
\item{}
The {\sl main difference equation}: \eqn\mndfr{\eqalign{ \zz{x} + \zz{x -
{\e}_1 - {\e}_2} - \zz{x- {\e}_1} - & \zz{x-{\e}_2} = \cr & \lg{\Lambda\over
x}\cr}} {\sl Reflection:} \eqn\rflct{{\g}_{-{\e}_1, {\e}_2}(x;{\Lambda}) =
-\zz{x - {\e}_1} - \frac{2x+{\e}_1}{2{\e}_2} \lg{\Lambda}} For ${\n} \in
{\bf Q}$ the function $\zz{x}$ can be related to $\zu{x}$. Suppose
\eqn\rn{{\e}_1 = - \frac{\hbar}{p}, \ {\e}_2 =  \frac{\hbar}{q}, \qquad p,q
\in {\bN} .} Then \eqn\urov{\zz{x} = \sum_{i=0}^{p-1}\sum_{j=0}^{q-1} \zu{ x
+\frac{\hbar}{pq} \left( pj-qi \right)}}

\subsubsec{Trigonometric analogue of $\zz{x}$}

The natural generalization of \zzdf\ is the function: \eqn\zzzdf{\zz{x |
{\b}} = \frac1{2{\e}_1{\e}_2} \left( -\frac{\b}{6} \left( x + \half ({\e}_1
+ {\e}_2)\right)^3 + x^2 \lg{\b\Lambda}\right)+ \sum_{n=1}^{\infty}
\frac1{n} \frac{e^{-{\b} n x}}{(e^{{\b}n {\e}_1}-1)(e^{{\b}n {\e}_2}-1)}  }

which obeys:
\item{} The {\sl main $Q$-difference equation}:
\eqn\mnqdf{\eqalign{ \zz{x | {\b}} + \zz{x -{\e}_1 - {\e}_2 | {\b}} & -
\zz{x-{\e}_1 | {\b}} - \zz{x-{\e}_2 | {\b}} = \cr & - \lg{\frac{2}{\b
\Lambda}{\rm sinh}\frac{{\b}x}{2}} \cr}} We shall only use in this paper the
special case ${\e}_2 = - {\e}_1 =  {\hbar}$, where we get the function
 \eqn\fdzfn{\eqalign{\zu{x ; {\b}} = &\frac{\b x^3}{12\hbar^2} - \frac{x^2}{2{\hbar}^2}
{\lg{\b\Lambda}} - \frac{\b x}{24} + \sum_{n=1}^{\infty} \frac1{n}
\frac{e^{-{\b} n x}}{(e^{-{\b}n {\hbar}}-1)(e^{{\b}n {\hbar}}-1)} \cr & =
\sum_{g=0}^{\infty} {\g}_{g} ( x ; {\b}) {\hbar}^{2g-2}\cr & {\g}_{0}(x;
{\b}) = -\frac{x^2}{2} {\lg{{\b}\Lambda}} + {\b}\frac{x^3}{12} -
\frac1{{\b}^2} \ {\rm Li}_{3}\left(e^{-{\b}x}\right) \cr & {\g}_{1}(x; {\b})
= - \frac1{12} \lg{2 {\rm sinh}\frac{\b x}{2} } \cr & {\g}_{g}(x; {\b}) =
\frac{B_{2g} {\b}^{2g-2}}{2g(2g-2)}\ {\rm Li}_{3-2g} \left( e^{-\b x}
\right) \cr }} Note that up to the terms of instanton degree zero the
function $\zu{x | {\b}}$ coincides with the all-genus free energy of the
type A topological string on the resolved conifold, with ${\b}x$ being the
K\"ahler class of the ${\bP}^1$, and ${\b\hbar}$ the string coupling
\vafaavg. Another Gromov-Witten interpretation is via the local ${\bF}_{1}$,
with the K\"ahler class of the base ${\bP}^1$ being $\lg{\b\Lambda}$
(considered to be big), and the fiber ${\bP}^1$ with the K\"ahler class
${\b}x$ \vafaengine.

\appendix{B}{Partitions, charges, colours}

\item{}{\it Partition}:  a nonincreasing sequence of
nonnegative integers, stabilizing at some point at zero: $$ {\bk}: k_1 \geq
k_2 \geq \ldots \geq k_{n} > 0 = k_{n+1}= k_{n+2} = \ldots $$
\item{}$n \equiv {\ell}({\bk})$ is called the {\it length} of ${\bk}$,

\item{}$\vert {\bk} \vert = \sum_i k_i$,
is called the {\it size}  of the partition, $k_i$'s are called the {\it
parts} of the partition.
\Figy{4}{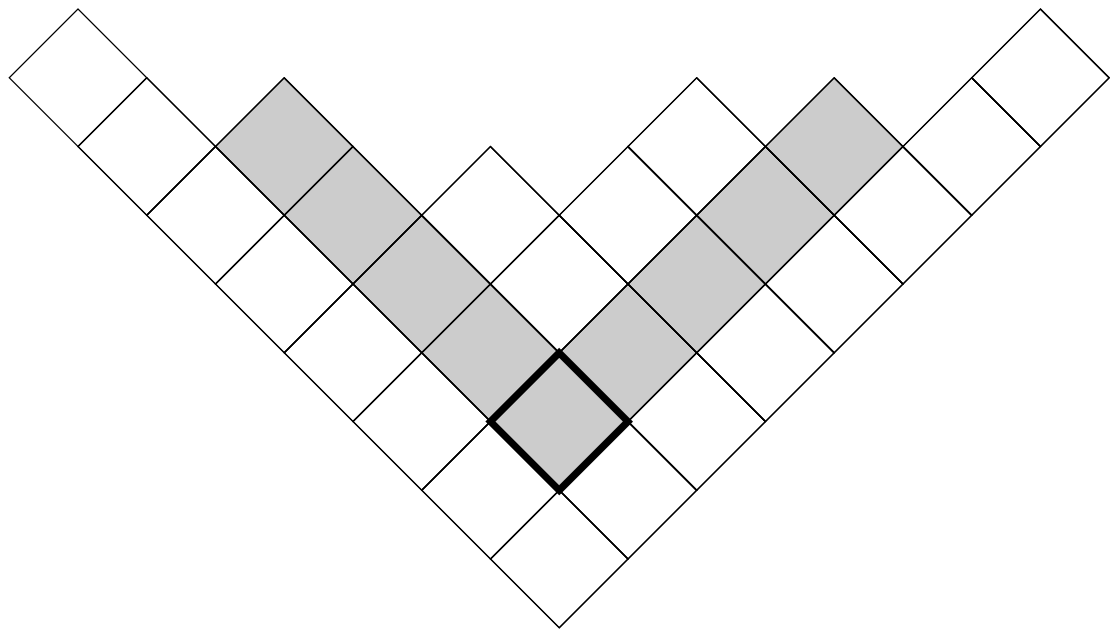}{Box,  hook, ....}
The parts of a partition are labeled by $i,j = 1,
2, \ldots , $
\item{}For the partition ${\bk}$: $${\square} = (i,j) \in {\bk} \Longleftrightarrow \ 1 \leq i, \quad 1
\leq j \leq k_i $$
\item{}{\it Dual partition:} ${\tilde\bk}$,
$$ (i,j) \in {\tilde\bk} \Longleftrightarrow (j,i) \in {\bk} $$ i.e.
${\tilde k}_i  = \# \{ j \vert i \leq k_j \}$
\item{}{\it Hook-length} $h_{i,j} = h ({\square})$ of the $(i,j)$ box in the
Young diagram of the partition ${\bk}$: $$ h_{i,j}  = {\tilde k}_j + k_i - i
- j +1 $$
\item{}{\it Coloured partition:} ${\vec \bk}$, the $N$-tuple of partitions:
$${\vec\bk} = ({\bk}_1, \ldots, {\bk}_{N}), $$ individual partitions are
denoted as: $${\bk}_l = (k_{l,1} \geq k_{l,2} \geq \ldots \geq k_{l,n_l} >
k_{l, n_{l} +1} = 0 = \ldots ),$$ $$ |{\vec\bk}| =  \sum_{l,i}k_{li} $$
\item{}{\it Charged partition:} $(p;{\bk})$,  the set of nonincreasing integers
${\k}_i = k_i +p$, where $${\bk} = ( k_1 \geq k_2 \geq \ldots)$$is a
partition, and $p \in \bZ$. The limit ${\k}_{\infty} \equiv p$ is called the
{\it charge}.
\item{}{\it Blending of coloured partitions:} given a vector ${\vec p} =
(p_1, \ldots, p_N)$, with $$ \sum_l p_l = 0 $$ and an $N$-tuple of
partitions ${\vec \bk}$, we define the {\it blended} partition $K$, as
follows: \eqn\blnd{\{ K_i - i \ \vert \ i \in {\bN} \} = \{  N ( k_{li} - i
+ p_l) + l-1 \ \vert \ l = 1, \ldots, N, \ i \in {\bN} \ \}}

\subsubsec{Power-sums} To the charged partition $(p;{\bk})$  it is useful to
associate the following shifted-symmetric generating function,  analytic in
$t$, with a single pole at $t=0$, defined for ${\Re}t > 0$ by the series:
\eqn\chnr{{\bp}_{p; \bk}(t) = \sum_{i=1}^{\infty} e^{t(p+ k_i - i+{\half})}}
The expansion of ${\bp}_{p;\bk}(t)$ near $t=0$ contains information about
the charge, the size of $\bk$, etc: \eqn\expn{{\bp}_{p;\bk}(t) = \frac{1}{t}
+ p + t \left( \frac{p^2}{2} + \vert \bk \vert - \frac{1}{24} \right) +
\ldots} \mdp Given $N$ partitions ${\bk}_l$, $l=1, \ldots, N$, and the
charges $p_l \in {\bZ}$, we associate to them the generating function:
\eqn\nchnr{{\bp}_{\vec p; \vec\bk}(t) = \sum_{l,i} e^{t \left( N ( p_l +
k_{l,i} - i ) + l - {\half} \right)} = \sum_{l} \ e^{N t {\r}_l} \
{\bp}_{p_l, {\bk}_l} (N t) } which corresponds to the blended partition
${\bf K}$ of charge: \eqn\chrgss{p = \sum_l p_l = \sum_l {\tilde p}_l} and
size: \eqn\szzz{\vert {\bf K} \vert = \sum_I K_I = N  \sum_l \left( {\half}
{\tilde p}_l^2 + \vert {\bk}_l \vert \right)
-
{\half} p^2  - \frac{N^2-1}{24}} where $$ {\tilde p}_l = p_l + {\r}_l $$
\mdp The function $f_{p;\bk}(x)$ is related to ${\bp}_{p;\bk}(t)$ by the
integral transformation: \eqn\intrr{f_{p;\bk}^{''}(x) = - {1\over {\pi i}}
\int_{\bR} \ {\rm d}t \ e^{-i t x} \ \sin{\frac{t\hbar}{2}} \
{\bp}_{p;\bk}(i t{\hbar}) } The  resolvent \rslvnt\ is
related to ${\bp}_{p;\bk}(t)$ by another integral transformation:
\eqn\intrri{R(z | {\e}_1, {\e}_2 ) = - {1\over {\pi i}} \int_{0}^{\infty} \ {\rm
d}t \ e^{- t z} \ \sinh{\frac{t\hbar}{2}} \ {\bp}_{p;\bk}( t{\hbar}) }

\appendix{C}{Theta-function}

For completeness we list here the relevant formulae for the odd theta
function, which we use in \EcmLax.

\eqn\thfnc{\eqalign{& {\t}_{11}({\varpi};{\tau}) = \sum_{n \in {\bZ}}
e^{{\pi}i {\tau} ( n + {\half})^2 + 2{\pi} i ( {\varpi} + {\half} ) ( n +
{\half} ) } \cr
& {\t}_{11}({\varpi} +1; {\tau}) = -  {\t}_{11}({\varpi}; {\tau}) \cr
& {\t}_{11}({\varpi} +{\tau}; {\tau}) = - e^{- {\pi} i ( 2 {\varpi} +  {\tau})}  {\t}_{11}({\varpi}; {\tau}) \cr
}}
From these formulae one easily concludes that the Lax
operator $L({\varpi})$ \EcmLax\ is the meromorphic Higgs operator in the
rank $N$ vector bundle over the elliptic curve $E_{\tau}$, twisted by the
one dimensional affine bundle, which makes its spectrum to live in the
affine bundle as well.

In \trcc\ we use Dedekind eta-function:
\eqn\ddknd{{\eta}(q) = q^{1\over 24} \prod_{n=1}^{\infty} ( 1- q^n), \qquad
q = e^{2\pi i {\tau}}}
\footatend\vfill\supereject\immediate\closeout\rfile\writestoppt
\baselineskip=14pt\centerline{{\bf References}}\bigskip{\frenchspacing%
\parindent=20pt\escapechar=` \input refs.tmp\vfill\eject}\nonfrenchspacing

\bye